# Assessment of Human Behavior in Virtual Reality by Eye Tracking

**Dissertation**

der Mathematisch-Naturwissenschaftlichen Fakultät

der Eberhard Karls Universität Tübingen

zur Erlangung des Grades eines

Doktors der Naturwissenschaften

(Dr. rer. nat.)

vorgelegt von

M. Sc. Hong GAO

aus Tangshan, China

Tübingen

2022

*"The Journey Is the Reward"*
"过程就是收获"

– Steve Jobs
– 史蒂夫·乔布斯

To my family and friends

# Acknowledgments

Many people I have to thank during my doctoral studies, from the first day I joined the Human-Computer Interaction Research Group to the last day I wrote this dissertation.

First of all, I would like to express my deepest gratitude to my supervisor Prof. Dr. Enkelejda Kasneci, who supervised me from various aspects throughout my doctoral studies. I am also very thankful to the late Prof. Dr. Rosenstiel for his supervision. I would also like to thank Prof. Dr. Richard Göllner for his consistent support during the BigFish and VirATec projects. I am also grateful to Jun.-Prof. Dr. Lisa Bardach, Prof. Dr. Andreas Schilling, Jun.-Prof. Dr. Jonathan Brachthäuser for evaluating my work.

I would like to acknowledge my collaborators Dr. Efe Bozikir, Dr. Lisa Hasenbein, Dr. Patricia Goldberg, Philipp Stark, Prof. Dr. Gerrit Meixner, and Prof. Dr. Jens-Uwe Hahn, Lasse Frommelt for their support with the projects. They have been very supportive during design of VR user studies, VR data collection, analysis of eye-tracking data, technical discussions, and writing of research papers.

I thank my colleagues in the Human-Computer Interaction research group. Dr. Thomas Kübler, Dr. Wolfgang Fuhl, Dr. Tobias Appel, and Dr. Shahram Eivazi gave me great guidance in my research. Thank you for your kindness and patience in answering my research questions and for your suggestions on my new research ideas. I also grateful for the rest of the group, Babette Bühler, Dr. Nora Castner, Dr. Benedikt Hosp, Yao Rong, Daniel Weber, Björn Severitt, Chen Xin, Ruikun Hou, and others.

Finally, I am extremely grateful to all my family. I thank my parents for their unwavering and endless support throughout my academic life. Without their support, I would not have made it. To my boyfriend Yapeng, I thank you very much for your support and encouragement in my studies and life. Thank you for being with me through the ups and downs. Especially, since I could not return to China due to the COVID-19 pandemic, I am also grateful to all my relatives who helped my parents in their time of need.

Hong GAO

# Abstract

Virtual reality (VR) is not a new technology but has been in development for decades, driven by advances in computer technology such as computer graphics, simulation, visualization, hardware and software, and human-computer interaction. Currently, VR technology is increasingly being used in applications to enable immersive, yet controlled research settings. Education and entertainment are two important application areas, where VR has been considered a key enabler of immersive experiences and their further advancement. At the same time, the study of human behavior in such innovative environments is expected to contribute to a better design of VR applications. Therefore, modern VR devices are consistently equipped with eye-tracking technology, enabling thus further studies of human behavior through the collection of process data. In particular, eye-tracking technology in combination with machine learning techniques and explainable models can provide new insights for a deeper understanding of human behavior during immersion in virtual environments.

In this work, a systematic computational framework based on eye-tracking and behavioral user data and state-of-the-art machine learning approaches is proposed to understand human behavior and individual differences in VR contexts. This computational framework is then employed in three user studies across two different domains, namely education, and entertainment. In the educational domain, the exploration of human behavior during educational activities is a timely and challenging question that can only be addressed in an interdisciplinary setting, to which educational VR platforms such as immersive VR classrooms can contribute. In this way, two different immersive VR classrooms were created where students can learn computational thinking skills and teachers can train in classroom management. Students' and teachers' visual perception and cognitive processing behaviors are investigated using eye-tracking data and machine learning techniques in combination with explainable models. Results show that eye movements reveal different human behaviors as well as individual differences during immersion in VR, providing important insights for immersive and effective VR classroom design. In terms of VR entertainment, eye movements open a new avenue to evaluate VR locomotion techniques from the perspective of user cognitive load and user experience using machine learning methods. Research in two domains demonstrates the effectiveness of eye movements as a proxy for evaluating human behavior in educational and entertainment VR contexts. In summary, this work paves the way for assessing human behavior in VR scenarios and provides profound insights into the way of designing, evaluating, and improving interactive VR systems. In particular, more effective and customizable virtual environments can be created to provide users with tailored experiences.

**Abstract**

# Zusammenfassung

Die virtuelle Realität (VR) ist keine neue Technologie, sondern befindet sich seit Jahrzehnten in der kontinuierlichen Entwicklung, angetrieben von Fortschritten in der Computertechnologie wie Computergrafik, Simulation, Visualisierung, Hard- und Software und Mensch-Computer-Interaktion. Gegenwärtig wird VR-Technologie in zunehmendem Maße in Anwendungen eingesetzt, die immersive, aber kontrollierte Forschungsumgebungen ermöglichen. Bildung und Unterhaltung sind dabei zwei wichtige Anwendungsbereiche, in denen VR als Schlüssel zu immersiven Erfahrungen und deren Weiterentwicklung angesehen wird. Gleichzeitig wird erwartet, dass die Untersuchung des menschlichen Verhaltens in solchen innovativen Umgebungen zu einer besseren Gestaltung von VR-Anwendungen beitragen wird. Daher sind moderne VR-Geräte durchweg mit Eye-Tracking-Technologie ausgestattet und ermöglichen so weitere Studien des menschlichen Verhaltens durch die Sammlung von Prozessdaten. Insbesondere kann die Eye-Tracking-Technologie in Kombination mit Techniken des maschinellen Lernens und erklärbaren Modellen neue Erkenntnisse für ein tieferes Verständnis des menschlichen Verhaltens während der Immersion in virtuellen Umgebungen liefern.

In dieser Arbeit wird eine systematische Methodik vorgestellt, die auf Eye-Tracking- und Verhaltensdaten von Nutzern und modernen maschinellen Lernansätzen basiert, um menschliches Verhalten und individuelle Unterschiede in VR-Kontexten zu untersuchen. Dieser Ansatz wird dann im Rahmen von drei Nutzerstudien in zwei verschiedenen Anwendungsbereichen, nämlich Bildung und Unterhaltung, eingesetzt. Im Bildungsbereich ist die Erforschung des menschlichen Verhaltens während Lern- und Lehraktivitäten eine aktuelle und herausfordernde Frage, die nur in einem interdisziplinären Rahmen angegangen werden kann, wozu pädagogische VR-Plattformen wie immersive VR-Klassenzimmer einen erheblichen Beitrag leisten können. Auf dieser Weise wurden zwei verschiedene immersive VR-Klassenzimmer entwickelt, in denen Lernende computergestützte Denkfähigkeiten erlernen und Lehrende das Klassenraummanagement trainieren können. Die visuelle Wahrnehmung und das kognitive Verarbeitungsverhalten von Lernenden und Lehrenden wurden dabei mithilfe von Eye-Tracking-Daten und maschinellen Lernverfahren in Kombination mit erklärbaren Modellen erforscht. Die Ergebnisse zeigen, dass Augenbewegungen verschiedene menschliche Verhaltensweisen sowie individuelle Unterschiede während der Immersion in VR aufzeigen und somit wichtige Erkenntnisse für die immersive und effektive Gestaltung von VR-Klassenräumen liefern. Im Hinblick auf VR-Unterhaltung eröffnen Augenbewegungen einen neuen Weg, um VR-Lokomotionstechniken aus der Perspektive der kognitiven Belastung des Nutzers und der Nutzererfahrung mit Methoden des maschinellen Lernens zu bewerten. Die im Rahmen der

## Zusammenfassung

Dissertation durchgeführte Forschung in diesen zwei Anwendungsbereichen zeigt die Effektivität von Augenbewegungen als Proxy für die Bewertung des menschlichen Verhaltens in VR-Bildungs- und Unterhaltungskontexten. Zusammenfassend ebnet diese Arbeit somit den Weg für die Bewertung des menschlichen Verhaltens in VR-Szenarien und liefert Erkenntnisse über die Gestaltung, Bewertung und Verbesserung interaktiver VR-Systeme. Auf dieser Basis können effektivere und personalisierbare virtuelle Umgebungen für maßgeschneiderte und immersive Erfahrungen geschaffen werden.

# Contents

# List of Figures

# List of Tables

# 1 List of Publications

**Publications Relevant to This Thesis**

1. **Hong Gao**, Efe Bozkir, Lisa Hasenbein, Jens-Uwe Hahn, Richard Göllner, and Enkelejda Kasneci. Digital transformations of classrooms in virtual reality. In *Proceedings of the 2021 CHI Conference on Human Factors in Computing Systems (CHI'21)*. Yokohama, Japan. 2021. doi: 10.1145/3411764.3445596.

2. **Hong Gao**, Lisa Hasenbein, Efe Bozkir, Richard Göllner, and Enkelejda Kasneci. Evaluating the Effects of Virtual Human Animation on Students in an Immersive VR Classroom Using Eye Movements. In *Proceedings of the 28th ACM Symposium on Virtual Reality Software and Technology (VRST'22)*. Tsukuba, Japan. 2022. doi: 10.1145/3562939.3565623.

3. **Hong Gao**, Lisa Hasenbein, Efe Bozkir, Richard Göllner, and Enkelejda Kasneci. Exploring Gender Differences in Computational Thinking Learning in a VR Classroom: Developing Machine Learning Models Using Eye-Tracking Data and Explaining the Models. In *International Journal of Artificial Intelligence in Education (IJAIED)*. 2022. doi: 10.1007/s40593-022-00316-z.

4. **Hong Gao**, Efe Bozkir, Philipp Stark, Patricia Goldberg, Gerrit Meixner, Enkelejda Kasneci, and Richard Göllner. Predicting Teacher Expertise Based on Fused Sensor Data from an Immersive VR Classroom and Explainable Machine Learning Models. In *Proceedings of the 2023 CHI Conference on Human Factors in Computing Systems (CHI'23)*. Hamburg, Germany. 2022. *Work under revision for CHI'23.*

5. **Hong Gao**, Lasse Frommelt, and Enkelejda Kasneci. The Evaluation of Gait-Free Locomotion Methods with Eye Movement in Virtual Reality. In *2022 IEEE International Symposium on Mixed and Augmented Reality (ISMAR-Adjunct'22)*. Singapore. 2022. doi:10.1109/ISMAR-Adjunct57072.2022.00112.

6. **Hong Gao** and Enkelejda Kasneci. Eye-Tracking-Based Prediction of User Experience in VR Locomotion Using Machine Learning. In *Computer Graphics Forum*. 2022. doi: 10.1111/cgf.14703.

## Further Contributions

7. **Hong Gao**, Zijian Lu, Vera Demberg, and Enkelejda Kasneci. The Index of Cognitive Activity Predicts Cognitive Processing Load in Linguistic Task. In *ACM CHI'21 Workshop on Eye Movements as an Interface to Cognitive State (EMICS'21)*. Yokohama, Japan. 2021.

8. Efe Bozkir, Philipp Stark, **Hong Gao**, Lisa Hasenbein, Jens-Uwe Hahn, Enkelejda Kasneci, and Richard Göllner. Exploiting object-of-interest information to understand attention in VR classrooms. In *2021 IEEE Virtual Reality and 3D User Interfaces (VR'21)*. Virtual. 2021. doi: 10.1109/VR50410.2021.00085.

9. Wolfgang Fuhl, **Hong Gao**, and Enkelejda Kasneci. Tiny convolution, decision tree, and binary neuronal networks for robust and real time pupil outline estimation. In *Proceedings of the 2020 ACM Symposium on Eye Tracking Research & Applications (ETRA'20)*. Stuttgart, Germany. 2020. doi:10.1145/3379156.3391347.

10. Wolfgang Fuhl, **Hong Gao**, and Enkelejda Kasneci. Neural networks for optical vector and eye ball parameter estimation. In *Proceedings of the 2020 ACM Symposium on Eye Tracking Research & Applications (ETRA'20)*. Stuttgart, Germany. 2020. doi:10.1145/3379156.3391346.

## 1.1   Scientific Contribution

This thesis provides a state-of-the-art investigation of human behavior in virtual reality (VR) using eye-tracking technologies in combination with machine learning techniques to shed light on the adaptive and effective design of VR applications for education and entertainment. Contributions include the assessment of various human behaviors during VR learning and entertainment experiences using eye-tracking, machine learning approaches to identify individual differences based on eye-tracking, and post-hoc model explanation approaches for deeper investigations of interpretable model outputs.

Chapter 1 lists the main publications during the doctoral studies (including papers used in this thesis and papers that contribute to fundamental aspects in eye gaze analysis) and provides an overview of their scientific contributions.

Chapter 2 introduces the background and fundamentals of this thesis, including the application of VR in education and entertainment and eye tracking as an evaluation tool to assess various human behavior in VR scenarios.

Chapter 3 presents and discusses the major contributions based on the core publications during the dissertation project, mainly from three VR user studies in education and entertainment domains. To investigate student behavior while learning in VR, an immersive VR classroom was created to provide students with an immersive learning experience. The contributions of this study include insights into designing adaptive VR classrooms by analyzing students' eye gaze behaviors, uncovering the impact of animated virtual avatars, particularly virtual peer learners' social interaction behaviors, on students, and identifying gender differences in the learning process through machine learning based on eye-tracking. In addition, teachers' behavior during VR immersions is also investigated. For this purpose, an immersive VR classroom was created to assess teachers' expertise in professional vision using eye-tracking and machine learning. This study mainly contributes to a viable and practical avenue of assessing teacher expertise with machine learning methods. The study of human behavior in entertainment VR applications provides insights to improve the design of VR systems. To this end, an VR locomotion study was proposed to assess human behavior in VR entertainment using eye tracking in combination with machine learning. This study provides a feasible way to evaluate locomotion methods from the perspective of cognitive load and user experience based on eye movements.

Chapter 4 discusses the main findings of the dissertation and provides an outlook on the assessment of human behavior using eye-tracking in VR-based systems.

# 2 Introduction

Virtual reality (VR) can be defined as an approach to a user-computer interface that involves a real-time simulation of an environment, scenario, or activity that enables user interaction through multiple sensory channels [1]. VR is not a recent technology but has been in development for decades, driven by advances in computer technology such as computer graphics, computer simulation, visualization, human-computer interaction technology, and software and hardware. For several decades, however, the applications of VR remained mainly limited to individuals. In recent years, as global events accelerated existing trends toward distance education and virtual experiences, many major technology companies embraced virtual and augmented reality and are investing massively. This has facilitated the development of the latest consumer-grade VR head-mounted displays (HMDs) (e.g., Oculus Rift [1] and HTC Vive [2]) and the proliferation of these technologies in people's daily lives [2, 3], such as Horizon Workrooms [3] for virtual business meetings launched by Facebook for Oculus Quest 2, or OneBonsai's Virtual Instructor Platform [4] for safety, medical, and industrial training.

VR is considered a highly promising and prominent educational aid of the 21st century, as it has the advantage of providing users with authentic and realistic learning experiences by mimicking real educational scenarios or those scenarios that are impossible in the real world, thus further improving users' learning performance. In particular, VR has been identified as an excellent educational tool for learning and training in STEM subjects (Science, Technology, Engineering, and Mathematics) such as physics, math, computer science, manufacturing, language, health, and medicine, e.g., [4, 5, 6, 7, 8, 9], as well as in non-STEM subjects such as business and arts, such as [10, 11]. VR not only provides an authentic environment that mimics the real world but can make the impossible possible. For example, some contexts such as hostile working environments, learning life-threatening procedures, and high-cost training, environments that are difficult to assess, make hands-on training and learning far too impossible and highly risky. In these cases, VR can provide "risk-free" experiences at a

---

reasonable cost.

Educational VR applications support both online and offline learning, and with the development of VR technology, the integration of VR into education may become more prevalent in the near future. In fact, a variety of VR-based learning platforms have been developed and widely deployed, such as Engage [5] and Mozilla Hubs [6]. Such VR platforms not only allow learners to learn flexibly but also provide a high level of immersion and enhanced social engagement compared to traditional non-VR web-based online learning. The development of VR-based online learning platforms has recently been further driven by advanced VR HMDs and especially under the special circumstances of the COVID-19 pandemic. Many universities and schools had to switch to online teaching, which further promoted the development of remote interactive online learning. Many researchers have made efforts to develop more efficient and interactive VR platforms for online education. In addition to online learning platforms, VR is also used to develop offline learning platforms where real social interactions with real people can be replaced by pre-programmed avatars in virtual environments, such as immersive VR classrooms [12]. Such VR-based offline learning platforms can provide learners with more flexibility, adaptability, and personalization. By adapting the design paradigms of VR applications, self-directed learning can be achieved and individualized support can be provided to learners.

VR-based educational platforms not only provide learners with a learning context but also facilitates researchers to study the effects of certain features of learning environments on learners, which can provide further opportunities to improve learning outcomes. Studying learner behavior during real-world learning is challenging due to the difficulty of fully controlling experimental conditions and preparing real-world learning experiments is time-consuming and expensive. For example, two of the most important factors that can shape learners' learning behavior and that need to be investigated are the teachers who deliver the instructional content to the class and the peer learners who provide social interaction information to the learners. In real-world contexts, it is difficult to fully control these two factors, and especially to repeat the experiments many times. VR, on the other hand, can overcome this limitation by manipulating the virtual teacher and virtual peer learners with virtual avatars programmed with real-world animations.

In addition to the benefits of VR for education, VR is also becoming increasingly popular in entertainment, enriching people's daily entertainment life. Designers and researchers are inspired to integrate VR into entertainment in various and innovative ways, such as the VR museum [13] used by many world-famous museums, virtual theme and amusement parks on Oculus Rift and Steam, VR music experiences, and numerous VR games.

Behind the booming adoption of VR in education and entertainment is the question of how to develop the optimal applications for specific purposes. That is, how to design a VR application that best achieves the goals and meets the needs, for example, an educational VR application that not only serves as a simple learning platform for learners but also promotes the best learning behaviors, whereas an entertainment VR application provides the best experience for users. Therefore, it is essential to study the behavior of users while they are immersed in virtual environment, as this will provide deep insights to improve VR systems and thus provide better VR experiences for users. In the literature, it was recognized already two decades ago that VR systems have the potential to facilitate the study of human behavior in psychology [14]. In VR systems that display 3D visual stimuli, users interact with virtual objects so that their behavioral responses to these visual stimuli can be measured. Thanks to the development of VR technology that enables tracking of various human behaviors (e.g., body tracking, head tracking, gaze tracking) and software such as Unity [7] and Unreal Engine [8] in recent years, VR has now proven to be viable for studying various human behaviors in different domains. In particular, eye-tracking technology built into HMDs makes it easy to obtain human gaze data, which contains extensive information about how users process visual scenes and whether a cognitive processing load is simultaneously triggered during information processing. Eye movements are the movements of the human eyes that help capture, track, and fixate visual stimuli and can reveal people's deep subconscious behaviors and thoughts about the systems they interact with.

---

[5]https://engagevr.io/
[6]https://hubs.mozilla.com/

However, studies on eye-tracking in VR are still limited. Several technical limitations account for the lack of eye-tracking data exploration in human behavior studies in the VR domain. First, it was difficult to obtain eye-tracking data when VR headsets are worn, but this is not a major issue nowadays as some advanced HMDs are equipped with built-in eye trackers. Second, is the lack of state-of-art tools or software to analyze eye-tracking data. For traditional remote and mobile eye trackers, the professional eye-tracking analysis software is well supported by the respective eye tracker manufacturer. However, for VR HMDs, there is no such software support despite eye-tracking integration. In this context, the 3D visual stimuli and head movements in VR environments make it challenging to analyze eye-tracking data with standard algorithms used for non-VR contexts. Therefore, researchers' attention should be drawn to overcoming these limitations, as the benefits of VR systems as tools for studying human behavior should not be annihilated by such limitations.

Based on eye-tracking technology and the data derived from such integration, users' visual attention and cognitive behavior can be easily assessed during immersion in the virtual environment, which provides further insights for improving VR systems. Specifically, in the field of VR education, previous literature has mostly evaluated learner behavior using post-hoc questionnaires that captured learning outcomes or perceptions of the VR environment, such as presence, immersion, and engagement [12, 15]. However, learner behavior during the learning process, which provides information about learning improvement, has rarely been studied and can be investigated using eye-tracking. This method is also applicable in the field of VR entertainment to investigate users' temporal visual perception behavior and provide

---

further guidance on designing VR applications that provide users with a better experience.

In summary, this chapter focuses on how eye-tracking can facilitate the development and application of VR in education and entertainment. Section 2.1 discusses in detail the application of VR in education, i.e., how VR has been used to facilitate education. Section 2.2 discusses the application of VR in entertainment. This section primarily discusses, from a technical perspective, how navigation in VR applications can be achieved. Building on this, Section 2.3 discusses the effectiveness of eye-tracking for detecting various human visual and cognitive behaviors and its potential for facilitating studies of VR applications in education and entertainment.

## 2.1 Virtual Reality in Education

Recently, digital technologies have gained significant popularity in education at all academic levels, from elementary school to postgraduate education. VR, in particular, has received substantial attention due to its ability to provide educators with tremendous opportunities to incorporate advanced technologies into their educational activities to provide learners with an improved learning motivation and experience, and thus better learning performance, e.g., [16]. Moreover, the current availability of consumer-grade HMDs that enable the creation of immersive experiences at a reasonable cost makes it possible to introduce immersive VR experiences in education in the near future [17]. Due to the many advantages of VR, such as first-hand active experiences, distraction-free, learner-centered, and entertaining, VR has been used to support various types of learning and teaching, such as virtual field trips [18], virtual labs [19], virtual classrooms [15, 20], virtual language immersion [21], and for training [22]. Such VR applications based on advanced technologies have the advantage of enabling remote and online education with increased engagement and immersion [23].

VR as an evolving technology facilitates education from several aspects. First, virtual learning and training scenarios can be easily and cost-effectively created to provide users with realistic experiences. In particular, VR is an optimal solution for educational activities that are difficult or not cost-effective to achieve in real-world situations. For example, VR offers solutions for science education that requires learners to perform experiments, such as learning chemistry concepts and performing chemistry experiments that are costly and dangerous [24], simulating physics experiments [4], and engineering labs [25]. By immersing learners in such virtual environments, they can learn complex science concepts or theories through active experience and increased interest. VR also facilitates training by simulating actual events for various organizations and businesses [26]. Training is the teaching or developing of any skills and knowledge in a person that are related to specific useful competencies in specific domains, which is a much more specialized form of education. Real-world training usually has several limitations, including the high cost of time to set up real-world training systems and travel to the site, the lack of intuitive visual hints such as 3D animations for concept illustrations, and the high cost and difficulty of preparing real-world training materials. VR technology can overcome these limitations and enable real-world training through virtual environments [27, 28]. Both teachers and learners can benefit from such training systems to develop specific skills [29, 30], for example training pre-service teachers with professional vision skills (i.e., teachers' ability to rapidly notice information in the class) [22, 31], training students in various fields such as medical training [32] and programming skill training [7].

Remote and offline education can be supported with advanced VR technology, which has been further promoted by the COVID-19 pandemic. Compared to real-world learning that requires students to be physically present, such as attending a lecture in a classroom, online remote learning with VR platforms can provide learners with an interactive and immersive learning experience without physical constraints [33], such as the social VR platform Mozilla Hubs. In such remote learning platforms, learners attend remote instruction by immersing themselves in a virtual environment and interacting with the instructor on the other side of the internet in real time [23]. With the accompaniment of the real person represented by virtual avatars, learners gain a high level of presence, immersion, and engagement. An example of this is remote classroom instruction [33]. Remote VR classrooms create a classroom that mimics a traditional classroom with the same features and configurations. Moreover, teachers and students represented by virtual avatars connect to the virtual classroom simultaneously. Thus, real-time distance learning can be achieved. A key feature of such distance learning classrooms is that they provide learners with a social accompaniment, similar to that of a real classroom, especially through peer learners. Learners can interact with both teachers and peers to get social feedback that is highly related to their learning experience.

VR-based online distance learning provides learners with immersive and interactive learning experiences by enabling real-time interactions with others who are simultaneously immersed in the virtual environment. Another form of education is offline learning, which can also be well supported by VR technology. Unlike online distance learning, learners do not interact with real people in offline learning, but can also have an immersive and authentic learning experience with VR. One advantage of offline learning is that there is no time synchronization issue and thus more flexibility. Apart from that, all the experiences that learners can obtain from a remote VR learning platform can also be obtained in offline VR applications. An immersive experience can still be provided by programming the specific features of the virtual environment, such as preprogrammed avatars with animations [12, 34], to provide a realistic-like interaction. In such VR-based offline applications, students can have the self-paced immersive learning experiences [35].

Regardless of the form of learning (online or offline), VR technology has shown its significance and promotion for education. Among the various types of VR applications, (immersive) VR classroom, where a teacher teaches students face-to-face, is the most widely used form of education, not only in K-12 but also in higher education. Therefore, it is very important and necessary to study the different aspects of teacher and student behavior to improve education [36, 37, 38]. VR classrooms provide learners with a learning experience like a real classroom by mimicking a traditional classroom environment, including classroom layout and virtual avatars representing the teacher and classmates [12, 39, 40]. Especially with the social information provided by animated avatars, especially avatar classmates, learners can learn in a more immersive and engaging way [20]. This raises the question of how to design effective VR classrooms. Some previous work has made significant progress in this research domain. For example, an online class that enables real-time interactions with teachers and classmates was developed based on a standard web conferencing system (i.e., Microsoft Teams) and a VR social platform (i.e., Mozilla Hubs) [33]. The teacher delivers a lecture to the students who attend the class simultaneously. Students perceived a high level of presence while immersed in such an online class. However, the virtual avatars representing real people were not fully featured with animations, but only audio information of the real people behind them was provided. This may reduce the engagement and authenticity of the learning experience. Some previous work has targeted more interactive avatars to increase authenticity. For example, a more immersive VR classroom system was designed to support online virtual teaching and learning [20]. In this system, the teacher and students were visualized with lifelike avatars and can interact with each other. Social information from classmates is one of the important aspects that influence students' learning behavior [40, 41]. Therefore, this should also be considered when designing an immersive and authentic VR classroom. To this end, several previous studies have created immersive VR classrooms that aim to provide learners with more peer accompaniment and social interaction by embodying historical messages of previous learners in the classroom during learning [12] and programming virtual classmates with social interactive behaviors (e.g., hand-raising [40]).

The aforementioned work has mainly discussed VR-based educational systems for learning purposes, especially from the students' perspective. VR might also be a powerful tool to facilitate teacher education, e.g., training teacher with special pedagogical skills [42]. For example, a VR-based learning platform incorporating a Kinect-enabled sensormotor interface has been developed for teaching training [43]. It was found that such a VR learning system supporting a wide range of teaching tasks with avatar-embodied live gestures can provide teachers with an enhanced sense of presence. Similar to other VR classrooms designed for students, active avatars were also created in some related works. The teacher can practice teaching in such a VR classroom with real students represented by avatars who provide real interaction feedback. In addition, immersive VR classrooms have been designed for pre-service teachers to practice teaching [44] and classroom management [22, 31]. In all these immersive VR classrooms, a realistic classroom was designed by programming animated avatars (including teachers and students), providing teachers with authentic teaching experiences.

## 2.2 Virtual Reality in Entertainment

Virtual reality was originally developed for entertainment. With the advances in computer graphics, user interfaces, and visual simulations, it has been used in various fields and also in our daily lives [45]. Especially in recent years, consumer-grade HMDs equipped with sensory accessories such as controllers, headsets, hand controllers, and treadmills have led to the development of various types of VR entertainment (e.g., games, exhibitions, concerts, and museums [13, 46, 47]), as well as education through gaming (edutainment) [48, 49]. Rather than spending hours sitting on the site or walking around a room (e.g., a museum), VR entertainment enables users to be physically active and explore the virtual environment while having immersive experiences without having to physically exert themselves too much. This is one of the essential features of VR entertainment applications, i.e., an unlimited virtual world can be provided, and users can explore this environment in a small real-world place with little or no physical movement.

Navigation in virtual environments requires that users can explore large virtual environments efficiently and infinitely while remaining confined within a room-scale real-world environment. This is achieved through VR locomotion techniques, which is a technology that enables avatar movement from one location to another within a virtual reality environment. Especially with the increasing integration of VR into entertainment and education [2, 3], VR locomotion techniques are increasingly being studied by developers and researchers. A variety of locomotion techniques have been developed and employed in VR applications, which can be classified into different categories according to the interaction type, VR motion type, and VR interaction space [50]: 1) Motion-based locomotion requires some kind of physical movement to enable locomotion (e.g., arm swinging [51], walking-in-place [52]), 2) room scale-based locomotion is special motion-based locomotion that is limited by the size of the real environment (e.g., real-walking [53]), 3) controller-based locomotion provides continuous movements based on controller inputs (e.g., joystick-based [54]), 4) teleportation-based locomotion supports non-continuous movement by instantaneously moving the user from one location to another in the virtual environment (e.g., teleportation [55]). These different VR locomotion techniques are unique, have their own benefits and drawbacks, and are designed for different applications. For example, point-and-click teleportation and teleportation-like locomotion techniques (e.g., grappling) are now very popular and are used and integrated into a variety of commercial VR systems such as the Oculus Rift and the HTC Vive [55]. Compared to other continuous locomotion techniques, teleportation-based locomotion enables faster and more effective movement, lower risk of motion sickness, ease of use for the user, and ease of implementation [56, 57]. However, such teleportation techniques have the drawbacks of causing spatial disorientation and interrupting presence [58], which is not the case with continuous locomotion techniques. This raises the question of how to select or, in other words, evaluate locomotion techniques for different purposes.

A number of previous studies provide a systematic literature review of locomotion techniques [59]. Many of these studies introduce new VR locomotion techniques and compare them to well-established and popular locomotion techniques, such as joystick, teleportation, and walking-in-place. In these studies, assessments such as user experience (usability of the locomotion technique), sense of presence, motion sickness, and other post-hoc surveys such as user preferences were usually quantified with questionnaires [60, 61]. In addition, some previous studies have examined the cognitive load of users while they experienced different 3D travel techniques (e.g., real walking, redirected walking, steering, and joystick) using either a post-task or dual-task paradigm to measure cognitive load [62, 63]. Such an assessment provides insight into the improvement of VR locomotion techniques and may provide further inspiration for the development of new techniques.

## 2.3  Studying Human Behavior in VR Environments based on Eye Tracking

Human gaze reveals a wealth of information about how we perceive the world. Eye movements are the voluntary and involuntary movements of the eyes controlled by six extraocular muscles. Eye movements such as saccade, fixation, and smooth pursuit assist humans in obtaining, fixating, and tracking visual stimuli [64]. Eye tracking has long been used as an intuitive modality for studying various conscious and unconscious human perception behaviors [65], e.g., visual attention in solving science problems [66], visual search [67], emotional and cognitive processes [68], as a modality in various human-computer interaction (HCI) tasks (e.g, multimedia learning [69], web search [70], medical image reading [71, 72, 73, 74, 75], driving [76, 77, 78, 79, 80, 81, 82], teaching [83], programming [84], software development [85], human-robot interaction [86, 87, 88]), and in psychology study [89]). Especially with the advances in hardware, various types of fancy eye trackers offer new opportunities and potential for human behavioral research. In the field of HCI, human behavior, which reflects how people interact with a system, is extensively studied to provide significant insights for system design: The more we understand about human behavior, the better we can design HCI interfaces that meet people's needs [90].

Eye-tracking technology is an excellent tool for investigating human visual perception and cognitive processing load during interaction tasks [91]. For example, it is necessary and important to investigate human behavior in entertainment applications, as such investigation provides direct evidence for design improvement [92]. In particular, the study of human behavior in different educational domains helps to better understand the learning and teaching process and thus provides profound insights for improving education effectiveness and thus performance [93], such as in computing [94], medicine [95], mathematics [96], linguistics [97], and training [98]. In these studies, eye-tracking technology has been applied to reveal human behavior in terms of visual attention, decision-making processes, and cognitive processing. Data obtained with eye-tracking technology (e.g., screen-based eye trackers and eye-tracking glasses) provide important information about users' behavioral patterns, which can inform how the learning process evolves. From the eye-tracking data, eye movements such as fixation and saccade can be calculated [99, 100, 101], as well as pupil diameter, which is often considered also an indicator of cognitive load [91]. Fixations are periods of time when the gaze is maintained in a single location, whereas saccades are the rapid shift of the eye from one fixation to another. Scanpaths are the combination of fixations and saccades [102, 103, 104]. On this basis, various eye-tracking measures can be conceptualized in different approaches, usually on three scales, i.e., temporal (e.g., fixation duration and saccade duration), spatial (i.e., fixation sequence, saccade length, and scanpath pattern), and counting (e.g., number of fixations and number of saccades). In previous work, these measures were mostly used to reveal different human behaviors. For example, fixations have been used to assess the visual attention allocation of novice programmers during reading source code [105]. Scanpaths have been used to explore students' visual strategies used in reading medical images (e.g., [71]), or in problem solving (e.g., [106]).

### 2.3.1 Eye Tracking in Virtual Reality

The integration of eye tracking systems into HMDs greatly advances VR research, as it provides informative data to reveal subconscious human behavior in virtual environments [107] and enables foveated rendering to reduce rendering workload by greatly reducing the image quality in the peripheral vision [108]. With such combined technologies, users' eye gaze data can be accurately and unobtrusively recorded during the VR experience with an eye tracker seamlessly built into the HMDs [109]. It should be noted that such eye-tracking data collection is not constrained by head movements. Such a combination breaks the limitations of investigating human perception and behavior in the virtual world. With the rapid development of VR technologies, many modern consumer-grade HMDs can support integrated eye tracking, e.g., HTC Vive Pro Eye and Varjo VR-3. Especially with the increasing popularity of VR in education and entertainment, easy access to eye-tracking data opens up many possibilities for VR research, providing profound insights into the design of effective VR applications that can be used not only as interactive interfaces but also as tools for studying human behavior in VR.

One aspect of the application of eye-tracking technology in VR is that it can serve as a tool for touchfree human interaction with the virtual environment. Gaze-based interaction in VR offers several advantages, such as minimized physical effort (e.g., using body movement for interaction), and easy interaction with distant objects in the virtual environment. This frees up the hands or body used for interaction with gestures or controllers. Based on the fovea region traced by the eye tracker inside the HMD, the 3D ray of the gaze vector can be generated towards the focal point of sight, allowing interaction with the virtual environment. Researchers have already embarked on this line of research, e.g. [110, 111]. Most commonly, gaze-based interactions are used in VR to point at objects and make selections. For example, in one comparison, gaze-based interaction was found to be faster than conventional 3D pointing, especially for distance objects [112]. A Gaze + pinch interaction technique that combines eye and freehand input for 3D interaction in VR has been proposed to provide enhanced effective interaction in a large virtual environment [113]. Several novel eye-gaze-based interaction techniques were proposed and evaluated [114]. In addition to object selection, eye tracking input was also used for motion control [115], VR navigation [116], and interaction with virtual environments such as gaze typing [117].

Another important aspect of eye tracking in VR is providing unique insights into human perceptual behavior. While immersed in various VR applications designed for different purposes (e.g., as an educational tool, for entertainment, or to explore human behavior), users perceive visual scenes with their eyes. This perception of the virtual environment contains important information about their potential human behaviors. It is important that this information is obtained in a non-intrusive way without affecting the natural behavior of the users. Although eye-tracking technology has been used extensively in education to study user learning behaviors and thus improve education [98, 118], the introduction of eye-tracking in VR for educational purposes is a fairly new topic that has only emerged in recent years. Many educational researches have used VR as a tool to create an authentic educational virtual environment that provides users with a more realistic experience (e.g., VR classroom [31, 44], VR lab [119]), however the evaluation methods used in these studies are mostly post-hoc questionnaires that assess either the effectiveness of VR systems or students' learning behaviors and performance. The lack of analysis of eye tracking data in VR is due to some technical limitations. Compared to traditional eye-tracking in the real world, which use 2D screen-based stimuli, 3D stimuli are presented on 2D screen lenses in VR HMDs only a few centimeters in front of the users' eye and the users can freely move their head. This makes eye tracking in VR challenging.

Advanced VR headsets with integrated eye trackers have solved the eye tracking challenge by providing original sensor data of the eyes. However, research on identifying eye movements (e.g., fixations, saccades, etc), which are commonly used to reveal various human perceptual behaviors in virtual contexts, is still at an early stages. Unlike eye trackers used for 2D stimuli in real world, for which there are professional data analysis tools for business use (e.g., Tobii Pro Lab [120]), there is no standard method or software for analyzing eye-tracking data collected in VR. In particular, due to the more complex eye-head coordination patterns, the fine-grained methods developed for monitor-based eye-tracking with 2D stimuli are no longer applicable. Some previous works have attempted to fill this research gap by proposing proprietary identification algorithms based on already established algorithms (e.g., I-VT: velocity threshold identification; I-DT: dispersion threshold identification; [121]) used for 2D stimuli. For example, a two-stage pipeline for the detection of fixations and saccades (in the first stage) and of more complex but not very frequently used eye movements (e.g., smooth pursuits, Vestibulo-ocular reflex; in the second stage) was proposed for eye tracking data collected in a 360-degree video context [122]. This detection method is based on the I-VT algorithm considering head movement. In another work [123], a fixation detection algorithm based on the I-DT algorithm was adapted to VR, with validation of the thresholds used. In [124], a modified hybrid I-VDT (velocity- & dispersion-threshold identification algorithm) was proposed for fixation and saccade detection, with optimal parameters tuned to achieve the best performance. These proposed algorithms have greatly facilitated and increased confidence in the study of human behavior in VR using eye-tracking [125]. Other methods known from eye-tracking research, such as bayesian identification of eye movements [101], a rule-based machine learning method [126], and gazeNet with deep neural networks [127], have not yet found their application and adaptation to eye movement detection in VR.

### 2.3.2 Eye Tracking with Machine Learning

Eye tracking has been used extensively to uncover users' visual attention in human-computer interaction tasks, providing insights into a wide variety of subconscious human behaviors, such as visual perceptual patterns [71, 103, 104], cognitive processing load [128, 74], and visual search dynamics [129]. Eye tracking combined with machine learning techniques offers more opportunities to study various aspects of human behavior in which gaze plays a role, including but not limited to human personality traits [130, 131], gender [132, 133], intelligence [134], skills and abilities [71, 73, 135], learning gains [136], task performance in driving [82, 79, 137, 138], etc. Eye-tracking data with timing information contain patterns of how people visually perceive the stimuli throughout the task. Visual patterns are usually post-experimentally extracted as various eye-tracking measures, e.g., fixations, saccades, blinks, scanpaths, and attention heatmaps, which are interpreted as various human behaviors. However, these patterns are usually considered separately and examined either in an average manner or in specific time windows when specific stimuli are presented. In this way, the gaze patterns in the eye-tracking data, reflected in different eye movements, are not considered collectively, although they would provide potential insight into a deeper understanding of human behavior.

Machine learning is the perfect solution for such multimodal data analysis, where hidden patterns or data groupings are automatically discovered from large volumes of data. Initially, first machine learning algorithms employed in the eye-tracking domain were related to improving the robustness of the eye-tracking signal, such as methods for high-quality pupil detection (e.g., [139, 140, 141, 142]), slippage compensation (e.g., [143, 144]), or eye-movement recognition (e.g., [145, 146]). Beyond this technical contributions, there are several ways to integrate multimodal eye-tracking data into machine learning model at the behavioral level. One way is to build machine learning models based on unprocessed original sensor data, such as raw gaze positions and gaze velocities [136]. In a data-driven approach based on state-of-the-art black-box models, a larger data set can typically contribute to better model performance, e.g., using original sensor data as features. However, this comes at the price of low model interpretation. When using eye tracking, understanding human behavior is usually the first goal. Therefore, even when combining eye tracking with machine learning, it is very important to provide interpretable findings. For this purpose, interpretable eye-tracking features have be extracted for machine learning models, providing thus deep insights into human behavior. This is also what previous works have done. Statistical metrics (e.g., mean, minimum, maximum, and sum) of eye tracking measurements are typically extracted from time series data as features [130, 147, 134, 148]. To explain the machine learning model outputs, often post-hoc explanation models are applied, e.g., [149]. Such model explanation offers several advantages for eye-tracking studies. First, it reveals which features are informative and which are not in the model for a specific prediction task. Understanding the contribution of eye-tracking features to the models can not only increase confidence in the validity of the models, but also provide clues for improving model performance by knowing exactly what to fine-tune and optimize. In addition, the model explainability approach uncovers the underlying re-lationships between eye movements and model targets. This provides further insights into understanding underlying human behaviors during tasks. In particular, VR technologies that enable the acquisition of eye-tracking data in combination with machine learning techniques and the model explainability approach offer a new avenue for deeper understanding of human behavior during various VR activities, such as in education and entertainment. This will greatly facilitate the development of more effective VR applications and promote VR as an outstanding tool for studying human behavior.

Overall, this chapter provided a discussion of current work on studies of VR-based environments applied in two significant domains, education and entertainment, and the limitations of studying human behavior using eye-tracking in such VR contexts. This provides the background and foundation for the comprehensive investigation of various human behaviors in such VR environments using eye-tracking technology (i.e., collection of eye-tracking data and detection and analysis of pupil and eye-movement events) and machine learning methods, which are presented in 6 papers in Chapter 3.

# 3 Major Contributions

This chapter provided a summary of the main contributions of this dissertation, based on several papers published in prestigious conferences and journals. A deep understanding of human behaviors through the use of multimodal eye-tracking data in various VR contexts, including VR in education and VR in entertainment, was provided based on several user studies. The main contributions and results are presented below. The relevant publications are listed in Chapter 1.

## 3.1  Human Behavior in IVR Classrooms

### 3.1.1  Student Gaze Behavior in IVR Classrooms with Different Configurations

**Hong Gao**, Efe Bozkir, Lisa Hasenbein, Jens-Uwe Hahn, Richard Göllner, and Enkelejda Kasneci. Digital transformations of classrooms in virtual reality. In *Proceedings of the 2021 CHI Conference on Human Factors in Computing Systems (CHI'21)*.

**Motivation**

With the development of advanced consumer-grade HMDs and computer graphics, VR is increasingly being applied in education to provide users with a more immersive and authentic distance or offline learning experience. Largely due to the COVID-19 pandemic, many universities and schools have switched to online teaching and are using various VR applications such as Mozilla Hubs [150]. VR holds great potential for facilitating education by providing learners with a more flexible and immersive learning experience at a low cost: no physical distance constraints, no need for time synchronization with other learners or instructors, high immersion and authenticity. Among the various types of VR-based educational applications, immersive VR (IVR) classrooms, which mimic traditional classroom environments and enable authentic social interactions and engagement, play an important role in VR education, especially in K-12 education. Compared to other VR learning systems, VR classrooms provide learners with a more immersive learning experience because they not only offer an authentic classroom environment, but also provide accompaniment of virtual instructors and peer learners, which in turn may positively impact learning outcomes. This leads to the research question of how to design an immersive and effective VR classroom.

Several key aspects can be important and need to be considered when designing an IVR classroom, such as the seating arrangement and seating position of learners in the VR classroom, the visualization style of social counterparts including virtual peer learners and the virtual instructor, and the accompaniment of peer learners. These factors are important for several reasons. First, educational psychology research has repeatedly demonstrated that peer learners (classmates) substantially shape student learning [151], especially peer learners' hand-raising behavior to ask or answer questions during class [40]. Second, as the Uncanny Valley effect explains [152], more human-like avatars are not always conducive to making virtual avatars palatable to users. Lastly, where learners sit in the classroom also matters, as it affects learner motivation and learning outcomes [153]. When learners are immersed in such IVR classrooms with different configurations, their behavior is what we are interested in. Therefore, the main motivation of this paper is to investigate how the various factors mentioned above for IVR classroom design affect learners' behaviors as seen through eye movements. Learners' perception of the configured IVR classroom can provide insight into the improvement of specific design features, and such perceptual behavior can be demonstrated as visual attention, visual search, and cognitive processing behavior, which can be measured with various eye-tracking features (e.g., fixation, saccade, pupil diameter). With modern VR headsets equipped with eye trackers, learners' behavioral data can be easily tracked, i.e., head movements, pupil size, and gaze vectors, and eye movements can be extracted from these sensor data. In this way, learners' visual perceptual and cognitive behaviors can be determined during immersion in the IVR classroom.

**Methods**

To investigate the effects of classroom configuration on learners, an IVR classroom was created considering three design features, namely, learners' seating position in the IVR classroom (front row vs. back row), the visualization style of avatar peer learners and the teacher (cartoon vs. realistic), and performance level of virtual peer learners (20%, 35%, 65%, or 80% of virtual peer learners with preprogrammed hand-raising behaviors). A user study was conducted in a between-subjects design that included a total of $2 \times 2 \times 4$ different conditions. 381 sixth-grade students participated in our study by listening to a virtual lesson delivered by a virtual teacher in an IVR classroom. Participants' raw head-tracking and eye-tracking data were recorded. Due to noisy sensor readings and blinking behavior, we smoothed [154] and normalized [155] the pupil diameter following the standard pipeline.

A modified velocity-threshold identification (I-VT) algorithm applied to the VR context that takes head movement into account [122] was applied in our study to extract eye movement features, with parameters adapted to our study. Several variables related to fixation, saccade, and pupil diameter were calculated. A full-factorial ANOVA was applied to compare all eye-tracking variables between conditions. The Aligned Rank Transform (ART) [156] ANOVA was used as a nonparametric factorial ANOVA. The Tukey-Kramer test was used as a post-hoc test. The significance level was set at $\alpha = 0.05$ for all tests. The overview of the designed IVR classroom is shown in Figure 3.1.

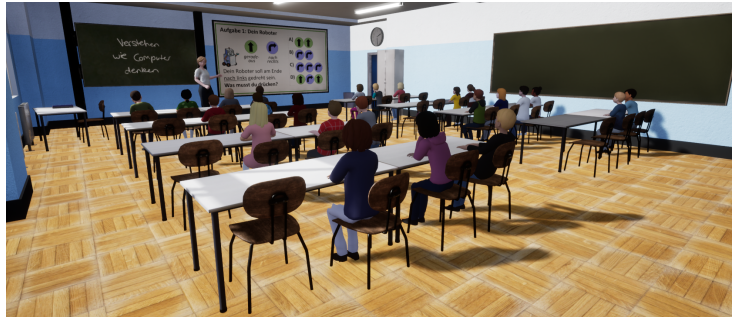

Figure 3.1: Immersive VR classroom for students developing computational thinking skills.

**Main Results**

Variables including the number of fixations, fixation and saccade duration, saccade amplitude, and pupil diameter were compared between conditions.

The results showed that students' seating position in the IVR classroom had significant effects on their visual attention and visual search behavior, as indicated by the fixation and saccade metrics. Compared to students seated in the front row of the classroom, students seated in the back row had statistically significant longer fixation durations, shorter saccade durations, and shorter saccade amplitudes. This suggests that students sitting in the back row had significantly longer information processing time and shorter visual search behavior. When students were seated in the back row, they had a wider view of the classroom and therefore were able to perceive more visual information (especially virtual peer learners) with fewer visual search activities.

The visualization style of virtual persons in the IVR classroom was found to have significant effects on students' visual behavior as well as cognitive load as indicated by the metrics related to fixation, saccade, and pupil diameter. Students in the realistic condition, in which all virtual persons were visualized with a realistic avatar style, had significantly shorter fixation duration, longer saccade duration, and larger pupil diameter than students in the cartoon condition. This suggests that students in the cartoon condition had longer visual processing time for the cartoon-style avatars, shorter visual search activities, and yet lower cognitive processing load.

Compared to the other two design factors, the performance level of peer learners (i.e., the percentage of virtual peer learners raising their hands) had significant effects on students, but to a less degree. Results showed that students had a significantly larger mean pupil diameter when they were immersed in a VR classroom with 80% of virtual peer learners raising their hands than in the 35% condition. This suggests that students had a higher cognitive load when they perceived more hand-raising information than when they perceived less hand-raising information. In addition, the results showed that students had significantly more fixations in the 65% condition than in the 80% condition. Overall, these results offer profound implications for designing an effective VR classroom using eye movement information.

### 3.1.2 Effects of Avatar Animations on Students' Visual Attention and Cognitive Behavior

**Hong Gao**, Lisa Hasenbein, Efe Bozkir, Richard Göllner, and Enkelejda Kasneci. Evaluating the Effects of Virtual Human Animation on Students in an Immersive VR Classroom Using Eye Movements. In *Proceedings of the 28th ACM Symposium on Virtual Reality Software and Technology (VRST'22).*

**Motivation**

VR classrooms that mimic traditional classroom environments provide learners with authentic learning experiences that further increase learner motivation, persistence, and interest, ultimately leading to better learning outcomes. To achieve an interactive and effective VR system, one of the key factors is the implementation of virtual avatars (e.g., the virtual teacher and, most importantly, virtual peer learners) that exhibit social interaction behaviors normally occurring in real classrooms. However, the current VR classrooms typically provide preprogrammed but limited social-interactive information from the virtual teacher or virtual peer learners. The effects of these animated avatars on learners' behavior, especially on learners' cognitive and visual attention behaviors, have rarely been studied, although this is very important for the design of virtual animated avatars in VR scenarios. First, it is important to know that the implemented animated avatars actually have an impact on learners, i.e., learners are expected to have perceived such avatar animations while immersed in VR, thus achieving the goal of increasing immersion and authenticity. Second, these effects of animated avatars should be moderate, i.e., learners' attention would not be diverted too much from the instructional content to the animated avatars, and their cognitive load would not be affected too much. To address this research gap, an IVR classroom was created with preprogrammed animated virtual avatars exhibiting interactive social behaviors that can increase immersion and authenticity. This study aims to investigate in depth how learners respond cognitively and visually to the animated avatars, especially to the most salient avatar animations, i.e., peer learners' hand-raising behavior.

**Methods**

The virtual avatars in the IVR classroom were preprogrammed with interactive social animations. These animations were created based on motion capture from real classrooms and therefore authentically mimic physical movements similar to those of real people in real classrooms. The virtual teacher walks around the podium while delivering a virtual lecture, proposes questions, and calls on the virtual peer learners with body gestures (e.g., hand gestures). The virtual peer learners turn around, think, and interact with the virtual teacher by raising their hands to answer questions. The avatar animations are shown in Figure 3.2.

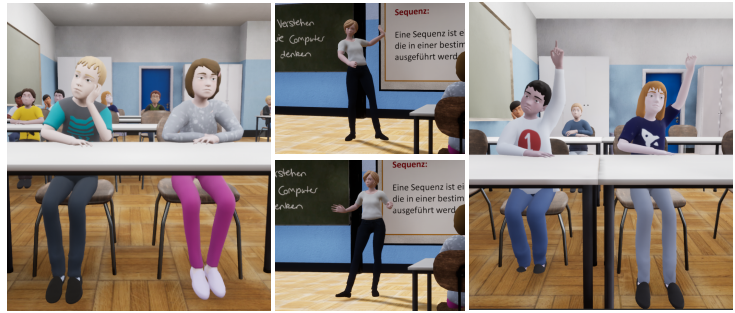

Figure 3.2: The animated virtual teacher and virtual peer learners in the IVR classroom.

Since the main objective of this study is to investigate the effects of the social interaction animations between the virtual teacher and the peer learners (i.e., the virtual peer learners' hand-raising behavior) on the students' cognitive responses and visual attention behaviors, the students' eye movements and pupil measurements before and after the hand-raising were mainly analyzed. Students' gaze behavior was examined in time windows of $2.5s$ before and after the hand-raising animation. Dependent variables including pupil diameter, fixation, saccade, dwell time, and the number of virtual peers on which students fixated, were extracted. First, to examine the effects of peer learners' hand-raising behavior on students, t-tests were conducted to compare these dependent variables before and after hand-raising animation. Second, whether the effects of hand-raising animation on students were related to the amount of hand-raising (20%, 35%, 65%, and 80%) was examined. The magnitude of these effects in the four hand-raising conditions was compared. That is, the change values of the above dependent variables before and after the hand-raising animation were calculated as new variables for comparison. Data collected in Section 3.1.1 were used, and detailed data processing and detection of eye movement events are described in the Methods of Section 3.1.1.

### Main Results

It was found that the avatar animations (i.e., the virtual peer learners' hand-raising behavior) significantly affected students' cognitive load, as indicated by pupil diameter. Students exhibited a significantly larger mean normalized pupil diameter after perceiving hand-raising animations than before the animation. Furthermore, the results showed that the increase in cognitive load caused by hand-raising animation differed significantly between different hand-raising conditions. The increase in pupil diameter was significantly greater in the 80% condition than in the other three hand-raising conditions (20%, 35%, 65%).

Results showed that the hand-raising behavior of peer learners had significant effects on students' visual attention and visual search behavior, as indicated by fixation and saccade. Specifically, it was found that students had a significantly longer mean fixation duration after the onset of the hand-raising animation than before. However, this effect was not found to be affected by the amount of hand-raising animation. Moreover, the results showed that students' saccadic behavior was significantly affected by the hand-raising animations: students showed significantly longer mean saccade duration, mean saccade amplitude, and more saccades after perceiving hand-raising behavior than before. And this effect was affected by the amount of animation, with students showing a greater increase in saccade amplitude in the 20% condition than in the other three conditions (35%, 65%, 80%). This suggests that students show a greater increase in their visual search behavior when the amount of animations is low.

In addition to the cognitive and visual responses elicited by the avatar animations, it was found that students exhibited significantly different behaviors toward the salient object-of-interests (OOI; i.e., the virtual teacher, the virtual peer learners, and the screen displaying the instructional content) in the VR classroom. First, it was found that after the hand-raising animations began, students initially focused their attention more on the virtual peer learners than on the instructional content. In addition, students' TTFF (i.e., the time to the first fixation) was significantly shorter for the peer learners than for the instructional content. Second, it was found that students' dwelling behavior on these three OOIs was significantly affected by the hand-raising animation. Specifically, students showed significantly longer dwell time on the peer learners after the hand-raising animation began than before. Accordingly, it was found that students' dwell time on the instructional content decreased significantly after virtual peer learners raised their hands. In addition, the magnitude of change in dwelling behavior towards instructional content was found to be influenced by the number of hand-raising animations, with students showing the greatest decrease in their attention towards instructional OOIs in the 80% condition than in the other three conditions (20%, 35%, and 65%). Last, the number of virtual peer learners perceived by students was significantly greater after the onset of hand-raising animation than before, and this increase was more significant in the 80% condition than in the other three conditions.

### 3.1.3 Gender Differences in Students' Eye Movements During CT Development in an IVR Classroom

**Hong Gao**, Lisa Hasenbein, Efe Bozkir, Richard Göllner, and Enkelejda Kasneci. Exploring Gender Differences in Computational Thinking Learning in a VR Classroom: Developing Machine Learning Models Using Eye-Tracking Data and Explaining the Models. In *International Journal of Artificial Intelligence in Education (IJAIED)*. 2022.

**Motivation**

With the growing popularity of user-friendly and open-source programming languages such as Python, Computational Thinking (CT) has already been incorporated into K-12 education to equip students with this 21st-century skill. However, gender differences in CT development continue to be observed, such as in students' interests and attitudes toward CT and in their CT skills as measured by student self-reports and learning outcomes. Measures related to the learning process that could shed light on how gender differences emerge during CT skill development have been little studied and should be investigated as they offer insights to improve CT development. Therefore, it is very important to understand learners' behaviors that may be associated with gender differences during CT development. VR classroom as an excellent tool to facilitate education can also be used as a tool to support students in developing CT skills. With advanced VR technologies, learners' behavioral data, such as head-tracking and eye-tracking during VR immersion, can be easily captured by VR devices, and such data can help explore underlying gender differences that are difficult to detect through post-hoc self-reported measures.

To understand how different gender groups differ in their behavior during CT development, a machine learning approach was proposed to predict gender information based solely on eye-tracking data alone in an IVR classroom. This offers advantages in several ways. First, the feasibility of the machine learning approach for gender prediction and the discriminative power of eye-tracking features for the machine learning models were investigated. Second, the underlying relationships between eye movements, which characterize students' visual attention and cognitive behavior during learning, and gender can be revealed through a state-of-art model explanation approach. This provides profound insights for improving CT development of specific genders to reduce gender disparities, and thus offers clues for designing personalized VR-based education tutoring systems.

**Methods**

In this study, the same data as in Section 3.1.1 were used, and detailed data processing and eye movement events detection are described in the Methods of Section 3.1.1. To build machine learning models for gender prediction based solely on eye-tracking data, features characterizing learners' behavior during the learning process were first extracted. A sliding window approach was used for feature extraction, using a range of window sizes from 10 seconds to 100 seconds with a step of 10 seconds. The window size was considered as a parameter and tuned during model training. A total of forty-three sensor features were extracted based on the processed tracking data, including features related to head movements, eye movements (including fixation and saccade), pupil diameter, and OOIs (i.e., the virtual teacher, the virtual peer learners, and the screen displaying the instructional content).

Five machine learning models were created for comparison in a binary classification task, including Support Vector Machine (SVM), Logistic Regression, and three ensemble machine learning models, namely Random Forest, eXtreme Gradient Boosting (XGBoost), and Light Gradient Boosting Machine (LightGBM). A nested cross-validation approach was used to optimize the hyperparameters of the models. For each iteration, the dataset was split into a training (80% × 80%), a validation (80% × 20%), and a test set (20%), using a subject-dependent and stratified manner. In the training process, 5-fold cross-validation was applied for training. To avoid overfitting the model, all data splits were participant-dependent, meaning that all data samples from the same participant were to remain in one dataset (i.e., either the training, validation, or test set). In addition, cross-validation was performed five times, each time randomly selecting different groups of participants as the test set to eliminate the participant-group effect and thus increase the generalizability of the model. For the final model performance evaluation, the models were re-trained on the unified training and validation set and tested on the test set to generalize the models to unseen data. Test results of accuracy and F1 score were reported. To investigate which particular features contribute to the machine learning models and to uncover the underlying relationships between eye-tracking features and genders, the SHapley Additive exPlanations (SHAP) approach was applied for post-hoc model explanation. The machine learning approach to predicting gender is shown in Figure 3.3.

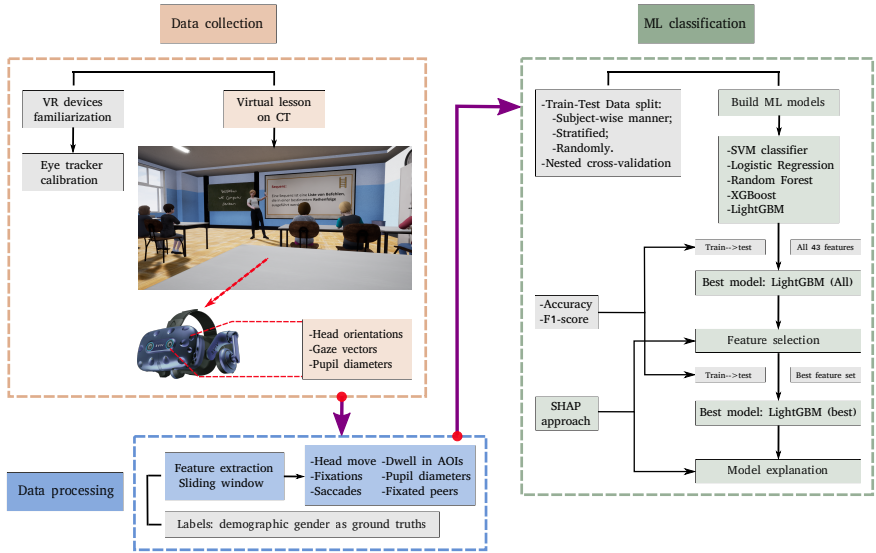

Figure 3.3: The machine learning approach to predicting gender in CT development in the IVR classroom.

**Main Results**

**Model performance in gender prediction**    All five classification models trained with eye-tracking features extracted with different time window sizes performed better than the chance level (50%). In particular, a window size of 60 seconds provides an optimal tradeoff for gender classification compared to time windows of other lengths. Of all five classification models, LightGBM trained on all features performed best with an average accuracy of 65.5%, followed by SVM with an average accuracy of 63.8% and XGBoost with an average accuracy of 63.5%. To improve the performance of the model (LightGBM) in gender classification, feature selection was performed based on the SHAP feature importances. The LightGBM models were trained with a set of selected features: Each time, the least important features according to SHAP were removed. The results showed that the LightGBM model trained with the 24 most important features performed the best with an average accuracy of 70.8%. With the feature selection approach, the model improved accuracy by more than 5%.

**Model explainability**    The SHAP results showed that the features of head movement rate and mean fixation duration on the virtual teacher provide the most information for the gender classification model, followed by the maximum fixation duration on the screen displaying instructional content, the number of virtual peer learners perceived by the students, and the dwell time on the virtual peer learners. In contrast to head movement and fixation (especially fixation on OOIs), features related to pupils and saccades, which are correlated with cognitive load and visual search behavior, were found to contribute less to the classification model. In addition, the SHAP approach showed how each feature contributes to the model outputs. Among the features that contribute the most, head movement rate was found to have a positive impact on the model outputs, i.e., a higher value of the head movement rate than its respective average value drives the classification into the model output of class-male. Conversely, the features of fixation duration on the virtual teacher and the screen, which were found to be informative, were found to have a negative impact on the model outputs, i.e., a higher value of those fixation features than their respective average value drives the classification into the model output of class-female. Similar to the head movement rate, features such as the number of peer learners perceived by the students and the dwell time on the peer learners were also found to have a positive impact on the model outputs.

### 3.1.4   A Machine Learning Approach to Predicting Teacher Expertise based on Sensor Data from an IVR Classroom

**Hong Gao**, Efe Bozkir, Philipp Stark, Patricia Goldberg, Gerrit Meixner, Enkelejda Kasneci, and Richard Göllner. Predicting Teacher Expertise Based on Fused Sensor Data from an Immersive VR Classroom and Explainable Machine Learning Models.  In *Proceedings of the 2023 CHI Conference on Human Factors in Computing Systems (CHI'23). Work under revision for CHI'23.*

**Motivation**

Teachers' professional vision in the classroom is considered one of the most important skills for remaining competent.  It requires teachers to visually perceive relevant events in the classroom, such as disruptive student behavior, and to intervene effectively.  Eye-tracking, which provides an intuitive interface for examining human behavioral patterns and cognitive processes, offers a viable and practical avenue to assess teachers' professional vision in the classroom. Researchers have made significant progress in assessing teachers' perceptual behavior using eye-tracking technology. They found significant differences between expert and novice teachers, with expert teachers generally processing visual information in the classroom more quickly than novice teachers and being more accurate and effective at identifying disruptive student behavior, as indicated by eye movements of the teachers.

Investigating teachers' expertise in professional vision can lead to a deeper understanding of how effective classroom management is developed, and thus provide further insight for designing gaze-based training systems to assist novice teachers to develop their expertise. Most previous works in this research area have used pre-recorded classroom videos from the teacher's perspective as experimental stimuli, which could lead to unnatural and unrealistic teacher visual behavior since teachers are not actually teaching. VR has several advantages that provide an excellent opportunity to overcome these limitations, such as the ability to manipulate certain experimental conditions by preprogramming the virtual avatars and virtual environment configurations, and to overcome significant privacy concerns as no real videos of the students whom the teacher delivers a lecture to are recorded. Using novel technological approaches to collecting a comprehensive set of teacher behavioral data that could indicate teacher expertise in classroom management, including eye-tracking, head-tracking, and controller-tracking, the teachers' recognition and handling of disruptive events in the classroom are taken into account. As no previous research has investigated teacher expertise in professional vision in a VR context using eye-tracking technology, this work is encouraged as it can pave the way for evaluating teacher expertise in an interactive virtual environment using eye tracking and provide further valuable insights for developing teacher expertise in classroom management in VR-based systems.

### Methods

To explore how experts and novices differ in their ways of detecting disruptive events in the classroom, 25 novice teachers and 17 expert teachers were recruited to participate in the experiment of teaching in VR. An IVR classroom was set up to mimic a traditional classroom. Since the main goal is to investigate how teachers visual perceive the classroom, visual student avatars were pre-programmed with different attentive behavior and off-task behaviors. Teachers were instructed prior to the experiment to detect disruptive student behaviors while giving a presentation in the IVR classroom. During the VR immersion, teachers' behavioral data were recorded simultaneously, including their detection of disruptive student behavior and their handling of disruptive student behavior. On this basis, a machine learning approach was proposed for predicting teacher expertise in professional vision. In contrast to previous studies that used a limited number of eye movements based on an average measure [157, 158], this study used extensive eye-tracking data along with controller-tracking data throughout VR instruction. Specifically, teachers' pupil diameter, fixations, saccades, and relevant eye movements to object-of-interests (OOIs; i.e., virtual students) were used to capture teachers' visual attention, visual search, as well as underlying cognitive processing behavior [159] during the recognition and handling of disruptive student behavior.

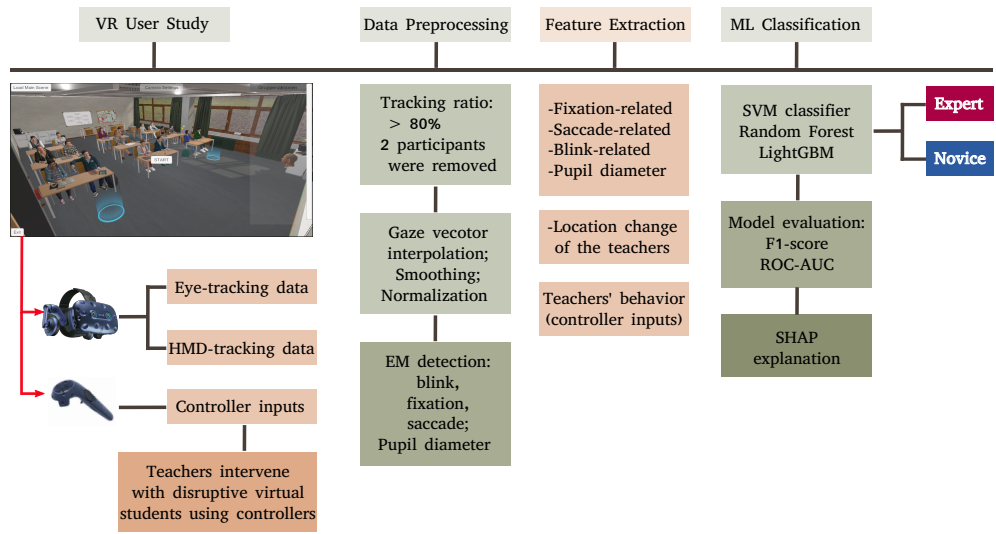

Figure 3.4: The machine learning approach to predicting teacher expertise in professional vision in the IVR classroom.

A large set of sensor features (mainly eye-tracking) were extracted from the teachers' behavioral data. The data samples were randomly split with a 80 (training):20 (test) ratio. For tuning the hyperparameters of the model, 5-fold cross-validation was performed. Strategies to avoid overfitting were performed as in the Methods of Section 3.1.3, including participant-dependent wise data split without regard to participants' identity. Three different classification models, SVM, random forest, and decision trees, were built and their performances were compared using the F1-score and the AUC score. In addition, statistical tests on teachers' behavioral features were performed to examine differences between novice and expert teachers, using the Mann-Whitney U test as a nonparametric test. Furthermore, the SHAP approach was applied for post-hoc model explanability. The most contributing features and how they contribute to the models can be revealed by SHAP. The machine learning approach was presented step by step in Figure 3.4.

### Main Results

**Differences in teachers' professional vision**    Several statistical differences in various behavioral aspects of teachers during classroom management were found between novice and expert teachers. Expert teachers showed significantly fewer fixations and shorter fixation duration throughout the classroom instruction, as well as shorter fixation duration in the OOIs. Teachers' saccadic behavior also differed significantly between experts and novices, with expert teachers exhibiting lower and smaller visual search behavior than novices, as evidenced by fewer saccades, shorter saccade duration and amplitude, and lower saccade peak velocity. In addition, experts were found to exhibit more blinks with shorter blink duration, and smaller pupil diameter. In addition, it was also found that teachers differed between expert and novice teachers in dealing with disruptive students in the classroom. Specifically, expert teachers detected more disruptive students than novice teachers, as indicated by controller input, and they had a longer visual information processing time for these disruptive students, as evidenced by a greater number of fixations and a longer fixation duration on C-AOIs (i.e., those disruptive students that teachers successfully identified with the controller).

**Model performance in expertise prediction**  All three models, including SVM, random forest, and decision trees built on fused sensor data were trained and compared for predicting teacher expertise. F1-score, ROC-AUC score, and ROC curve were used for model performance evaluation. Results showed that all three models can successfully predict teacher expertise. Random forest performed best with an F1-score of 0.78 and a ROC-AUC score of 0.77. The SVM classifier and LightGBM achieved slightly lower but still predictive performance, with an F1-score of 0.74 and a ROC-AUC score of 0.74, and an F1-score of 0.76 and a ROC-AUC score of 0.75, respectively.

**Model explainability**  The post-hoc model explanation results showed that all types of teacher behavioral features provide discriminative information for the model to distinguish teacher expertise, and their contributions are different. Specifically, teachers' visual attention behavior toward disruptive virtual students detected by teachers contributed the most to the classification model, including features such as the number of fixations and fixation durations (sum and maximum) on C-AOIs. These features are the three most important features. In contrast, teachers' general visual attention across the whole classroom did not contribute as much, for example, the general averaged number of fixations and fixation durations. Pupil diameter, which indicates teachers' cognitive processing load during classroom management, was also found to contribute strongly to the model, i.e., mean pupil diameter is the fourth most important feature. Moreover, teachers' direct reactions to the disruptive student behaviors, represented by the number of clicks they made on these disruptive behaviors, also contributed considerably to the model, but to a lesser extent than attention behaviors on the AOIs. On the other hand, saccadic features were found to contribute less compared to other eye-tracking features such as fixation and pupil diameter.

Teachers' behavioral features contributed differently to the model for distinguishing expert and novice teachers. Of the features that contributed greatly (i.e., fixations on the C-AOIs), a positive impact of these features on the model was found. In addition, the feature representing teachers' direct reaction to disruptive students, i.e., the number of clicks on the controller performed by teachers, was also found to have a positive impact on the model. This means that for these features, a feature value that is higher than the feature average contributes to the model output of class 1 (expert) and vice versa. Conversely, features such as pupil diameter and blink duration had negative impact on the model outputs, i.e., a feature value that is higher than the feature average contributes to the model output of class 0 (novice) and vice versa.

### 3.1.5   Conclusion

The works presented so far demonstrate the feasibility and effectiveness of eye-tracking as tool to study human behavior in VR-based educational scenarios (i.e., IVR classrooms) from two aspects, namely student learning and teacher training. On the one hand, students' learning behavior while immersed in the IVR classroom was studied from various aspects with eye-tracking technology in combination with machine learning techniques and explainable models. First, from the perspective of VR learning environment design, several classroom configurations that are relevant to student learning were investigated, namely, the seating position of students in the IVR classroom, visualization style of avatars (i.e., the virtual teacher and peer learners), and peer learners' social interaction behavior with the virtual teacher (i.e., hand-raising). Students' eye movements, including fixation and saccade, as well as pupil diameter were examined. The results showed that such configurations of the classroom significantly affected students in various aspects, especially in terms of students' visual attention and visual search behavior throughout the classroom. Such findings provide guidance for future studies on the design of effective VR classrooms by considering and evaluating different configurations that may affect students' behavior using eye-tracking.

The first work is a general investigation of students' eye-tracking behavior throughout the learning in the IVR classroom. Based on this, it is necessary and natural to move to the next step: a deeper understanding of how avatar animations influence student behavior during learning activities in VR, especially the hand-raising animation of virtual peer learners, which is a salient classroom event. The following second paper specifically investigated how students visually and cognitively respond to the social interactive activities between virtual peer learners and the virtual teacher (i.e., hand-raising) during learning in VR by using various eye-tracking metrics. Results showed that students' attention was drawn from the instructional content (i.e., the virtual teacher and the screen displaying instructional material) to hand-raised virtual peer learners, which was associated with increased cognitive load. Moreover, these attentional shifts differed between conditions with varying amounts of hand-raising animation. These results highlight the importance of considering the effects of avatar animations on users during learning when designing VR learning environments.

The immersive VR classroom was designed as a learning tool for students to learn computational thinking (CT), a subject in which gender disparity exists. Therefore, it is interesting to examine gender differences in CT skill development in a VR context, particularly to explore whether gender differences can be revealed by students' gaze information. This is therefore the main research goal of the third paper. Such an investigation could provide insights into the design of customized VR-based systems to support students' CT skill development. A machine learning approach for gender prediction based solely on eye-tracking data was proposed. A comprehensive set of eye-tracking features was extracted, including fixation, saccade, and pupil diameter. The trained models in the binary classification tasks achieved the best performance with an accuracy of over 0.7, suggesting that the eye-tracking data contain discriminative information about gender differences in CT skill development. Furthermore, post-hoc model explainability approach was conducted with SHAP. The most important features contributing to the model for classifying gender and their relationships with gender information were uncovered. This sheds light on what gaze behaviors specific gender groups exhibit during learning in VR, which further provides clues for tailoring assisting systems for specific gender groups.

Inspired by these previous studies, the research question of studying human behavior in virtual environments from the teachers' perspective is proposed. In the fourth paper, teachers' visual perceptual behaviors during classroom management were studied. Specifically, this study examined how expert and novice teachers differ in the way of they manage classroom events, based on their visual behaviors (i.e., eye movements) and their intervention actions toward disruptive students (i.e., teachers' ability to detect disruptive students as indicated by the number of clicks by the controller). Similar to the previous work on gender prediction, a machine learning approach was also proposed for teacher expertise prediction using features extracted from various teacher behavioral data during classroom management, including eye-tracking and controller-tracking data. The trained models showed the predictability of teacher expertise in classroom management, with random forest achieving the best performance of over 0.78. In addition to the differences in extracted features between experts and novices revealed by statistical tests, the post-hoc model explanation approach showed that these informative features contributed differently to the model outputs with different importance. This uncovered the question of how experts managed the class, such as how they direct their visual attention to disruptive students and what visual strategies they employ. These findings provide guidance for the design and development of VR-based training systems that utilize gaze-based interfaces to help novices develop visual strategies that experts employ during classroom management.

## 3.2 Human Behavior in VR Locomotion

### 3.2.1 The Impact of VR Locomotion Techniques on Eye Movements

**Hong Gao**, Lasse Frommelt, and Enkelejda Kasneci. The Evaluation of Gait-Free Locomotion Methods with Eye Movement in Virtual Reality. In *2022 IEEE International Symposium on Mixed and Augmented Reality (ISMAR-Adjunct'22)*.

**Motivation**

With the proliferation of commercially available VR head-mounted displays (HMDs), VR is becoming increasingly prevalent in the entertainment industry, driving the design and development of VR entertainment applications. Navigation is considered an important feature of VR applications (especially VR games) that enables users to move efficiently and indefinitely in large virtual environments and can be achieved through the use of VR locomotion techniques. A good locomotion technique should not only perform its intended function (i.e., enable the users' effective movements in the virtual environment) but also not interfere with the users' main task in VR by providing a poor experience or causing an additional cognitive load. Therefore, it is important to evaluate these aspects of locomotion techniques before they are deployed. Previous work has addressed the evaluation of VR locomotion techniques from various perspectives, including but not limited to user experience (e.g, the usability of the locomotion technique), presence, motion sickness, and other post-hoc surveys such as user preferences. However, previous evaluation methods have mostly used post-hoc questionnaires for surveys or quantification, while real-time user behavior data (e.g., visual attention and cognitive load) have rarely been assessed, which may reveal deeper subconscious thoughts of users that are difficult to capture with users' self-reports. Eye-tracking is an excellent tool for this purpose. In particular, eye movements have been found to correlate with cognitive load, which could provide a viable avenue for exploring users' cognitive load behavior during VR locomotion. To address this research gap, an approach to assess five gait-free (i.e. no gait movement is required) locomotion techniques using eye movements was proposed.

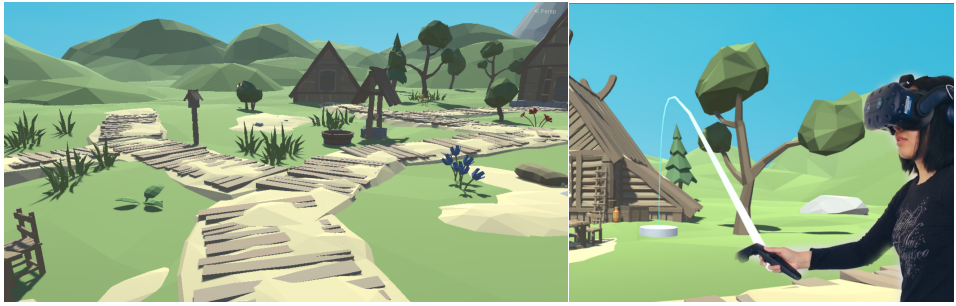

Figure 3.5: The virtual environment designed for VR locomotion.

**Methods**

The purpose of this study is to investigate the feasibility and effectiveness of using eye-tracking technology to measure users' cognitive processing load and experiences during VR locomotion and to explore whether and how eye movements are related to user experiences. Rather than proposing new locomotion techniques, five of the most commonly used and representative gait-free locomotion techniques were selected as test subjects, including arm swinging, dash, grappling, joystick, and teleportation. All five locomotion techniques were implemented using the Unity engine and the scripts were written in C#. A VR locomotion user study was conducted in a within-subjects design in which users navigated a designed virtual environment using five different locomotion techniques to complete a simple and easy search & collect task. Fifteen users took part in the experiment. During VR locomotion, users' eye-tracking data were recorded with the eye tracker integrated into the HMD. The virtual environment designed for VR locomotion is shown in Figure 3.5. The eye-tracking data preprocessing and algorithms used for eye movement event detection suitable for a VR context are the same as described in Section 3.1.1, with parameters adjusted accordingly. Various eye movement metrics related to cognitive load were used as dependent variables, including fixation, saccade, and blink. To explore whether the effects of locomotion techniques on users' cognitive load were reflected in eye movements, statistical tests of the these eye movements metrics were performed to compare among the five locomotion conditions.

**Main Results**

Three types of eye movements that correlate with cognitive load were examined, including blinks, fixations, and saccades. The effects of locomotion techniques on users' cognitive load, as reflected by eye movements, were found to vary.

Users' blinking behavior was found to be significantly affected by locomotion techniques. It was found that users exhibited a significantly higher blink rate in the joystick condition than in the arm swing and dash conditions. In addition, a significantly higher blink rate was also observed in the teleportation condition than in the dash condition. No significant differences were observed between grappling and the other locomotion conditions. Two fixation-related measures were evaluated, namely fixation rate and fixation duration. It was found that users in the arm swing condition had significantly lower fixation rates than in the joystick and teleportation conditions. Conversely, fixation duration was found to be significantly higher in the arm swing condition than in the joystick and teleportation conditions. Saccadic measurements, i.e., saccade amplitude, also differed significantly between some locomotion conditions. Users showed significantly greater saccade amplitude in the joystick condition that in the arm swinging condition. Moreover, a significantly greater saccade amplitude in the teleportation condition than in the arm swinging and grappling conditions were observed.

### 3.2.2 Predicting User Experience with VR Locomotion Based on Eye Movements

**Hong Gao** and Enkelejda Kasneci. Eye-Tracking-Based Prediction of User Experience in VR Locomotion Using Machine Learning. In *Computer Graphics Forum*. 2022.

**Motivation**

Previous literature has primarily focused on the development of new locomotion techniques to meet the requirements of various VR applications. However, the evaluation of locomotion techniques is also very important, which provides deep insights for improving VR locomotion. In particular, user experience is highly valued in this research area because it provides direct feedback on the usability of the locomotion method and how users feel using the locomotion technique. To date, most studies on VR locomotion (or on the design of VR-based systems) have examined user experience using methods such as observational data during VR locomo-tion (or during interaction with VR systems), post-hoc surveys, and questionnaires. However, such methods may lack deeper insights into users' unconscious thoughts and behaviors during locomotion experiences.

With this in mind, alternative methods for evaluating user experience with VR locomotion should be explored, to which eye-tracking technology can contribute. As discussed in Section 3.2.1, users' eye movements have been found to reveal the underlying cognitive processing behavior of users during VR locomotion and are further correlated with user experience. These earlier findings inspire the idea of investigating the feasibility of predicting user experience with VR locomotion using eye-tracking data alone. Moreover, post-hoc model explanations may further reveal the underlying correlations between eye-tracking metrics and the model outputs, i.e., user experience level. Such an approach provides a potential avenue for future studies that evaluate user experience using eye-tracking technology not only in VR locomotion, but also in other VR contexts where user experience needs to be evaluated to improve VR system design. It can also provide deep insights into predicting real-time user experiences in interactive VR systems for educational or entertainment purposes. With real-time feedback on user experiences, interactive systems can provide users with tailored and optimal experiences by adjusting system settings accordingly.

**Methods**

A new machine learning approach was proposed for predicting user experience with VR locomotion based on eye-tracking alone. Data from the VR user study presented in Section 3.2.1 were used. The eye-tracking data samples of all the trials were labeled L-EU (i.e., low user experience) and H-UE (i.e., high user experience) based on the ground truth of user experiences obtained from the post-hoc user experience questionnaire. Various eye-tracking features were extracted from time-series sensor data using a sliding-window approach similar to that used in Section 3.1.3, including pupil diameter, fixation, and saccade. Based on the extracted features, random forest classification models were built to solve a binary classification problem. Dataset was randomly divided into a training set (80%) and a test set (20%). For model training, 5-fold cross-validation was performed to tune the hyperparameters of the random forest models on the training set. Strategies to avoid overfitting were performed as in the Methods of Section 3.1.1, including participant-dependent wise data split without regard to participants' identify.

In addition, the SHAP approach was applied for post-hoc model explanation. First, SHAP can help gain insights into the feature importance and identify the most informative eye-tracking features for the machine learning models to predict user experience. Second, SHAP can reveal underlying relationships between eye movement and user experience level (i.e, L-UE, H-UE) by demonstrating what impact an individual feature has on model outputs. To further validate the SHAP explanation results, the statistical tests for comparison between the two classes were performed on the feature variables that can be interpreted as associated with different human behaviors (e.g., cognitive load, visual attention, and visual search) during VR locomotion.

**Main Results**

**Model performance in user experience prediction** Classification results showed that the random forest models trained based on eye-tracking features extracted with different window sizes were able to classify users into L-UE and H-UE categories with average accuracies above 0.62. In addition, the model trained on features extracted with a $20s$ time window performed best with an average accuracy of 0.71, precision of 0.72, recall of 0.71, and F1-score of 0.71.

**Model explainability** The SHAP results showed that three pupil-related features contributed the most to the classification model for user experience level prediction. It is worth mentioning that these three most important features all had a positive impact on the model, meaning that the feature value higher than the feature average drives the classification into the prediction output of class-1 (i.e., L-UE). The feature maximum saccade amplitude was found to have a large contribution and a negative impact on the model, meaning that the feature value higher than the feature average drives the classification into the prediction output of class-0 (i.e., H-UE). Compared to pupil- and saccade-related features, fixation-related features tend to be less informative in classifying user experience levels.

**Statistical findings of eye-tracking features** The results showed some significant differences between the two groups in the most important features, i.e., pupil-related features. In particular, both the mean pupil diameter and the mean pupil diameter during fixations were significantly larger in the L-UE group than in the H-UE group, indicating a higher cognitive load of users with low user experience. In addition, several significant differences were found in saccadic metrics, including saccade rate, mean of saccade duration, amplitude, and peak velocity. These saccadic feature variables had significantly higher values in the H-UE group than in the L-UE group.

### 3.2.3 Conclusion

In Section 3.2, the study of human gaze behavior during VR locomotion was presented based on two works. First, it was investigated how locomotion techniques affect users' visual perception and cognitive behaviors during navigation in the virtual environment using eye-tracking technology. A locomotion study was designed for users to experience five common gait-free locomotion techniques. Eye movements indicating users' visual attention and cognitive load were analyzed, including fixation, saccade, and blink. The different locomotion techniques were found to affect users' cognitive behavior significantly differently, with the joystick eliciting the least cognitive load in users. This study demonstrates the effectiveness and feasibility of evaluating locomotion techniques using objective eye-tracking data as a proxy to study human behavior during VR locomotion. Eye-tracking revealed the underlying unconscious human behavior that is not easily captured by questionnaires. This study provides a feasible and practical way to evaluate VR locomotion techniques or even other VR-based applications where it is important to provide a good experience for users.

Since eye movements have been found to be indicative of users' cognitive load, which is interrelated with the user's experience during VR locomotion, it is interesting to further investigate whether such such eye-tracking data is informative of the user experience level. This is the research goal of the second paper in Section 3.2. A machine learning approach was proposed for predicting user experience based on a variety of eye-tracking features. The trained classification models achieved a performance of over 0.7, indicating the informative power of eye-tracking features in determining the user experience level.

In summary, this work provides a viable user experience assessment tool for future studies, especially when new locomotion techniques are proposed, and can be extended to other VR research that aims to provide a good experience for users or involves system assessment and improvement. The ultimate goal would be a system that can detect the user experience level in real-time to offer users a tailored and optimal experience.

## 3.3 Conclusion

The insights from the education and entertainment domains are interrelated and not separate. On the one hand, the findings on individual differences in the educational domain can also be applied to the entertainment domain to enable better design of VR entertainment applications that provide tailored experiences to different users with different needs. On the other hand, VR locomotion techniques can also be integrated into VR educational applications to provide more opportunities for interactive and innovative learning and teaching. For example, teleportation can be a promising method for VR educational applications that require users to move around to perform various learning and teaching activities, such as free group discussion in class or effectively assisting students by teachers effectively moving around to students. Thus, this work makes an important contribution to the field of VR applications used on a daily basis.

# 4 Discussion & Outlook

This chapter discusses the main findings of the papers included in this dissertation to explain how human behavior in VR is assessed by eye-tracking, following the structure of the major contributions Section (Section 3). Section 4.1 presents the discussion of the main findings from the paper presented in Section 3.1. More specifically, it discusses the assessment of learners' eye gaze that reveals different human behaviors in IVR classrooms, including significantly different student responses to IVR classrooms with different configurations (Section 4.1.1), the effects of avatar animations on students' visual attention and cognitive behavior (Section 4.1.2), gender differences in CT development in the IVR classroom (Section 4.1.3), and the assessment of teacher expertise in classroom management in an IVR classroom using eye tracking and machine learning (Section 4.1.4). Section 4.2 discusses the main findings from the work presented in Section 3.2, i.e., the assessment of users' eye-tracking behavior during VR locomotion. Specifically, it discusses how eye movements indicate cognitive load and user experience during VR locomotion (Section 4.2.1) and how informative eye movements are in machine learning models for predicting user experience with VR locomotion (Section 4.2.2). An outlook on how eye tracking facilitates the human behavior research in both educational and entertainment VR scenarios is provided in Section 4.3.

## 4.1 Eye Tracking Assessment in IVR Classrooms

### 4.1.1 Student Behavior Differs in IVR Classrooms With Different Configurations

IVR classrooms that mimic traditional classrooms offer learners authentic and realistic learning experiences, in particular, the inclusion of the accompaniment of virtual teachers and virtual peer learners further enhances learners' immersion and engagement [12, 160]. This work conducted a systemic investigation of various aspects of the design factors of an IVR classroom: the sitting position of students, the visualization style of avatars, and the performance level of social counterparts (i.e., the hand-raising behavior of virtual peer learners). In the field of VR research, eye movement analysis is still a challenge, even though advanced VR HMDs provide easy-to-acquire eye tracking data recorded with the integrated eye tracker. Only raw eye-tracking data is available and there are no well-established algorithms for detecting eye movement events such as fixation and saccade. In this work, an algorithm suitable for VR contexts was proposed for post-experimentally detection of eye movement events. Thus, students' subconscious visual perceptual and cognitive processing behaviors in IVR classrooms with different configurations was assessed by calculating various eye-tracking metrics such as pupil diameter, fixation, and saccade.

Students' position in the classroom affects their visual perception behavior, as indicated by various eye movement variables. Students who sat in the back row in the IVR classroom had longer fixation durations than students who sat in the front row, indicating that students in the back row needed more time to process visual information. This could be due to the fact that students in the back row had a smaller field of view and more visual information. Therefore, students in the back rows have difficulty processing the instructional content, resulting in a longer fixation time. When sitting in the back row, they are also exposed to more visual peer learners exhibiting social interactive behaviors (i.e., hand-raising), which affects students' learning experience [40]. Such visual information from visual social counterparts diverted students' attention from the instructional content. Students' visual attention behavior (indicated by fixations) is interrelated with their visual search behavior (indicated by saccade). Students in the back row exhibited smaller saccade amplitude and thus shorter saccade duration than those in the front row. This is due to the fact that in the back row, the size of the instructional content in students' view is smaller due to the long distance than in the front row, so students shift their attention less to perceive instructional information. These findings have important implications for future studies using IVR classrooms as educational tools: Sitting position in the VR classroom influences students' behavior and should be considered by educators depending on the learning objective. If students attend a teacher-centered lecture where attention to the instructional content is required, they should be seated in the front row of the VR classroom so that they can easily perceive the instructional information. On the other hand, if interactive information from social counterparts is a very important factor during learning, e.g., in a discussion-dominated lecture where virtual peer learners raising their hands play an important role, it is better to seat students in the relatively back row in the VR classroom.

Avatar visualization styles are not a factor in the real classroom, but a special factor in VR that affects users' experience [161]. Theoretically, it was expected that learners immersed in an IVR classroom with cartoon-style avatars would exhibit shorter fixation and longer saccade. First, the larger size of the avatar makes it easier for students to perceive social information from virtual peer learners, resulting in a shorter visual processing time, i.e., shorter fixations. Second, the larger avatars located in front of students' gaze may block students' view of the instructional content to some extent, resulting in less focus on the instructional content thus shorter fixations than when students are seated in the frontal position of the classroom. However, an opposite visual behavior of students was observed. Students showed longer visual processing time and slower visual search behavior when immersed in the IVR classroom with cartoon-style avatars than with realistic-style avatars. These findings are surprising, yet reasonable and explainable. In this study, since the students who participated in the VR lecture were six-graders, they might be more engaged with the cartoon-style avatars, which are more interesting to them. This explanation can be further supported by the findings on pupil diameter: Students showed significantly larger pupil diameter in the realistic avatar condition than in the cartoon avatar condition, which may indicate that students had a higher cognitive load when surrounded by avatars visualized in the realistic style. These results suggest that an optimal trade-off between different avatar styles should be made when designing VR learning environments.

The accompaniment of peer learners is one of the most important features of classroom education. Not only can peer learners act as companions, but their social interactive behavior with the teacher (i.e., peer learners' hand raising) is an important factor that affects students' behavior during classroom learning. And these effects depend on the amount of social information. Results showed that students responded differently under different hand-raising conditions. It was found that students exhibited a higher cognitive load when the performance and attendance levels of peer learners were relatively high. In addition, students showed more fixations in the condition in which 65% of peer learners raised their hands than in the 80% condition. This is not theoretically expected, as it was assumed that students would show more extreme gaze behavior in the extreme hand-raising condition. Regarding the hand-raising of peer learners, it can be concluded that such social behavior had an impact on students and should be considered in the design of the VR classroom, however, no logic was found and the extent of the impact of this manipulation is relatively mixed and should be further investigated.

Overall, these findings demonstrate the effectiveness of eye-tracking technology for studying underlying human behavior that is difficult to capture with questionnaires. From a VR classroom design perspective, these findings show that the classroom configurations discussed in this study affect students' learning experiences differently and should be considered when designing effective VR classrooms.

### 4.1.2 Avatar Animations Affect Students' Visual Attention and Cognitive Behavior

Avatars in IVR classrooms are rendered with human-like animations to increase authenticity and immersion. In developing VR-based educational applications where avatar animations are intended to influence learners' learning experience, it is necessary to investigate how learners respond to these animated virtual persons, especially learners' visual and cognitive responses at the moment when the avatar animations occur. Such an investigation provides cues for a better and more effective IVR classroom. Hand-raising is a very salient behavior of students that contains important information about a student's (in this study, the virtual peer learner) performance and motivation that is also perceived by peers (in this study, the student participants) and substantially shapes peers learning experience in the classroom. Therefore, it is important to determine whether peer learners' hand-raising really works as it should in a VR context, especially considering that student's field of view in VR is slightly different from that in the real classroom, which may lead to differences in how students perceive hand-raising behavior in two different contexts. In this study, the effects of social interaction animations, i.e., hand-raising, of virtual peer learners on students' behavior were assessed using eye-tracking. It was found that students exhibited higher cognitive load after being exposed to the hand-raising animations, as reflected in pupil diameter. Students showed significantly longer visual information processing time AND more extensive exploratory behaviors after perceiving the hand-raising animations than before, as evidenced by longer fixation durations, more saccades, longer saccade durations, and larger saccade amplitudes. In addition, students' attention shifted significantly from the instructional content to the social interactive peer learners thus resulted in perceiving more hand-raised peer learners. Not only were students' responses to avatar animations noted, but it was also found that these student responses were related to the amount of animation.

The investigation in this study shows that students are indeed affected by peer learners' animations in different aspects. These findings imply that the effects of avatar animation should be considered by developers when presenting animated virtual humans to enhance immersion or for interaction purposes in VR scenarios. Overall, these findings have important implications for future studies aiming to create more effective, interactive, and authentic VR-based (learning) environments through the manipulation of avatar animations.

### 4.1.3 Eye Movements Reveal Gender Differences in CT Development in an IVR Classroom

VR-based educational applications facilitate skills training at low cost [27], such as the development of computational thinking (CT) skills [162]. When it comes to learning, one of the important questions is how to improve learning, especially in CT skill development where gender gaps are still observed [163]. Advanced HMDs record ecologically valid user behavioral data (e.g., eye tracking and head tracking) that reveal users' unconscious behaviors during the VR experience. The use of eye tracking technology in combination with machine learning techniques and explainable models can provide a pathway to understanding gender differences in the development of CT skills in VR and provide further insights into reducing gender gaps and improving learning.

Machine learning models based on sensor features can successfully predict gender differences with an average accuracy of over 0.7. A large set of features, including fixation, saccade, head movement, and pupil diameter, indicative of various human behaviors were extracted from student behavioral data recorded by VR devices using a sliding-window approach. These features captured students' visual attention, visual search, and cognitive processing behaviors during learning in the IVR classroom. In particular, students' behavior toward OOIs (i.e., the virtual teacher and screen, and virtual peer learners) that is highly associated with their learning activities were considered. The predictability of gender differences during CT development in an immersive VR classroom based on eye-tracking data demonstrates the effectiveness of eye-tracking data in revealing underlying gender differences.

Moreover, the post-hoc model explanation provided further information for understanding gender differences in CT development. The features that contribute most to machine learning models for gender prediction and the effects of these features on the model outputs were identified. Different gender groups were found to exhibit different attentional behaviors during CT development. The results suggest that the way students attribute their visual attention to the different classroom content (i.e., the virtual teacher and the screen are instructional content, and the virtual peer learners are social comparison information) reflects their gender information. For example, the model explanation approach reveals that students' head movements are the most informative features for predicting gender. It was observed that head movements had a positive impact on the classification model, suggesting that this feature may correlates with gender. With regards to the design of CT learning environments from a pedagogical perspective, this finding suggests that different gender groups may need different guidance in exploring the VR classroom. In addition, SHAP revealed that different gender groups differed in the way they allocated their attention to instructional content and social counterparts. These findings provide implications for offering tailored experiences for students to support their skill development in VR-based learning systems. For example, instructional guidance from the teacher is likely to support girls' CT learning, as they tend to focus more on the teacher than on virtual peer learners. The model explanation findings provide insights for future studies examining gender differences in a VR context in response to more specific aspects of the learning experience (e.g., critical time points in the lesson) and in other subjects, particularly those that typically exhibit gender differences. From a technical perspective, this approach offers a viable way to assess student behavior during learning and skill development in VR-based setups using eye-tracking technology in combination with machine learning techniques and model explainable approaches.

### 4.1.4 Teacher Expertise in Classroom Management Is Predictable Based on Sensor data

The three previous works extensively examined human behavior while immersed in a virtual educational environment to participate in learning activities from the student's perspective. However, the behavior of another important role in education, namely the teacher, has not been fully investigated, especially not with the help of eye-tracking. One of the areas of research is the assessment of teacher expertise, where novel technological approaches to collecting process data, especially eye-tracking, in combination with machine learning techniques and model explainability can provide new insights into ways to train novice teachers and develop expertise, and hence better exploit human resources.

Teacher expertise in classroom management was investigated using an IVR classroom design from the teacher's perspective. Teacher's behavioral data during immersive in such IVR classroom were recorded using VR devices, including eye movements and response behaviors toward disruptive virtual students (i.e., controller clicks on disruptive students). By fusing two types of sensor data into machine learning models for predicting teacher expertise, teachers' recognition and handling of disruptive events in the classroom are considered, filling the research gap that exists in studying teachers' classroom management with only limited gaze information. The novice-expert classification models based on different classifiers were found to be predictive of teacher expertise, with random forest performing the best at 0.77. These results show that predicting expertise based on sensor data collected in a virtual environment, particularly eye-tracking data, is feasible in the educational domain. Expertise detection has been investigated in various domains, but not yet in the domain of teachers' classroom management in a VR context. This study fills this research gap and provides further evidence for the discriminative power of eye-tracking in detecting expertise.

Previous work has employed the SHAP model explanation in the machine learning approach for predicting gender information during skill development in an IVR classroom, providing profound insights for a deeper understanding of gender differences as reflected in various eye movements, and thus providing specific guidance for improving the learning of specific gender groups (see section 3.1.3). Such a post-hoc model explanation is also necessary in the research on teacher expertise recognition, as it provides more information about how expert and novice teachers differ in the way they perceive classroom events. The results showed that both gaze behavior and teachers' handling of disruptive events provide discriminative information for the novice-expert model, but to varying degrees. The SHAP approach revealed that teachers' visual attention, especially towards disruptive students as indicated by fixations, contributes the most to the novice-expert model. In addition, the high contribution of pupil diameter suggests that teachers' expertise in classroom management may also be strongly reflected in their cognitive processing behavior, which is second only to the importance of teachers' attentional behavior toward disruptive students. It is noteworthy that the feature related to teachers' handling of disruptive classroom events in the classroom, as indicated by controller input, also proved informative of teacher expertise, although its informativeness was lower than that of visual perceptual behavior. Not only was the features' importance determined, but also their impact on the novice-expert model. Several features were found to verily affect the model outputs, which revealed underlying relationships between teacher behaviors (i.e., eye movements and actions toward disruptive students) and teacher expertise. These relationships were further supported by statistical analysis of these behavioral features.

Overall, the computational data-centric methods based on machine learning techniques and model explainability approaches demonstrate the predictability of educational expertise based on a rich set of sensor features, particularly eye movements, in VR settings. These findings provide valuable insights for future studies assessing teacher expertise in VR-based systems and offer further insight into the design of gaze-based training interfaces. With a better understanding of how expert teachers in the classroom perceive and intervene in disruptive student behaviors, it may be possible to support and train novice teachers to develop expertise by teaching them exhibiting expert behaviors.

## 4.2 Eye Tracking Assessment in VR Locomotion

### 4.2.1 Eye Movements Are an Indicator of Cognitive Behavior and User Experience during VR Locomotion

Researchers continue to bring new locomotion techniques for improving the navigation experience. This experience includes not only effective movement in virtual environments but also small additional effects for users, e.g., a lower cognitive processing load. Previous studies on locomotion have mostly evaluated locomotion techniques in terms of users' perceived experience, e.g., presence, usability and effectiveness of locomotion techniques, using post-hoc questionnaires or surveys. With state-of-the-art VR devices, users' real-time visual attention and cognitive processing behavior during locomotion in VR can be assessed using eye-tracking data provided by integrated eye trackers.

Various eye movements were found to be related to the human cognitive processing load. With this in mind, a new approach was proposed to assess common gait-free locomotion techniques using eye-tracking, including arm swing, dash, grapple, joystick, and teleportation. Eye movements, including blinks, fixations, and saccades were assessed as measures of users' cognitive processing load during VR locomotion. Significant differences in eye movements were found between the five locomotion conditions; in particular, joystick and teleportation had significantly different impacts on user behavior compared to the other three locomotion techniques. These two locomotion techniques elicited low cognitive load in users, as indicated by higher blink rate, lower fixation duration, higher fixation rate, and longer saccade amplitude. These findings are supported by previous work showing that joystick and teleportation provide a better experience for users.

These findings demonstrate the effectiveness and feasibility of eye movements as a proxy for studying human cognitive behavior during VR locomotion. Such an investigation provides several contributions and implications for the VR locomotion research field. It provides a viable way of evaluating locomotion techniques with regards to users' cognitive load using eye-tracking, filling the research gap of VR locomotion technique evaluation that relies primarily on subjective post-hoc reports from users. Eye-tracking revealing the underlying human behavior that occurs during VR locomotion provides important clues for the improvement of locomotion techniques by taking into account the user's cognitive load. It also offers further implications for future studies in the VR domain, where the user experience needs to be assessed, to use eye movements as a superior objective measure of human behavior during VR experience.

### 4.2.2 User Experience with VR locomotion Is Predictable Based on Eye Movements

In Section 4.2.1, it was discussed how eye movements can be used as a proxy for assessing user behavior, such as cognitive processing load, during VR locomotion. This previous work has shown that eye movements reveal human subconscious behavior that is also related to the user experience during VR navigation with five different locomotion techniques. This inspires a machine learning approach to predicting user experience with VR locomotion based on eye-tracking data. When developing a VR application, user experience is usually at the forefront of system evaluation, as it provides explicit insights into the usability of the system and users' subjective thoughts about it. This is also true for VR locomotion techniques. It is necessary and important to measure users' experience during VR locomotion to gain insights into the improvement of VR locomotion techniques. Previous work has investigated the assessment of user cognitive load during VR locomotion using eye movements (see Section 3.2.2). This inspires the idea of evaluating the user experience based on eye movements, since the user's experience is highly correlated with cognitive load. On this basis, a machine learning approach combined with the model explainability approach was proposed for predicting user experience with VR locomotion based on eye-tracking data collected with advanced VR devices in a VR locomotion user study.

The classification models for user experience level detection are predictive with an accuracy of over 0.7. This result demonstrates the discriminative power of eye-tracking features in distinguishing user experience levels. The success of the proposed machine learning approach provides an opportunity to evaluate locomotion techniques in terms of user experience, which provides further insights to improve locomotion techniques. This approach also provides the opportunity to evaluate user experience in VR-based interactive systems where the user experience needs to be assessed. In addition, the SHAP model explanation approach provides further information about how eye movements reveal the user experience level, and this is further supported by statistical analysis. The most informative features for the machine learning model to predict the user experience with VR locomotion were identified. Among all types of eye-tracking features, pupil diameter was found to be the most informative, having a positive impact on the classification model, i.e., a pupil diameter feature value higher than the feature average contributes to the model output of class-1 (i.e., low experience level). These findings suggest that users' experiences with VR locomotion are highly reflected in their cognitive processing behavior. Saccade features related to users' visual search behavior were also found to discriminate users into low and high experience groups. These saccade measures were found to have a negative impact on the classification model, i.e., a higher value of these features than the feature average contributes to the model output of class-0 (i.e., high experience level). These results are consistent with pupil diameters, as saccades were also found to be related to cognitive load, in a negative relationship. Overall, the effectiveness of eye movements in determining user experience provides deep insights into predicting user experience in real-time, which may be relevant and important for interactive VR systems or intelligent user interfaces for educational and entertainment purposes. By obtaining real-time feedback from users regarding their experiences with the systems they interact with, it might be possible for such systems to provide users with tailored and optimal experiences by adjusting system settings accordingly.

## 4.3 Outlook

VR technologies in combination with eye-tracking technologies are advancing human-computer interaction research by providing users with more natural and efficient ways to interact with virtual environments and thus improve user interaction performance. With advanced VR technologies, immersive and authentic yet controlled virtual environments can be created and used as research environments. Since state-of-the-art VR devices are consistently equipped with eye-tracking technology, fields of research that deal with the evaluation of human behavior and VR systems can be supported through the collection of process data, e.g., VR education and entertainment. In this dissertation, these two main directions of VR application were investigated together and provide profound insights for future work aimed at exploring human behaviors in virtual environments and providing effective VR applications.

In terms of VR education, this work used IVR classrooms designed for different types of users (i.e., learners and instructors) as examples to illustrate how VR supports classroom teaching and learning by examining users' eye-tracking behaviors during VR immersions. First, with VR technology, an IVR classroom that mimics traditional classrooms can be created for students as learning interfaces, with the animated virtual teachers and virtual peer learners rendered as social companions. While immersed in such a learning environment, it is important to evaluate student behavior to gain insights for improving classroom configurations. Eye tracking is an excellent tool for assessing various human behaviors during interaction with virtual environments. In this way, important classroom configurations related to student learning behaviors can be explored to achieve better system design. For example, it was found that students' visual attention and cognitive behavior differ significantly in different classroom configurations (i.e., students' sitting position in the classroom, visualization styles of avatars, and especially the social hand-raising information from the animated peer learners). This informs that these are important aspects that need to consider while designing an IVR classroom. Furthermore, machine learning techniques and explainable models provide further ways of exploring individual differences. Based on a large set of eye-tracking features that characterize various human behaviors, machine learning models can be trained for predicting individual differences, such as gender. The post-hoc model explanation approach further highlights the discriminative power of eye-tracking in detecting individual differences during learning in VR.

IVR classrooms can be designed not only for learning purposes but also for training purposes, e.g., to train teachers in classroom management. By manipulating different classroom configurations especially by rendering animated virtual students exhibiting disruptive behavior in IVR classroom, teachers' expertise in classroom management can be assessed. Since eye-tracking has already demonstrated its discriminative power in detecting individual differences, it offers great potential for assessing differences in teachers' professional vision during classroom management. Similarly, a machine learning approach can be performed in combination with model explanation techniques to gain a deeper understanding of teacher expertise as revealed by eye movements in VR settings. The predictive power of novice-expert models again demonstrates that eye-tracking contains discriminative information about how teachers manage classroom events during teaching in an IVR classroom scenario. In addition, the underlying relationships between eye movements and teacher expertise as revealed by model explanation approach offer further insights for teacher training, e.g., training novice teachers with expert visual perceptual strategies used in classroom management and thus developing their expertise.

In the field of VR entertainment, user experience is particularly important and should be the primary consideration when developing entertainment VR applications. Eye-tracking can contribute to this by revealing various subconscious human behaviors during VR immersion that are difficult to assess with post-hoc user reports (e.g., questionnaires and surveys). VR locomotion is one of the feature of a lot of VR gaming applications. With different locomotion techniques, users can effectively navigate virtual environments in different ways and for different navigation purposes. Since locomotion is generally not the primary task in VR, it is important to ensure that the user experience is not be compromised by locomotion techniques. The real-time visual and cognitive load of users in relation to their VR locomotion experience has not been previously studied and can be investigated with eye-tracking. In particular, eye movements have been used as an indicator of cognitive load, which opens a new avenue for assessing VR locomotion techniques from the perspective of user cognitive load. Moreover, user experience can be assessed based on a rich set of eye-tracking data using machine learning techniques. The post-hoc model explanation approach further demonstrates the effectiveness of eye-tracking in detecting user experience during VR locomotion. Such a machine learning approach makes important contributions to the research in VR entertainment. Similar to the educational field, eye-tracking can provide a novel avenue to assess human behavior during VR entertainment, thus providing insights for improving the VR entertainment application.

### 4.3.1 Ethical Considerations

In this work, all participants in each user study were informed with full transparent details about the experiments and the purpose of data use before data collection, which in this work is for research purposes only. Eye-tracking data were collected from participants with their consent and each study was approved by Ethics Boards of the University of Tübingen. All participants had the right to request deletion of their data from the study at any time. All data were recorded anonymously without recording the real names of the participants. Specifically, in this study, individual differences revealed by machine learning models, such as gender differences in the development of computational thinking skills, teachers' expertise in classroom management, and users' perceptual behavior during VR navigation, are fully protected and can only be analyzed by the researchers involved in the corresponding projects. No other persons can access the raw data without the consent of the participants.

However, with the increasing availability of sensing technologies integrated into VR devices, more data is being recorded by these VR devices, making it thus possible to infer individual information about users, such as preferences or skills. This was demonstrated throughout this dissertation based on the different applications covered in this work, which found individual differences from multiple perspectives, such as gender-related characteristics during learning, expertise skill during classroom management, and the level of expertise level during VR immersion. It is notable here, that eye-tracking data is an information source that can be used to reveal even more information about a user, most importantly biometric characteristics [164]. Therefore, special care must be taken with such data to not only avoid privacy invasion [165] but also to think of efficient means of user empowerment and self determination.

# A Human Behavior in IVR Classrooms

This chapter is based on the following publications:

1. **Hong Gao**, Efe Bozkir, Lisa Hasenbein, Jens-Uwe Hahn, Richard Göllner, and Enkelejda Kasneci. Digital transformations of classrooms in virtual reality. In *Proceedings of the 2021 CHI Conference on Human Factors in Computing Systems (CHI'21)*. Yokohama, Japan. 2021. doi: 10.1145/3411764.3445596.

2. **Hong Gao**, Lisa Hasenbein, Efe Bozkir, Richard Göllner, and Enkelejda Kasneci. Evaluating the Effects of Virtual Human Animation on Students in an Immersive VR Classroom Using Eye Movements. In *Proceedings of the 28th ACM Symposium on Virtual Reality Software and Technology (VRST'22)*. Tsukuba, Japan. 2022. doi: 10.1145/3562939.3565623.

3. **Hong Gao**, Lisa Hasenbein, Efe Bozkir, Richard Göllner, and Enkelejda Kasneci. Exploring Gender Differences in Computational Thinking Learning in a VR Classroom: Developing Machine Learning Models Using Eye-Tracking Data and Explaining the Models. In *International Journal of Artificial Intelligence in Education (IJAIED)*. 2022. doi: 10.1007/s40593-022-00316-z

4. **Hong Gao**, Efe Bozkir, Philipp Stark, Patricia Goldberg, Gerrit Meixner, Enkelejda Kasneci, Richard Göllner. Predicting Teacher Expertise Based on Fused Sensor Data from an Immersive VR Classroom and Explainable Machine Learning Models. In *Proceedings of the 2023 CHI Conference on Human Factors in Computing Systems (CHI'23)*. Hamburg, Germany. 2022. *Work under revision for CHI'23.*

## A.1 Digital Transformations of Classrooms in Virtual Reality

### A.1.1 Abstract

With rapid developments in consumer-level head-mounted displays and computer graphics, immersive VR has the potential to take online and remote learning closer to real-world settings. However, the effects of such digital transformations on learners, particularly for VR, have not been evaluated in depth. This work investigates the interaction-related effects of sitting positions of learners, visualization styles of peer-learners and teachers, and hand-raising behaviors of virtual peer-learners on learners in an immersive VR classroom, using eye tracking data. Our results indicate that learners sitting in the back of the virtual classroom may have difficulties extracting information. Additionally, we find indications that learners engage with lectures more efficiently if virtual avatars are visualized with realistic styles. Lastly, we find different eye movement behaviors towards different performance levels of virtual peer-learners, which should be investigated further. Our findings present an important baseline for design decisions for VR classrooms.

### A.1.2 Introduction

Recently, many universities and schools have switched to online teaching due to the COVID-19 pandemic. Online and remote learning may become more prevalent in the near future. However, one of the disadvantages of teaching and learning in such ways compared to conventional classroom-based settings is the limited social interaction with teachers and peer-learners. As this may demotivate learners in the long term, better social engagement providing solutions such as immersive virtual reality (IVR) can be used for teaching and learning. Next-generation VR platforms such as Engage[1] or Mozilla Hubs[2] may offer better social engagement for learners in the virtual environments; however, the effects of such environments on learners have to be better investigated. In addition to the opportunity to provide more efficient social engagement configurations, VR also enables building and evaluating situations that are difficult to set up in real life (e.g., due to the privacy-related concerns or current availability).

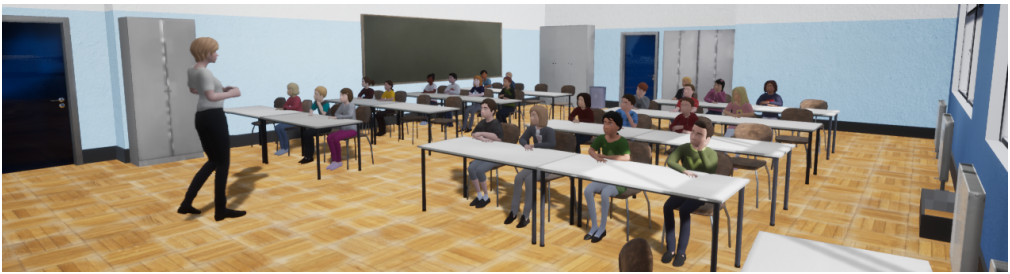

Figure A.1: Immersive virtual reality classroom.

---

[1]https://engagevr.io/

[2]https://hubs.mozilla.com/

While VR technology has a long history in the education domain [26, 166], the current availability of consumer-grade head-mounted displays (HMDs) allows for the creation of immersive experiences at a reasonable cost, making it possible to employ immersive personalized VR experiences in classrooms in the near future [11]. However, the digital transformations of classrooms reflect an important and critical step when developing VR environments for learning purposes and require further research. A unique opportunity to understand the gaze-based behavior, and consequently, attention distribution of learners in such VR settings is provided through the analysis of the eye movement of learners [167]. Since some of the high-end HMDs already consist of integrated eye trackers, it does not require extensive effort to extract eye movement patterns during simulations in VR. A thorough analysis of the eye movements allows us to infer information on the users going beyond the gaze position, for example stress [168], cognitive load [169], visual attention [76], evaluation and diagnosis of diseases [170], future gaze locations [171], or training evaluation [172]. In the virtual classroom, this rich source of information could even be combined with the virtual teachers' attention, similar to real-world classrooms [173, 174], to design more responsive and engaging learning environments.

In this study, we design an immersive VR classroom that is similar to a real classroom, enabling students to perceive an immersive virtual classroom experience. We focus on exploring the impact of the digital transformation from the classroom to immersive VR on learners by analyzing their eye movements. For this purpose, three design factors are studied, including sitting positions of the participating students, different visualization styles of the virtual peer-learners and teachers, and different performance levels of virtual peer-learners with different hand-raising behaviors. Figure A.1 shows the overall design of the virtual classroom. Consequently, our main contributions are as follows.

- We design an immersive VR classroom and conduct a user study to enable students to virtually perceive "interactive" learning.
- We analyze the effect of different sitting positions on learners, including sitting in the front and back. We find significantly different effects in fixation and saccade durations, and saccade amplitudes in relation to the sitting position.
- We evaluate the effect of different visualization styles of virtual avatars on learners including cartoon and realistic styles and find significantly different effects in fixation and saccade durations, and pupil diameters.
- We assess the effect of different performance levels of virtual peer-learners on learners by evaluating various hand-raising percentages, and find significant effects particularly in pupil diameters and number of eye fixations.

### A.1.3  Related Work

As head-mounted displays (HMDs) and related hardware become more accessible and affordable, VR technology may become an important factor in the educational domain, particularly given its provided immersion and potential for teaching [17, 175]. Various recent works on VR and education indicate that VR may offer significant advantages for learning and teaching. For instance, based on the post-session knowledge tests, both augmented and virtual reality (AR/VR) are found to promote intrinsic benefits such as increasing learners' immersion and engagement when used for learning structural anatomy [9]. In [176], the impact of VR systems on student achievements in engineering colleges was investigated by evaluating the results of post-quizzes and the results show that VR conditions present significant advantages when compared to no-VR conditions since students improve their performance, which indicates that VR can successfully support teaching engineering classes. Additionally, VR was also evaluated to help teachers develop specific skills that can be helpful in their teaching processes [177]. In addition to teaching and learning processes, another aspect under evaluation concerns the types of virtual environment configurations that are used not only for learning, but also for exploring immersion, motivation, and interaction. To this end, different types of VR setups have been studied. [11] introduced an immersive VR tool to support teaching and studying art history, which indicates, when used for high-school students, an increased motivation towards art history. [178] explored the possibility of using low-cost VR setups to improve daily classroom teaching by using a smartphone-based VR system. According to the evaluations using pre- and post tests, the proposed VR setup helps students perform better compared to traditional teaching using whiteboard and slides. Furthermore, HMD-based VR environment was studied in an elementary classroom for teachers to guide their students in exploring learning elements in immersive virtual field trips [119]. It has been concluded that students' motivation was enhanced after the virtual field trips. Overall, such works imply that while increasing motivation and engagement, different types of VR environments provide plenty of benefits and can be used to assist learning and teaching processes by providing users with immersive experiences.

One disadvantage of such VR and online learning tools is that learners' motivation and performance may be affected by lack of social interaction [179], peer accompaniment [180], or immersion [181]. Furthermore, realism in immersive environments can have various implications [182], related to both learning and interaction. To address these issues, several works have focused on how to provide more realistic and immersive environments. For example, [20] discusses the design of the VR environments for classrooms by replicating real learning conditions and enhancing learning through real-time interaction between learners and instructors. Furthermore, [12] constructed virtual classmates by synthesizing previous learners' time-anchored comments and indicates that when students are accompanied by a small number of virtual peer-learners built with prior learners' comments, their learning outcomes are improved. In addition to virtual peer-learners, the presence of virtual instructors may also have an impact on learning in VR. [183] investigated this and reports that learners engaged more with the environment and progressed further with the interaction prompts when a virtual instructor was provided. These works and findings indicate that the styles and types of virtual agents in the virtual environments may have several effects on students' attention and perception during immersion and should be taken into account. The evaluation of real-time visual attention towards similar configurations, which could be carried out using sensors such as eye trackers, may not only help to understand learning processes but also provide empirical insights about interactions during virtual classes for digital transformations of classrooms in VR.

From immersion and interaction point of view, video teleconferencing systems share similar goals with the VR classrooms as such systems enable people to experience highly immersive and interactive environments [184] and have been studied in the VR context as well. For example, [185] proposed a video teleconference experience using a VR headset and found that the sense of immersion and feeling of presence of a remote person increases with VR. Furthermore, different mixed reality (MR)-based 3D collaborative mediums were studied in terms of teleconference backgrounds and user visualization styles [186]. The real background scene and realistically constructed avatars promote a higher sense of co-presence. Low-cost setups were investigated also for real-time VR teleconferencing [187], as it was done for VR learning environments and it is found that it is possible to improve image quality using headsets in these setups. The possibility of having low-cost setups may become an important factor in the future when accessibility and extensive usage of everyday VR environments for learning [176] and interaction [20] are considered.

In general, while the visualization styles and rendering are considered to affect learners' perception and attention, in virtual learning environments particularly in IVR classrooms, other design factors are also important for attention-related tasks. For instance, [188] has studied the effect of being closer to the teacher, being in the teacher's field of view (FOV), and the availability of virtual co-learners in virtual classrooms. In particular, the authors found that students learn more if they are closer to the teacher and by being in the center of the teacher's FOV. In addition, when no co-learners or co-learners who have positive attitudes towards the lecture (e.g., looking at the teacher or taking notes) are available, students learn more information about the lecture instead of the virtual room. Gazing time was approximated according to the time students kept the virtual teacher in their FOVs; however, real-time gaze information was missing during the experiments. Exact gazing patterns and different eye movement events during learning are particularly needed for understanding moment-to-moment visual behaviors of students. In another work, [189] studied the effect of the sitting position on attention-deficit/hyperactivity disorder (ADHD) experiencing students in such classrooms and found indications that front-seated students are affected positively by this configuration in terms of learning. However, similar to [188], the authors did not have gaze information available but identified that the evaluation of eye movements may provide additional insights during learning, particularly in terms of real-time visual interaction, when learning and cognitive processes are taken into consideration. In addition, eye movements are also considered as choice of measurements to study visual perception during learning [64, 190]. [191] and [192] have studied attention measures and social interaction in similar setups using continuous performance tests and head movements, respectively. The latter work has used head movements as a proxy for visual attention and found that head movements shift between target and interaction partner. This finding partly supports the finding of [183] that the learners' engagement increases when a virtual instructor is presented. However, both works lack eye movement measurements. As also reported by [192], eye movements should be examined along with head movements to understand attention and interaction more in-depth, since eyes can move differently. In addition, [193] studied the relationship between performance, sense of presence, and cybersickness, whereas [194] examined attention, more particularly ADHD with continuous performance task in a virtual classroom. However, both works are more in the clinical domain, which are relatively different from an everyday classroom setup. [195] provides a general overview more from clinical perspective. Lastly, although has not been studied extensively in VR yet, peer-learners' engagement expressed by hand-raising behavior [41] may also affect the attention and visual behaviors of learners in the VR classrooms, which could be further studied.

In summary, while showing that VR could be a useful technology to support education, the aforementioned works primarily focused on the importance of used mediums and configurations, visualization styles, participant locations for visual attention, engagement, motivation, and learning of participants in VR classrooms. Yet, real-time and moment-to-moment interactions with the environment and visual behaviors of students in an everyday VR classroom setup were not studied in depth. Although obtaining such information in real-time is challenging, analyzing eye-gaze and eye movement features can provide valuable understanding into visual attention and interaction in a non-intrusive way, especially for designing such classroom configurations. For instance, long fixations can be related to the increased amount of cognitive process [196], whereas long saccadic behaviors are related to inefficient search behavior [197]. Furthermore, pupillometry is highly related to cognitive workload [128, 198]. Such information is also argued for consideration in IVR environments [199, 200]. In fact, when designing immersive VR environments for digital transformations of classrooms in virtual worlds, such features can be key to understand visual attention, cognitive processes, and visual interactions towards different classroom manipulations, which may also affect learning and teaching processes. To address this research gap, we study three configurations in an everyday VR classroom setup including different visualization styles of virtual avatars, sitting positions of participants, and hand-raising based performance levels of peer-learners by using eye movement features.

### A.1.4 Methodology

The main purpose of our study is to investigate the effects of digital transformations of the classrooms to VR settings on learners. Therefore, we designed a user-study to study these effects. In this section, we discuss the participant information, apparatus, experimental design, experiment procedure, measurements, data pre-processing steps, and our hypotheses. Our study and data collection were approved by the institutional ethics committee at the University of Tübingen (date of approval: 25/11/2019, file number: A2.5.4-106_aa) as well as the regional council responsible for educational affairs at the district of Tübingen.

### Participants

Participants were recruited from local academic track schools via e-mails and invitation letters. After obtaining written informed consent from both students and their parents or legal guardians, all students who indicated interest were admitted to the study. 381 volunteer sixth-grade students (179 female, 202 male), whose ages range from 10 to 13 ($M = 11.51$, $SD = 0.56$), were recruited to participate in the experiment. Due to hardware problems or incorrect calibration, data from 32 participants were removed. In addition, data from 61 participants were also removed due to eye tracker related issues including low eye tracking ratio (lower than 90%). Therefore, data from 288 participants (137 female, 151 male), whose ages range from 10 to 13 ($M = 11.47$, $SD = 0.51$), were used for evaluations. We had 16 different conditions in the experiment and the average number of participants for each condition was 18 ($SD = 5.3$). In addition to the actual study and data collection, we successfully piloted both our technical setup and the experimental workflow with 55 similar aged ($M = 11.35$, $SD = 0.52$) sixth-grade students (20 female, 35 male).

### Apparatus

In our experiments we employed HTC Vive Pro Eye devices with a refresh rate of 90 Hz and a field of view of 110°. The VR environment was designed and rendered using the Unreal Game Engine[3] v4.23.1. The screen resolution for each eye was set to 1440 × 1600. To collect eye movement data, we used the integrated Tobii eye tracker with a 120 Hz sampling rate and a default calibration with 0.5° − 1.1° accuracy.

### Experimental Design

The virtual classroom designed in our study has 4 rows and 2 columns of desks along with chairs, as well as other objects which typically exist in the conventional classrooms such as a board and display. In total, there are 24 virtual peer-learners sitting on the chairs. A virtual teacher standing in front of the classroom teaches a ≈ 15-minute virtual lecture to the students about computational thinking [201]. During the lecture, the virtual teacher walks around the podium. The virtual peer-learners and participants sit on the chairs throughout the lecture. The lecture has four phases including **(a) topic introduction** (≈ 3 minutes), **(b) knowledge input** (≈ 4.5 minutes), **(c) exercises** (≈ 5.5 minutes), and **(d) summary** (≈ 1.5 minutes). There are distracting behaviors from virtual peer-learners (e.g., raising hands, turning around) in the first, second, and third phases of the lecture.

In the beginning of the first phase, the teacher enters the classroom, stays in the classroom for a while, and then leaves for ≈ 20 seconds, giving participants the opportunity to look around and adjust to the virtual environment. The topic of the lecture is displayed on the board as *"Understanding how computers think"*. During the first phase, the teacher asks five simple questions to interact with the students. Some of the peer-learners raise their hands and answer the questions. In the second phase, the teacher explains two terms to the students, namely, the terms *"loop"* and *"sequence"*. These terms are also shown on the display. Then, the teacher asks four questions about each term and the peer-learners raise their hands to answer the questions. In the third phase, the teacher gives the students two exercises to evaluate whether or not they understand the terms correctly. For each exercise, the students have some time to think. Then, the teacher provides the answers for each exercise, and the peer-learners vote for the correct answer by raising their hands. In the last phase, the teacher stands in the middle of the classroom to summarize the lecture. No questions are asked in this phase; therefore, none of the peer-learners raise their hands.

---

[3]https://www.unrealengine.com/

Our study is in between-subjects design. The participants are located either in the front or back region of the virtual classroom. The participants that sit in the front of the virtual classroom have one row in front of them, whereas the participants that sit in the back have three rows in front of them. The visualization styles of the avatars have two levels as well, in particular cartoon and realistic. Lastly, the hand-raising percentages, which are intended to show the performance levels of the virtual peer-learners, have four different levels, including 20%, 35%, 65%, and 80%. Combining all, we have a 2 × 2 × 4 factorial design that forms 16 different conditions in total. Participants' views from back and front sitting positions, cartoon- and realistic-styled avatars are depicted in Figures A.2 (a), (b), (c), and (d), respectively.

**Procedure**

Each experimental session took ≈ 45 minutes including preparation time. We conducted the experiments in groups of ten participants by assigning each participant randomly to

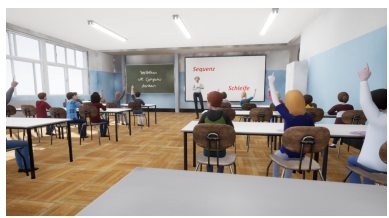

(a) Back sitting participant experiencing the VR classroom.

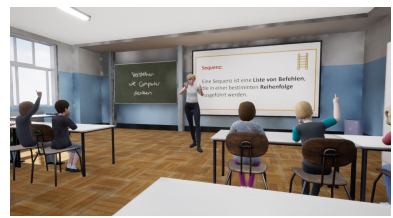

(b) Front sitting participant experiencing the VR classroom.

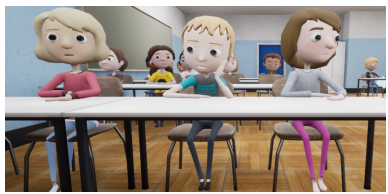

(c) Cartoon-styled avatars.

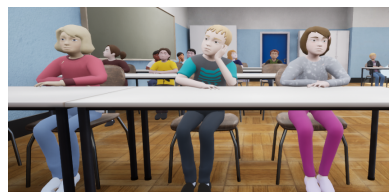

(d) Realistic-styled avatars.

Figure A.2: Views from the immersive virtual reality classroom.

one of the sixteen conditions. Before the data assessment took place at the participating schools, students were informed that they could drop out of the study at any time without consequences. After a brief introduction to the experiment and the data collection process, participants had the opportunity to acclimate with the hardware and the VR environment.

The experiment started with the eye tracker calibration. After calibration success, the experimenters pressed the "Enter" button to start the actual experiment and data collection process, wherein participants experienced the immersive virtual environment and the lecture. The experiments were supposed to be carried out in one session without breaks, mimicking thus a real classroom teaching session, lasting about 15 minutes. At the end of the experiment, the VR application displayed a message telling the participants to take off their HMDs. Lastly, participants filled out questionnaires about their experienced presence and perceived realism.

**Measurements**

For this work, our main focus was eye-gaze, head-pose, and pupil related activities of the participants as these are considered to be rich information sources, especially in VR. Fixations are the periods during which eyes are stationary within the head while fixated on an area of interest. Saccades, on the other hand, are the high-speed ballistic eye movements that shift eye-gaze from one fixation to another.

Using fixations, saccades, and pupil diameters, plenty of eye movement features are extracted. In this study, we extracted the number of fixations, fixation durations, saccade durations, saccade amplitudes, and normalized pupil diameters to analyze different conditions of the experiment. In the eye tracking literature, longer fixation durations correspond to engaging more with the object or increased cognitive process [196]. Fixation durations are mainly related to cognition and attention; however, it is argued that they are affected by the procedures that lead to learning and it is reported that fixation durations can be used to understand learning processes as well [202]. For instance, [203] has studied fixation patterns during learning in simulation- and microcomputer-based laboratory and found that simulation group had longer fixation duration, which means more attention and deeper cognitive processing. In addition to the fixations, longer saccade durations correspond to less efficient scanning or searching [197], whereas longer saccade amplitudes mean that attention is drawn from a distance [70]. Furthermore, a larger pupil diameter is related to higher cognitive load [204]. In addition, while being task dependent, [205] has indicated that pupil diameter measurements in high task load correlate with individual's performance. However, as pupil diameter values are also affected by the illumination, a controlled environment is needed to assess it. In our VR setup, the illumination is controlled across different conditions. Besides, a general overview of considering eye tracking as a tool to enhance learning with graphics is provided in [206].

Additionally, the self-reported presence and realism were assessed by questionnaires. The items in the questionnaires were based on the conceptualizations of [207] and [208] which were developed particularly to assess students' perception of the VR classroom situation. The Table A.1: Head and eye movement event identification thresholds.

| Event | Conditions for velocity ($v$) | Conditions for duration ($\Delta$) |
|---|---|---|
| Stationary HMD | $v_{head} < 7°/s$ | - |
| Fixation | $v_{head} < 7°/s$ and $v_{gaze} < 30°/s$ | $100\,ms < \Delta_{fixation} < 500\,ms$ |
| Saccade | $v_{gaze} > 60°/s$ | $30\,ms < \Delta_{saccade} < 80\,ms$ |

experienced presence and perceived realism were assessed via using a 4-point Likert scales ranging from 1 ("do not agree at all") to 4 ("completely agree") with nine (e.g., "I felt like I was sitting in the virtual classroom." or "I felt like the teacher in the virtual classroom really addressed me.") and six items (e.g., "What I experienced in the virtual classroom, could also happen in a real classroom." or "The students in the virtual classroom behaved similarly to real classmates."), respectively.

**Data Pre-processing**

As the raw eye tracking data collected from the VR device does not include fixations, saccades or similar eye movements, we first pre-processed the data to identify these events. Detecting different eye movements in the VR setup is a challenging task and different from the traditional eye tracking experiments that include equipment such as chin-rests, as participants have opportunity to move their heads freely in VR. In the eye tracking literature, Velocity-Threshold Identification (I-VT) method is used to classify fixations based on velocities [121]. In the VR context, [122] applied a similar method to detect eye movement events. We opted for a similar approach.

Before applying the I-VT, we first applied linear interpolation for the missing gaze vectors. After the interpolation, we identified the fixations when the HMD was stationary. However, the identification of saccades was not restricted by the HMD movement. The used velocity and duration thresholds for the HMD movement states, fixations, and saccades are depicted in Table A.1, where the velocities and durations are given as $v$ and $\Delta$, respectively. Unlike the fixations and saccades, the pupil diameter values are reported by the eye tracker. As raw pupil diameter values are affected by blinks and noisy sensor readings, we smoothed and normalized the pupil diameter readings using Savitzky-Golay filter [154] and divisive baseline correction using a baseline duration of $\approx 1$ seconds [155], respectively.

**Hypotheses**

We developed three hypotheses, each corresponds to one design factor.

- **Hypothesis-1 (H1)**: We hypothesize that the different sitting positions of the participants yield different effects on the eye movements. As the participants that sit in the front are closer to the board, displays, and the teacher, we assume that they can attend the virtual lecture more efficiently than participants in the back and have less difficulty extracting information about the lecture. However, as they have a narrower field of view, particularly towards the frontal part of the classroom, they need to shift their attention more than the participants sitting in the back.

- **Hypothesis-2 (H2)**: We hypothesize that different visualization styles of virtual avatars affect student visual behaviors differently. More particularly, as students are familiar with realistic styles in the conventional classrooms, we claim that compared to cartoon-styled visualization condition, they attend the scene shorter during fixations in the realistic-styled visualization setting as cartoon-styled avatars are more attractive to the students. Therefore, students engage with the environment more in the cartoon-styled visualization condition than in the realistic-styled condition.

- **Hypothesis-3 (H3)**: We hypothesize that different hand-raising percentages of virtual peer-learners can distinctively affect the behaviors of participants. Specifically, we anticipate that when relatively higher percentages of hand-raising levels are provided, such as 65% or 80%, the participant's cognitive load will be higher due to the fact that many of the peer-learners attend the lecture with a high focus. Similarly, participants have more fixations in the classroom in the higher hand-raising percentage conditions as a higher number of hand-raising percentage creates an opportunity for various attention and distraction points.

### A.1.5   Results

As we have three factors that form 16 different conditions, we applied 3-way full-factorial analysis of variance (ANOVA) by setting the level of significance to $\alpha = 0.05$ with Tukey-Kramer post-hoc test. For the non-parametric factorial analysis, we used the Aligned Rank Transform (ART) [156] before applying ANOVA procedures.

**Analysis on Different Sitting Positions**

Different sitting positions have an impact on the mean fixation and saccade durations, and mean saccade amplitudes. The mean fixation durations of the front and back sitting participants are illustrated in Figure A.3 (a). The participants that sit in the back have significantly longer mean fixation durations ($M = 222.6\,ms, SD = 14.57\,ms$) than the participants that sit in the front ($M = 218.75\,ms, SD = 13.11\,ms$), with $F(1, 272) = 6.7$, $p = .01$.

Both saccade durations and amplitudes are influenced by the sitting positions and are depicted in Figures A.3 (b) and (c), respectively. The results reveal significantly longer saccade durations in the front condition ($M = 50.23\,ms, SD = 1.7\,ms$) than in the back condition ($M = 47.9\,ms, SD = 2.62\,ms$), with $F(1, 272) = 73.76$, $p < .001$. Similarly, the mean saccade amplitude is significantly larger in the front condition ($M = 10.93°, SD = 1.54°$) than in the back condition ($M = 10.05°, SD = 1.38°$), with $F(1, 272) = 22.6$, $p < .001$.

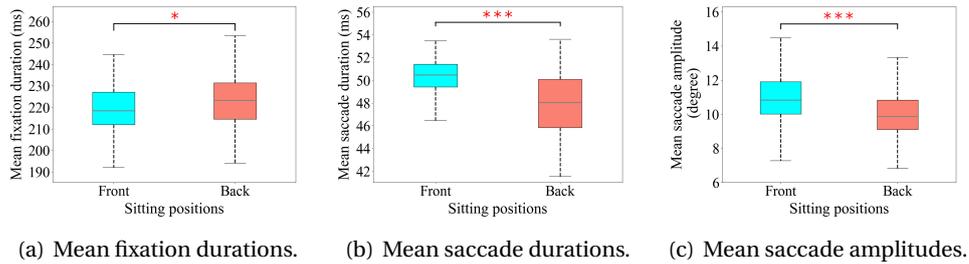

(a) Mean fixation durations.  (b) Mean saccade durations.  (c) Mean saccade amplitudes.

Figure A.3: Results for different sitting positions. Significant differences are highlighted with *
and *** for $p < .05$ and $p < .001$, respectively.

### Analysis on Different Avatar Styles

Different avatar visualization styles affect the mean fixation and saccade durations, and
pupil diameters. The results are depicted in Figures A.4 (a), (b), and (c), respectively. The
mean fixation durations are significantly longer in the cartoon-styled avatar condition ($M =
222.88ms, SD = 14.06ms$) than in the realistic-styled avatar condition ($M = 218.6ms, SD =
13.76ms$), with $F(1, 272) = 5.27$, $p = .022$. By contrast, the mean saccade durations are signifi-
cantly shorter in the cartoon-styled avatar condition ($M = 48.58ms, SD = 2.66ms$) than in the
realistic-styled condition ($M = 49.3ms, SD = 2.35ms$), with $F(1, 272) = 6.22$, $p = .013$.

The normalized mean pupil diameter, which reflects the cognitive load, is significantly
larger in the realistic-styled avatar condition ($M = 0.94, SD = 0.16$) than in the cartoon-styled
avatar condition ($M = 0.91, SD = 0.13$), with $F(1, 272) = 3.94$, $p = .048$.

### Analysis on Different Hand-raising Behaviors

The hand-raising behaviors of virtual peer-learners have significant impacts on the pupil
diameters and number of fixations as depicted in Figures A.5 (a) and (b), respectively. We
found significant effects on normalized mean pupil diameter values with $F(3, 272) = 4.78$,
$p = .003$. Particularly, mean pupil diameter in the 80% hand-raising condition ($M = 0.96, SD =$

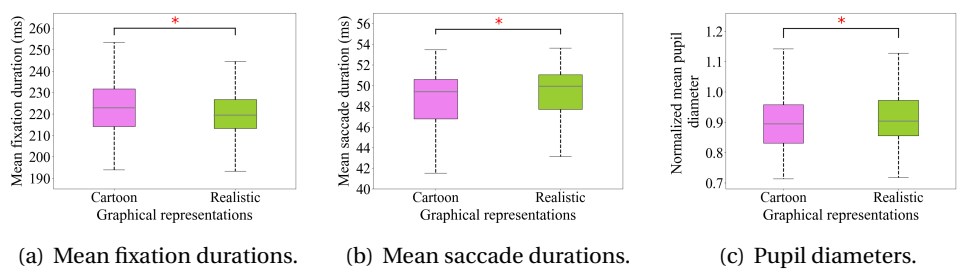

(a) Mean fixation durations.  (b) Mean saccade durations.  (c) Pupil diameters.

Figure A.4: Results for different avatar visualization styles. Significant differences are high-
lighted with * for $p < .05$.

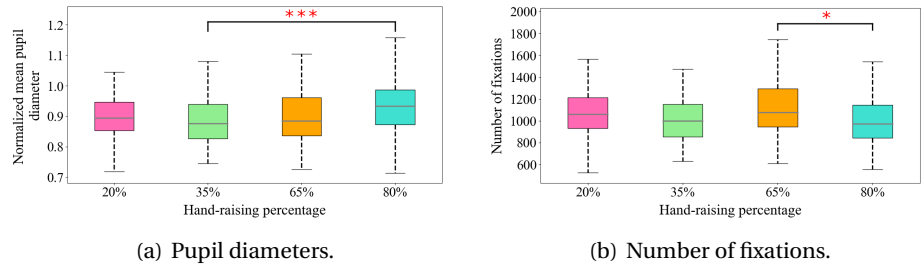

(a) Pupil diameters.

(b) Number of fixations.

Figure A.5: Results for different hand-raising percentages. Significant differences are highlighted with * and *** for $p < .05$ and $p < .001$, respectively.

0.16) is significantly larger than in the 35% hand-raising condition ($M = 0.9, SD = 0.12$), with $F(3, 272) = 4.78$, $p < .001$. In addition, we found significant effects on number of fixations with $F(3, 272) = 3.01$, $p = .03$. More specifically, there are notably more fixations in the 65% hand-raising condition ($M = 1112.92, SD = 245.07$) than in the 80% hand-raising condition ($M = 995.49, SD = 211.98$), with $F(3, 272) = 3.01$, $p = .028$.

### Analysis on Experienced Presence and Perceived Realism

We did not find significant effects of different experimental conditions on the self-reported experienced presence and perceived realism. Overall, the self-reported experienced presence and perceived realism values are in the vicinity of highest values with ($M = 2.91$, $SD = 0.55$) and ($M = 2.91$, $SD = 0.57$), respectively. These mean that even though we did not obtain statistically significant differences between conditions, the participants experienced high levels of presence and realism in the IVR classroom environment.

### A.1.6 Discussion

The results show that there are significant differences in the eye movement features between front and back sitting position conditions. Firstly, participants had longer fixations in the back sitting condition. This indicates that they had more processing time than the participants sitting in the front, which can be related to difficulty extracting information, similar to the relationship between task difficulty and mean fixation duration [209]. Secondly, the participants that sit in the front had longer saccade durations and amplitudes, which suggests that they needed to shift their attention more during the virtual lecture. While being located closer to the lecture content, longer saccade durations indicate that the participants sitting in the front had less efficient scanning behavior [197] during the lecture. We assume that this was due to the narrower field of view. These results support our **H1**. When designing virtual classes, these results should be taken into account, particularly when determining where students should be located in the classroom, depending on the context.

Our results show consequential effects in the eye movement features in different avatar style conditions. As mean fixation durations are longer in the cartoon-styled visualization condition, we assume participants found the cartoon-styled avatars more attractive and attention-grabbing. Therefore, their fixation behaviors were longer during the virtual lecture. On the contrary, the mean saccade durations are longer in realistic-styled conditions as the fixation durations are shorter, which is theoretically expected. Furthermore, the pupil diameters of the participants in the realistic-styled condition are larger, indicating that the cognitive load of these participants was significantly higher during the lecture, which is suggested by the previous work [204]. This is an indication that participants may have taken the lecture more seriously and in a more focused manner when the visualization was realistic. These findings support our **H2**. Rendering realistic-styled avatars may be computationally expensive depending on the configuration. Therefore, an optimal trade-off should be decided, taking the behavioral results into account while designing the virtual classrooms.

Furthermore, we observe significant effects in attention towards different hand-raising based performance levels of the peer-learners. Particularly, the pupil diameters of the participants in the 80% condition are significantly larger than the pupil diameters of the participants in the 35% condition. We interpret this to mean that when the performance and attendance level of peer-learners was relatively higher, the participants' cognitive load became higher, indicating that they might pay more attention to the lecture content. This partially supports our **H3**. In addition, a greater number of fixations are observed in the 65% condition than in the 80% condition. We claim that when almost all of the peer-learners participated in hand-raising behaviors during the lecture, participants acknowledged this information without significantly shifting their gaze. However, this claim requires further investigation. Manipulation of different hand-raising conditions may affect student self-concept [210], which should be further studied as well.

In our study, the interaction and perception in the immersive VR classroom were assessed mainly by using eye-gaze and head-pose information. However, while the virtual teacher and peer-learners talk in the simulations, no response or interaction by means of audio or gestures was expected from the participants. Combining visual perceptions and interactions with such data may provide additional insights particularly for better interaction design in VR classrooms. A future iteration can also evolve into an everyday virtual classroom platform where each virtual agent is actually connected to a real person, similar to in platforms such as Mozilla Hubs. To this end, further design settings such as optimal seating arrangement (e.g., U-shape, circle shape) in addition to the sitting positions should be investigated. Evaluation of similar configurations in online learning platforms such as Coursera[4], Udemy[5], or MOOCs[6] could provide additional implications for interaction modeling. Furthermore, gaze-based attention guidance can be considered for more interactive VR classroom experience and it can be achieved by fine-grained eye movement analysis focusing on short time windows instead of complete experiments. While being out of the scope of this paper, assessing learning outcomes and combining them with visual interaction and scanpath behaviors from immersive VR classroom could also offer insights for optimal VR classroom design.

---

[4]https://www.coursera.org/
[5]https://www.udemy.com/
[6]https://www.mooc.org/

### A.1.7   Conclusion

In this work, we evaluated three major design factors of immersive VR classrooms, namely different participant locations in the virtual classroom, different visualization styles of virtual peer-learners and teachers, including cartoon and realistic, and different hand-raising behaviors of peer-learners, particularly through the analysis of eye tracking data. Our results indicate that participants located in the back of the virtual classroom may have difficulty extracting information during the lecture. In addition, if the avatars in the classroom are visualized in realistic styles, participants may attend the lecture in a more focused manner instead of being distracted by the visualization styles of the avatars. These findings offer valuable insights about design decisions in the VR classroom environment. Few indicators were obtained from the evaluation of the different hand-raising behaviors of peer-learners, providing a general understanding of attention towards peer-learner performance. However, these indicators should be further investigated and remain a focus of future work.

### Acknowledgments

This research was partly supported by a grant to Richard Göllner funded by the Ministry of Science, Research and the Arts of the state of Baden-Württemberg and the University of Tübingen as part of the Promotion Program of Junior Researchers. Lisa Hasenbein is a doctoral candidate and supported by the LEAD Graduate School & Research Network, which is funded by the Ministry of Science, Research and the Arts of the state of Baden-Württemberg within the framework of the sustainability funding for the projects of the Excellence Initiative II. Authors thank Stephan Soller, Sandra Hahn, and Sophie Fink from the Hochschule der Medien Stuttgart for their work and support related to the immersive virtual reality classroom used in this study.

## A.2 Evaluating the Effects of Virtual Human Animation on Students in an Immersive VR Classroom Using Eye Movements

### A.2.1 Abstract

Virtual humans presented in VR learning environments have been suggested in previous research to increase immersion and further positively influence learning outcomes. However, how virtual human animations affect students' real-time behavior during VR learning has not yet been investigated. This work examines the effects of social animations (i.e., hand raising of virtual peer learners) on students' cognitive response and visual attention behavior during immersion in a VR classroom based on eye movement analysis. Our results show that animated peers that are designed to enhance immersion and provide companionship and social information elicit different responses in students (i.e., cognitive, visual attention, and visual search responses), as reflected in various eye movement metrics such as pupil diameter, fixations, saccades, and dwell times. Furthermore, our results show that the effects of animations on students differ significantly between conditions (20%, 35%, 65%, and 80% of virtual peer learners raising their hands). Our research provides a methodological foundation for investigating the effects of avatar animations on users, further suggesting that such effects should be considered by developers when implementing animated virtual humans in VR. Our findings have important implications for future works on the design of more effective, immersive, and authentic VR environments.

### A.2.2 Introduction

With the increasing availability of consumer-grade head-mounted displays (HMDs), virtual reality (VR) has been successfully deployed and is gaining immense popularity in the field of education [2]. Particularly, VR classrooms that mimic traditional classroom environments allow for authentic social interaction and engagement, which has been seen as crucial to increase learner motivation, persistence, and interest [12], ultimately leading to better learning outcomes [15]. However, in VR classrooms to date, virtual avatars (particularly virtual class-mates) that students can interact with or that provide them with social information (i.e., hand raising) to enhance immersion have not been fully developed and envisioned yet. Instead, VR classrooms are typically designed to provide either social interaction based on real-time interaction with another real person, meaning that individual learners need to participate synchronously with others in the same class [43], or provide pre-programmed but limited interactions with avatar teachers and classmates [12, 39]. Given that synchronous interaction in a VR classroom with a comparable number of students as in a real classroom (e.g., more than twenty students) is difficult to achieve, the development of VR environments that allow offline yet authentic social interaction with (pre-programmed) avatar teachers and classmates is a promising avenue for VR-based classroom learning, as such environments offer learners an authentic degree of flexibility with regard to the social information they process during learning.

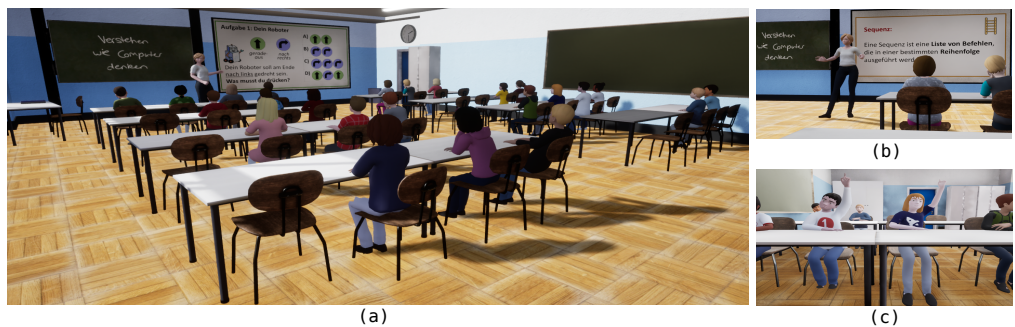

Figure A.6: Immersive virtual reality classroom: (a) overall view, (b) student's view, (c) virtual peer learners with hand raising animations.

It has been recently reported in [12] that presenting virtual humans pre-programmed with interactive animations in VR classrooms improved learners' motivation and immersion. However, this poses the question of how animated virtual humans affect learners while they are immersed in VR. Previous work has addressed this question and found that virtual human animations elicit a wide range of affective responses in humans, such as stress [211] and self-disclosure [212], and have effects on users' emotional responses [213, 214] and visual attention [215, 40]. These works highlight the importance and necessity of exploring the effect of avatar animations on various aspects of user behavior when presented in VR.

Against the background that students perceive their virtual peer's behavior in the VR classroom similarly to in a real-world learning scenario [216, 217], peer learners can be purposefully implemented in a pre-programmed VR classroom to promote student learning. Educational psychological research on so-called peer effects has repeatedly shown that the performance and motivation of peer learners influence individual learning trajectories regarding motivational as well as performance outcomes [151] Hand-raising (i.e., a students' self-motivated and clearly visible response to the teacher's questions) presents a highly salient behavior of students in a classroom setting, and contains important information about a student's performance and motivation that is perceived by peer learners and substantially shapes their learning experience in the classroom [41, 40]. Therefore, we consider it as one of the most important peer animations needed to be explored in VR classroom studies. In particular, studying the effect of (a crowd of) animated virtual humans, which typically convey social information (e.g., hand-raising of virtual peer learners), on students' cognitive and visual attention behaviors reflected in eye movements can provide valuable insights for the design features of VR-based learning environments (e.g., VR classroom) where social information is typically present to enhance immersion. Furthermore, a comprehensive understanding of the effects of avatar animations on learners is indispensable to creating well-functioning VR learning systems or more interactive online learning systems.

Eye movements as a non-intrusive and objective measurement have been proposed in previous works as a more intuitive method to investigate subjects' conscious and unconscious temporal behavior during tasks, such as cognitive load [204, 128, 74], visual attention [218, 219], and visual search behavior [219]. With this in mind, eye movements could provide a promising avenue to study users' cognitive and behavioral (e.g., visual attention and visual search) responses triggered by virtual human animations (e.g., social information) during VR experiences. And thanks to the state-of-the-art HMDs, eye-tracking data can be easily obtained via the built-in eye tracker, further facilitating the use of eye movements in VR studies.

In this work, we propose an authentic immersive VR classroom for offline learning and conduct an in-depth investigation of the effects of a crowd of physically animated virtual humans (particularly students' primary social counterparts: peer learners) on students' cognitive responses and visual attention behaviors while immersed in VR. Specifically, we designed an immersive VR classroom where students experience an immersive lesson that simulates a real classroom situation with a virtual teacher and a class of virtual peer learners pre-programmed with animations. The animations were created based on motion capture from real classrooms. To evaluate students' cognitive and visual attention behaviors before and after the onset of animation (i.e., hand raising of peers), we measured their eye movements and pupil measures, including pupil diameter, fixations, saccades, and dwell times. Furthermore, to examine whether the magnitude of the effects of animations was related to the amount of social information provided by the virtual peer learners, we purposefully implemented hand raising to mimic different variations of real classroom social interactions (between-subjects design: 20%, 35%, 65% or 80% of virtual peer learners are programmed with hand raising). We calculated the change scores of the eye movement variables before and after the onset of hand raising as new variables and use these new variables to represent the magnitude of the effects of animations on students' cognitive response and visual attention. A comparison was made between the four hand-raising conditions. Our study provides foundation for future work investigating the effects of a crowd of animated virtual peer learners on students' cognitive response and visual attention behavior during an immersive VR learning experience. Hence, it offers profound insights for optimizing the design of virtual learning environments from the perspective of virtual human animations. This improvement in system design will help to further improve learning, which is however beyond the scope of our current work.

### A.2.3   Related Work

**Immersive VR Classrooms**

The rapid development of affordable VR devices makes it easier to create VR learning environments for learners in various ways, such as VR labs [220, 221], VR field trips [222, 119], and VR classrooms [20, 219, 218]. Among these different VR learning contexts, VR classrooms that adopt traditional classroom settings, such as room layout and especially the presence of virtual teachers and virtual classmates exhibiting social interaction behaviors, have the advantage of providing a more authentic learning experience and higher engagement for students [12, 39, 43]. For instance, Liao et al. [12] suggested constructing virtual classmates as peer companions by synthesizing time-anchored comments from previous learners to reduce learners' sense of isolation and increase immersion and motivation during learning in the VR classroom. Learners were found to achieve better learning outcomes with the accompaniment of a few virtual classmates than without virtual classmates. Similarly, Liu et al. [39] developed a VR classroom with a virtual teacher and virtual classmates as an experimental platform to investigate the redundancy effect in learning. Interactive behaviors and animations of the virtual classmates, e.g., head movements (i.e., raising, lowering, and shaking the head), turning around, sneezing, and taking notes, were generated and found to have effects on learners and to be a decisive factor in the reverse redundancy effect. In addition to VR classrooms designed for learning, Ke et al. [43] developed a VR-based, Kinect-integrated learning environment for teaching training. Participants were teaching assistants at a university and were asked to play the role of a teacher and a student, respectively, in two sessions with corresponding avatars. With other avatars in the classroom played by peer trainees (whose body movements were projected onto their avatars in real time) and controlled by computers (with pre-programmed animations), participants maintained higher level of presence during immersion in VR-based training environments, regardless of whether they were acting as teachers or students.

The aforementioned works demonstrate the importance of the presence of animated virtual humans in VR environments, specifically classrooms, for improving learner engagement and immersion, as assessed by their subjective perception of the virtual human animations and learning outcomes. This inspired our study to develop an immersive VR classroom that mimics a real classroom scenario for offline learning, where the virtual teacher and virtual peer learners were presented with pre-programmed interactive animations to provide students with social information and companionship.

**Virtual Human Animations**

Researchers have studied various effects that virtual humans have on users during VR experiences. Robb et al. [223] examined the effects of the presence of a virtual human animated with verbal responses on students taking a prostate exam in a prostate exam simulation. They found that the virtual human elicited stress in students and led to increased engagement. In another study, Volonte et al. [215] examined how animation fidelity of the virtual human affected users' gaze behavior in a medical VR training simulation similar to [223]. It was found that the conversational and passive animations of the virtual human elicited visual attention responses from users, and users' visual attention shifted between the virtual human and goal directed activities. However, in these studies, users took part in VR simulations presented on a screen rather than in an immersive HMD-based VR environment, performing simple tasks and interacting with a limited number of (individual) virtual humans with limited animations. Closer to our work in terms of the number of animated virtual humans and VR environment settings, Volonte et al. [34] further investigated the effects of virtual human animations on users' affective and non-verbal behaviors while interacting with a crowd of virtual humans with animated emotions in an immersive HMD-based VR market scenario. It was found that virtual crowds with positive emotions elicited the highest scores on metrics related to interaction with the virtual agents.

However, it has not yet been investigated how animated virtual humans, which are intended to enhance immersion and serve as companions, in an immersive VR classroom affect learners' real-time and instantaneous and cognitive responses and visual attention behaviors reflected in eye movements during learning. Instead, such animated virtual humans in VR classrooms were found to influence learners in their overall experience but were not assessed by eye movements. [12, 39, 43]. A few previous works have investigated learners' eye movements during the VR experience. For instance, Gao et al. [219] investigated how different configurations of VR classrooms affect learners' gaze behavior while learning in VR. A range of eye movement features, including fixations and saccades, were extracted and analyzed with average measures. Their results showed that learners' gaze behavior differed significantly between different configurations of VR classrooms, suggesting that eye-tracking is indicative of learners' response to changes in classroom configurations. Following, Bozkir et al. [218] found that learners' visual attention switched between different virtual objects of interest (OOIs) while participating in a virtual lesson in the VR classroom. Different from the work of Gao et al. [219], which examined different classroom configurations, the study looked in more detail at learner attention to specific OOIs, such as instructional content and social information from animated peer learners, which have more to do with student learning behaviors. Although these studies examined an averaged gaze behavior across the entire duration of the VR experience, they provide evidence for our study to use eye movements as a metric to explore learners' real-time and instantaneous cognitive and visual attention responses to virtual human animations that last for a short period of time in VR classroom.

### A.2.4 Methods

**Participants and Apparatus**

In our study, 381 sixth-grade volunteer students (179 female, 202 male) with an average age of 11.5 years (10 to 13, $SD = 0.56$) were recruited from local schools to participate in our experiment. All volunteer participants and their guardians provided informed consent before the experiment. Our study was IRB-approved.

The VR classroom environment was rendered using the Unreal Game Engine [7] v4.23.1. The HTC Vive Pro Eye with a refresh rate of 90 Hz and a field of view of 110° was used. The integrated Tobii eye tracker with a sampling frequency of 120 Hz and a standard calibration accuracy of $0.5° - 1.1°$ was used to record the eye tracking data.

---

[7]https://www.unrealengine.com/

**Experimental Design**

To mimic a real classroom environment for sixth graders, we used the same configuration in the virtual classroom as in real classrooms (e.g., the classroom layout). In addition, a virtual teacher and twenty-four virtual peer learners were rendered with pre-programmed animations to enhance realism and authenticity. The virtual human animations were created based on motion capture from real classrooms and therefore authentically mimic physical movements similar to those of real people in real classrooms. We used recordings and motion captures from a sixth-grade classroom to ensure that the virtual peers' behaviors correspond to their virtual representation as well as the study participants' age. We used Xsens Motion Capture suits to record the authentic movements of the teacher and of six different students during a 15-min IVR lesson in a regular school setting. The students were asked to behave like they usually would in their classroom. The hand-raising was embedded in natural movements that showed students' motivation to be called on (e.g., leaning forward); overall, however, to avoid any confounding effects, any other distracting behaviors (like students moving around without reason and showing off-task behavior) were not included in the animations. Participants sat in such a VR classroom and listened to a virtual lesson ($\approx$ 15 minutes) on computational thinking (including basic concepts such as sequences and loops, practical exercises applied to them) delivered by the virtual teacher. During the lesson, the virtual teacher walks around the podium, asks simple questions (twenty-one in total), and calls on virtual peer learners with body gestures (e.g., hand gestures), while the virtual peer learners turn around, think, and interact with the virtual teacher by raising their hands to answer questions. To ensure controllability, all animations were pre-programmed. To mimic different variations of real-world social interactions in the classroom, a fixed group (between-subjects design: 20%, 35%, 65% or 80%) of virtual peer learners was programmed with hand raising behavior. We chose these percentages of hand-raising students (a) for pragmatic reasons (i.e., limiting the number of conditions to four so we could reach sufficient statistical power with the target sample size) and (b) for conceptual reasons (i.e., to examine differentiated effects based on an unambiguous picture of either a minority or majority of peers raising their hands; hence, a minority of 20% or 35% or a majority of 65% and 80%).

Note that the VR classroom had four rows of tables with virtual peer learners that showed the pre-programmed behavior without any randomizations. Participants were randomly either placed in the second (front position) or fourth row (back position). To ensure comparability across conditions, the peer learners in rows 1-2 and rows 3-4 were programmed in a similar manner to ensure that the participating students had similar experiences with regard to the position and proximity of the (hand-raising) peers. The hand-raising animations were identical across the conditions (with natural variations in authentic movements). Similarly, the audio file used was the same across conditions; one of the hand-raising peers was each called on and answered the teacher's question (based on audio recorded from the same sixth graders as the motion captures. The immersive VR classroom is shown in Figure A.6.

## A. Human Behavior in IVR Classrooms

### Procedure

After signing the informed consent form, participants were randomly assigned to one of the experimental conditions. Each experimental session lasted approximately 45 minutes and included a paper-based pre-test for demographic and learning background information, participation in the virtual lesson, and a paper-based post-test reporting on the VR experience. The post-test used a 4-point rating scale to assess participants' experienced level of spatial and social presence in the VR classroom (9 items based on REFS [208, 207]; e.g., "I felt like I was sitting in the virtual classroom" or "I felt like the teacher in the virtual classroom really addressed me") and their perceived realism of the VR classroom (6 items; e.g., "What I experienced in the virtual classroom could also happen in a real classroom" or "The students in the virtual classroom behaved similarly to real classmates"). The results indicated an overall authentic experience with mean levels of experienced presence and perceived realism ranging from 2.82 to 2.97 ($0.52 < SD < 0.62$) in all configuration conditions. Prior to the virtual lesson, a standard 5-point eye-tracking calibration routine was performed. All participants were informed before the experiment that they could drop out the experiment at any time without consequences.

### Data Preprocessing and Measures

Data including eye movements and head poses were collected from 381 participants. 40 participants who experienced hardware problems, invalid calibration, or synchronization issues in the VR environment were excluded. In addition, 61 participants with an eye-tracking ratio of lower than 90% (less than 90% of signal was recorded) were excluded. Therefore, 15 minutes of behavioral data from each of the 280 participants (140 female, 140 male) were used in our study. Since only raw tracking data, i.e., pupil size, gaze vectors, and head vectors, were collected, we performed data preprocessing that included normalization of pupil diameter and detection of eye movements for further analysis.

Pupil diameter has been suggested in previous studies as an indicator of cognitive load in human cognitive processes [204, 128, 74], so we used it as a measure in our study. Since pupillometry signals were affected by noisy sensor readings and blinks, we smoothed and normalized pupil diameters using the Savitzky-Golay filter [154] and divisive baseline correction with a baseline duration of $\approx 1$ second [155].

Fixations, i.e., the periods of time during which the eyes are stationary in the head, are generally considered an indicator of visual attention behavior [202, 203]. Saccades, i.e., rapid eye movement shifts between fixations, have been found to correlate strongly with visual search behavior [64, 197]. Before detecting such eye movements from the raw eye-tracking data, linear interpolation was performed for the missing gaze vectors. Fixations were detected using a modified velocity-threshold identification (I-VT) method applied to VR conditions that accounts for head movement [122, 219]. Specifically, fixations were detected under the stationary head condition (head velocity $< 7°/s$) [122] with a maximum gaze velocity threshold of $30°/s$. As saccade detection was not constrained by head movement, saccades were detected using the normal I-VT method, with a minimum gaze velocity threshold of $60°/s$. In addition, duration thresholds were applied, with a minimum duration of $100ms$ and a maximum duration of $500ms$ for fixation detection and a minimum duration of $30ms$ and a maximum duration of $80ms$ for saccade detection.

Dwell time, i.e., the total time spent in an area-of-interest (AOI), including all fixations and saccades as well as revisits, is a metric that conveys the level of interest and attentive behavior within a certain AOI of the stimuli [64]. In addition, the time to the first fixation (TTFF) and the first fixation duration (FFD) in an OOI are interpreted as visual scene priority [64]. Therefore, we computed these measures for the main objects of interest (OOIs) in the VR classroom, namely the virtual teacher, virtual peer learners, and the screen displaying the instructional content. The number of virtual peers that participants fixated on was also used as a measure.

### Hypotheses

Our aim is to investigate the effects of social interaction animations between the virtual teacher and peer learners (i.e., hand raising of peers) on participants' cognitive responses and visual attention behaviors by analyzing various eye movements and pupil measures. We tested the following hypotheses:

**H1** We hypothesized that the animated virtual peers will influence participants' cognitive responses. We expected participants' pupil diameter will increase after the hand raising activation compared to before.

**H2** We hypothesized that hand raising will affect participants' visual attention and visual search behavior, as reflected in fixations and saccades. We therefore expected to observe a significant effect of animation in these measures, i.e., longer fixation durations, more saccades, longer saccade durations, and larger saccade amplitudes.

**H3** We hypothesized that participants' visual attention to the virtual objects will change after the onset of hand raising, shifting attention from instructional content (i.e., virtual teacher and screen) to peers. Therefore, we expected an increase in dwell time on peers as well as the number of peers that participants fixated on, in other words, a decrease in the dwell time on instructional content.

**H4** We hypothesized not only that animated virtual peers will have effects on participants' behavior, but also that different numbers of peer animations (four hand-raising conditions) will affect participants differently.

### A.2.5 Results

To examine the effects of peer animations (i.e., hand raising) on participants' cognitive responses and visual attention, we measured eye movements and pupil diameters before and after the onset of hand raising. Previous studies have shown that task-evoked pupillary responses (TEPR), which index cognitive processing load, have a latency of several hundred milliseconds across tasks [204, 224]. Therefore, given the duration ($\approx 2$ seconds) of the hand raising animation, we assessed participants' gaze behavior within time windows of 2.5 seconds before and after the hand raising activation. Dependent variables including pupil diameter, fixations, saccades, dwell times, and the number of virtual peers participants fixated on were examined.

First, to investigate whether animated peer learners influenced participants, we compared these dependent variables before and after the onset of hand raising based on all experimental data. Variables used for statistical analysis were described as *V_before* and *V_after*. For this purpose, we used a within-subjects design with one-tailed t-tests. Specifically, a paired t-test was performed for normally distributed data; a Wilcoxon signed-rank test was performed for non-normally distributed data. Furthermore, to verify whether such effects existed in each hand-raising condition, we performed the same statistical analysis (paired t-test) separately for each condition. The statistical significance (see below) that we found based on all experimental data was also found in each hand-raising condition, meaning that the animated peer learners in the VR classroom influenced the participants regardless of the number of animations (20%, 35%, 65% or 80% of virtual peer learners are pre-programmed with hand raising). Given the length of the paper, we have included the detailed paired t-test results for each condition in the Appendix.

Second, since peer learner animations were found to have an effect on participants, as indicated by significant differences in the paired t-tests (see below), we were interested in examining how such effects differed between conditions. We compared the magnitude of these effects across the four hand-raising conditions. Specifically, we calculated the change scores of the above dependent variables before and after activation of hand raising as new variables, described as *V_change* (= *V_after* − *V_before*). For comparison of *V_change* between four groups, we performed a one-way between-subjects ANOVA and the Bonferroni-corrected Tukey-Kramer test as a post-hoc test for the pairwise comparisons. For non-normally distributed data, the Kruskal-Wallis H test was used as a non-parametric version of ANOVA and the Bonferroni-corrected Dunn's Test as a post-hoc test for pairwise comparisons.

All statistical analyses were performed using SciPy [225], an open-source Python[8] library. The significance level was set at $\alpha = 0.05$ for all tests. Asterisks in Figure A.8 and Figure A.9 indicate significant differences (*, **, *** and n.s. for $p < .05$, $p < .01$, $p < .001$, and no statistical significance, respectively).

### Pupil Diameter

To gain direct insight into the changes in pupil diameter during the virtual lesson, we plotted the normalized pupil diameter over time. As shown in Figure A.7, the blue dots represent the mean normalized pupil diameter of all participants at specific time points (every $100ms$), while the red rectangles each mark the 2.5-second time window after the onset of hand raising.

---

[8] https://www.python.org/

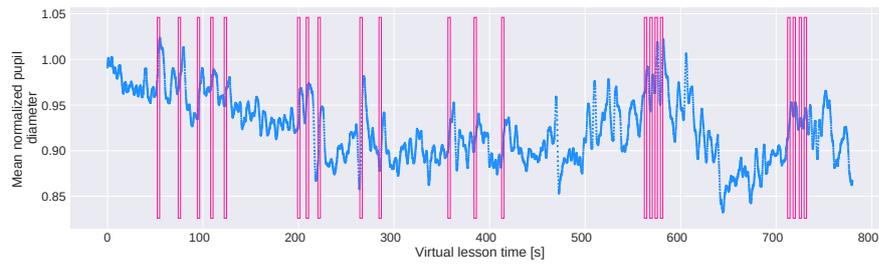

Figure A.7: Pupil diameter changes throughout the virtual lesson.

As shown, pupil diameter increases dramatically after the onset of hand raising compared to before peer learners raised their hands. The mean normalized pupil diameter before and after the onset of hand raising (i.e., *Pupil_before* and *Pupil_after*) was calculated for each participant. As shown in Figure A.8 (a), *Pupil_after* ($M = 0.95$, $SD = 0.17$) is significantly larger than *Pupil_before* ($M = 0.93$, $SD = 0.17$), with $t = 1,224$, $p < .001$.

Moreover, we calculated the change scores of mean normalized pupil diameter (i.e., *Pupil_change=Pupil_after−Pupil_before*) for each participant. A Kruskal-Wallis test revealed a statistically significant difference between groups in the change in mean normalized pupil diameter ($H(3) = 13.59$, $p < .01$). Notably, *Pupil_change* in the 80% condition ($M = 0.0243$, $SD = 0.011$) is significantly greater than in the 20% ($M = 0.0163$, $SD = 0.009$), 35% ($M = 0.0168$, $SD = 0.009$), and 65% ($M = 0.0168$, $SD = 0.011$) conditions, as shown in Figure A.9 (a).

**Fixation and Saccade**

The mean fixation duration before and after the onset of hand raising (i.e., *FixaDur_before* and *FixaDur_after*) was calculated for each participant. As shown in Figure A.8 (b), *FixaDur_after* ($M = 196.38ms$, $SD = 96ms$) is significantly longer than *FixaDur_before* ($M = 192.54ms$, $SD = 95ms$), with $t = 8,127$, $p < .05$.

In addition, we calculated the change scores of mean fixation duration (i.e., *FixaDur_change=FixaDur_after−FixaDur_before*) for each participant. However, no statistical differences in the change in mean fixation duration were found between groups, as shown in Figure A.9 (b).

We found a significant effect of hand raising on saccade measures, i.e., number of saccades (i.e., *SaccNum_before* and *SaccNum_after*), saccade duration (i.e., *SaccDur_before* and *SaccDur_after*), and saccade amplitude (i.e., *SaccAmpli_before* and *SaccAmpli_after*), as shown in Figure A.8 (c), (d), and (e), respectively. In particular, there are significantly more saccades after the onset of hand raising ($M = 3.29$, $SD = 1.75$) than before ($M = 2.99$, $SD = 1.73$), with $t = 1,184$, $p < .001$. *SaccDur_after* ($M = 42.35ms$, $SD = 9.95ms$) is significantly longer than *SaccDur_before* ($M = 39.97ms$, $SD = 10.08ms$), with $t = 6,597$, $p < .001$. Similarly, *SaccAmpli_after* ($M = 11.67deg$, $SD = 4.63deg$) is significantly greater than *SaccAmpli_before*

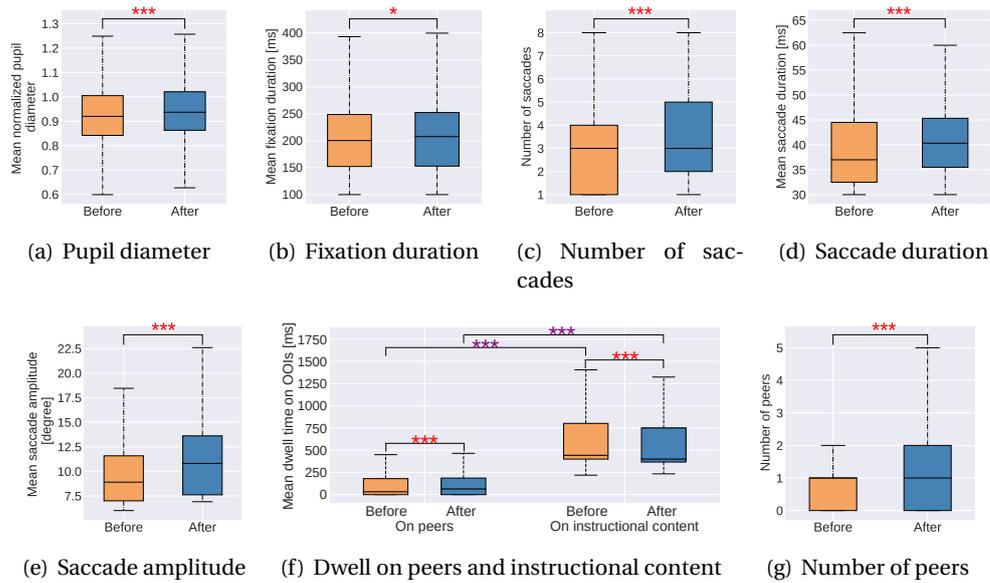

(a) Pupil diameter   (b) Fixation duration   (c) Number of saccades   (d) Saccade duration

(e) Saccade amplitude   (f) Dwell on peers and instructional content   (g) Number of peers

Figure A.8: Statistical comparison of the dependent variables between the conditions before and after the start of the hand raising animations. Dependent variables include (a) pupil diameter, (b) fixation duration, (c) number of saccades, (d) saccade duration, (e) saccade amplitude, (f) dwell on peers and instructional content, and (g) number of peers fixated by participants.

($M = 10.07 deg$, $SD = 3.91 deg$), with $t = 5,926$, $p < .001$.

Furthermore, we calculated the change scores of saccade variables (i.e., *SaccNum_change*, *SaccDur_change*, *SaccAmpli_change*, where *Sacc_change=Sacc_after-Sacc_before*) for each participant. A Kruskal-Wallis test revealed a statistical significance between groups in the change in mean saccade amplitude ($H(3) = 19.13$, $p < .001$). Notably, *SaccAmpli_change* in the 20% condition ($M = 2.13 deg$, $SD = 5.76 deg$) is significantly greater than in the 35% ($M = 1.37 deg$, $SD = 6.05 deg$), 65% ($M = 1.53 deg$, $SD = 5.76 deg$), and 80% ($M = 1.36 deg$, $SD = 5.56 deg$) conditions. However, no statistical significance was found between groups with respect to the change in saccade number and mean saccade duration. The results are depicted in Figure A.9 (c), (d), and (e), respectively.

**Eye Movements on OOIs**

We found that after the activation of hand raising, participants initially focused their attention more on the virtual peers OOIs than on the instructional content OOIs (i.e., virtual teacher and screen), as evidenced by the fact that 76% of their first fixations occurred in peer OOIs. Moreover, the time to the first fixation (TTFF) was significantly shorter in peer OOIs ($M = 0.45s$, $SD = 0.21s$) than in instructional content OOIs ($M = 0.93s$, $SD = 0.27s$), with $t = 4,847$, $p < .001$. Although no statistical difference in the first fixation duration (FFD) was found, a longer mean FFD was observed in peer OOIs ($M = 251.32\,ms$, $SD = 110.55\,ms$) than in instructional OOIs ($M = 242.56\,ms$, $SD = 102.09\,ms$).

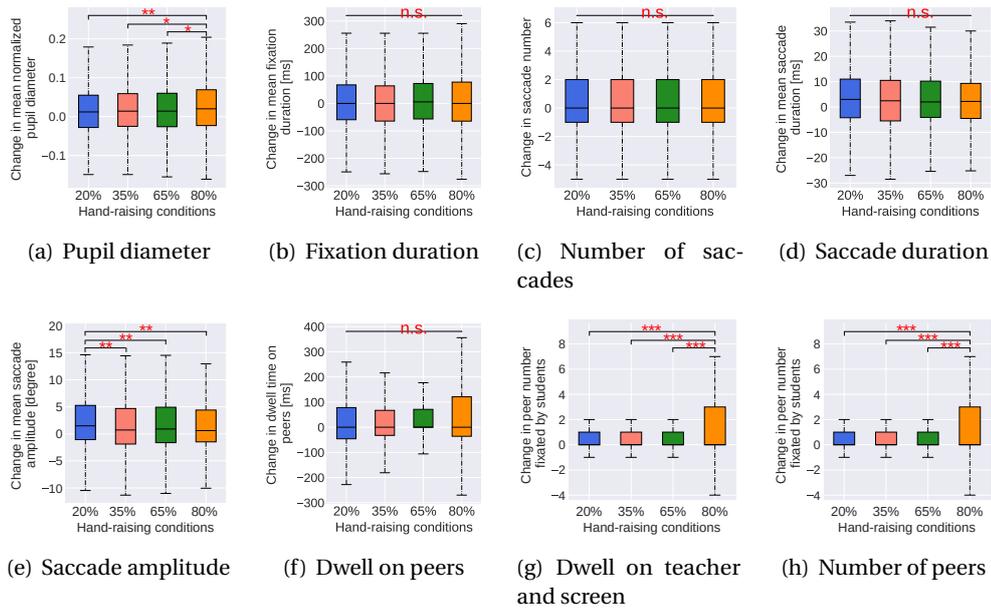

(a) Pupil diameter  (b) Fixation duration  (c) Number of saccades  (d) Saccade duration

(e) Saccade amplitude  (f) Dwell on peers  (g) Dwell on teacher and screen  (h) Number of peers

Figure A.9: Comparison of the change scores of pupil and eye movement variables before and after the onset of hand raising between four hand-raising groups, including change scores of (a) pupil diameter, (b) fixation duration, (c) number of saccades, (d) saccade duration, (e) saccade amplitude, (f) dwell on peers, (g) dwell on teacher and screen, and (h) number of peers fixated by participants.

A significant effect of hand raising on participants' dwelling behavior to peers (i.e., *DwellPeer_before* and *DwellPeer_after*) and instructional content (i.e., *DwellInstr_before* and *DwellInstr_after*) was found, as shown in Figure A.8 (f).

*DwellPeer_after* ($M = 146.08\,ms$, $SD = 59.07\,ms$) is significantly longer than *DwellPeer_before* ($M = 132.96\,ms$, $SD = 65.85\,ms$), with $t = 2,927$, $p < .001$. In contrast, participants were found to pay decreased attention to the instructional content after peers raised their hands, with *DwellInstr_after* ($M = 607.65\,ms$, $SD = 368.39\,ms$) being significantly shorter than *DwellInstr_before* ($M = 638.77\,ms$, $SD = 383.99\,ms$), with $t = 5,625$, $p < .001$. Notably, dwell time on instructional content was found to be significantly longer than dwell time on peers, regardless of whether or not there were hand raising animations. As seen in Figure A.8 (f), *DwellInstr_before* ($M = 638.77\,ms$, $SD = 383.99\,ms$) is significantly longer than *DwellPeer_before* ($M = 132.96\,ms$, $SD = 65.85\,ms$), with $t = 9,052$, $p < .001$; *DwellInstr_after* ($M = 607.65\,ms$, $SD = 368.39\,ms$) is significantly longer than *DwellPeer_after* ($M = 146.08\,ms$, $SD = 59.07\,ms$), with $t = 10,610$, $p < .001$.

In addition, we calculated the change scores of dwell time on peers (*DwellPeer_change=DwellPeer_after-DwellPeer_before*) as well as on instructional content (i.e., the virtual teacher and screen) (*DwellInstr_change=DwellInstr_after-DwellInstr_before*). A Kruskal-Wallis test revealed a significant difference between groups in the change in mean dwell time on instructional content ($H(3) = 56.96$, $p < .001$), as shown in Figure A.9 (g). Notably, *DwellInstr_change* (absolute value) is significantly greater in the 80% condition ($M = -48.44ms$, $SD = 38.93ms$) than in the 20% ($M = -22.51ms$, $SD = 47.74ms$), 35% ($M = -26.98ms$, $SD = 36.02ms$), and 65% ($M = -24.22ms$, $SD = 45.56ms$) conditions. However, no statistical differences were observed in the change in mean dwell time on peers between groups (see Figure A.9 (f)).

We calculated the number of virtual peers that participants fixated on before and after the onset of hand raising (i.e., *NumPeer_before* and *NumPeer_after*). As shown in Figure A.8 (g), we found a statistically significant increase in the number of peers that participants fixated on, with *NumPeer_after* ($M = 1.40$, $SD = 0.91$) being significantly larger than *NumPeer_before* ($M = 0.92$, $SD = 0.61$), with $t = 1,344$, $p < .001$.

Moreover, we calculated the change scores of the number of peers fixated by participants (*NumPeer_change=NumPeer_after-NumPeer_before*). A Kruskal-Wallis test revealed a statistically significant difference between groups in the change in the number of peers fixated by participants ($H(3) = 187.09$, $p < .001$). The *NumPeer_change* is significantly greater in the 80% condition ($M = 1.25$, $SD = 2.39$) than in 20% ($M = 0.11$, $SD = 1.51$), 35% ($M = 0.11$, $SD = 1.38$), and 65% ($M = 0.32$, $SD = 1.47$) conditions, as shown in Figure A.9 (h).

### A.2.6 Discussion

**Results Discussion**

In this section, we discuss the results based on the previously postulated hypotheses **H1**-**H4**. Overall, the results show that the pre-programmed social animation of virtual peers (i.e., hand raising in response to the virtual teacher's questions) had an effect on participants' cognitive responses and visual attention behaviors during the VR classroom experience, as indicated by various eye movements and pupil measures (see Figure A.8). The magnitude of this effect differed between four hand-raising groups (see Figure A.9).

Specifically, it was found that participants' pupil diameter increased significantly after the onset of hand raising (see Figure A.7 and Figure A.8 (a)). Since pupil diameter is often reported as an indicator of cognitive load [226, 227], this significant increase in pupil diameter suggests that participants have increased cognitive processing load and exert more cognitive effort when exposed to the social information that needs to be processed [228], which can be further confirmed by the results of other eye movement measures. Thus, our first hypothesis **H1** is confirmed.

Analysis of fixation and saccade measures allowed investigation of the effects of peer animations on participants' visual attention and visual search behavior. First, we found that participants exhibited longer fixation duration after hand raising was activated (see Figure A.8 (b)), suggesting that they had longer cognitive processing time and more attentive behavior [203, 196] while immersed in a VR classroom with animated virtual peers that were animated with more hand raising. This could be due to participants directing their attention to the raised hands around them. And this social information conveyed by their peers requires more processing time, which consequently implies additional cognitive effort [229, 230]. The increase in information processing time is consistent with the increase in cognitive processing load reflected in pupil diameter. Second, it was found that participants showed significantly different visual search behavior when exposed to hand raising animations, which was reflected in saccade measures (see Figure A.8 (c), (d), and (e)). Participants were found to exhibit more saccades after peers raised their hands, suggesting that their visual spatial attention was influenced and that they showed more visual search behavior [197]. In addition, saccades with greater amplitude were detected after the onset of hand raising, which is another indication that participants' attention might be drawn to more salient attraction cues from a distance around them [70]. These results support our second hypothesis **H2**.

Before the peer learners raise their hands, the virtual teacher and the screen displaying the learning content are the main focus of the participants' attention. After being confronted with social information from peers (i.e., hand raising), participants gave higher visual priority to the animated peers than to the instructional content to which they paid attention before hand raising, as indicated by their first-fixation behavior [64]. This visual priority is further evidenced by the time to the first fixation (TTFF) within OOIs. Furthermore, participants were found to shift their attention from instructional content to their peers (see Figure A.8 (f)). As a consequence, the number of peers participants fixated on was found to increase significantly after hand raising (see Figure A.8 (g)), further supporting our attention shift hypothesis. The change in visual priority is consistent with the shift in visual attention reflected in dwell time measures, and they can be parsed together. Taken together, all these results support our third hypothesis **H3**. Surprisingly, it is noteworthy that although participants' attention was captured by their peers' social animations, their attention remained primarily focused on the learning content (i.e., virtual teacher and screen) while the hand raising was occurring (see Figure A.8 (f)). This indicates that regardless of the animations, participants exhibited significantly longer attention time on the instructional content than on peers, which in turn implies that their learning activities were still ongoing when they were exposed to social information.

The attention shift results evidencing hypothesis **H3** (attention shifts from instructional content to peer learners) further support our hypothesis **H1** (cognitive load increases) and hypothesis **H2** (visual attention and visual search behavior change). When participants were exposed to peer animation (i.e., hand raising) intended to increase immersion and authenticity, their attention inevitably shifted to such animations, resulting in more focused attentional behavior and stronger visual search behavior, which ultimately also led to a change in cognitive processing load. Overall, our results support the first three hypotheses (**H1**-**H3**) that participants respond cognitively and attentionally to animated virtual companions while participating in a VR lesson, as reflected in various eye movements and pupil measures.

Furthermore, we measured the change in all variables after hand raising compared to before and considered it as a measure of the extent to which the animation affected participants. Results show that the hand raising animation affected participants differently across groups (see Figure A.9), and this difference is mainly between groups with lower (20%) and higher (80%) number of animations. Specifically, when participants were surrounded by a large number of animated peers (80%) from different directions, they were able to perceive the most animations, and thus this large amount of social information caused participants' cognitive load to increase the most [228]. This is evidenced by significant increase in pupil diameter and number of peers fixated by participants in the 80% condition than in the other conditions. Consequently, participants in groups with a higher (80%) number of animations showed a significantly greater decrease in their visual attention to the lesson content than in groups with a lower (20%) number of animations. This is because that their attention was more likely to shift from the instructional content to virtual peers when they were surrounded by a larger number of animated virtual peers. Therefore, participants in the 20% condition showed a significantly greater increase in saccade amplitude than in the other conditions. Since when participants were surrounded by a small number of animated peers (20%), they had to exert more effort to search for the raised hands, which were slightly more difficult to catch, than when they were surrounded by more animations, resulting in a greater increase in visual search behavior [197, 70]. However, no significant difference was found in the increase in fixation duration, saccade number, saccade duration, and dwell time on peers between groups. Thus, our fourth hypothesis **H4** is partially validated by these results. Our results indicate that participants' cognitive and visual attention responses elicited by virtual avatar animations are related to the number of animations, which could provide insights for designing VR environment with optimal virtual avatar animations.

In this study, it was found that animated peer learners significantly affect learners from several aspects, and that these effects are related to the number of animations provided. Such a finding further contributes to how and to what extent virtual peer learners' animations can be implemented in IVR classrooms to both enhance authenticity and control the distractions and additional cognitive load that the VR systems impose on learners to an acceptable level. Deciding on the number of social animations depends on the purpose of the designed VR applications, i.e., whether instructional content or social information is more important to learners. Our study does not aim to tell to what extent avatar animations should be provided, but is an exploratory study that provides bases for further studies that need to decide on the animations provided in their VR applications. Such findings provide important insights specifically for designing educational VR applications by implementing certain avatar animations related to users' learning behaviors (e.g., hand raising of peer learners), but also more generally for designing avatar animations in other VR-based systems, such as animated avatars in contexts like medical training simulation [34, 215]. Eye-tracking technology offers a viable way to investigate this research question by extracting users' temporal eye movements that are indicative of various human behaviors such as visual attention, visual search, as well as cognitive processing load.

**Limitations and Future Work**

Our immersive VR classroom was built with animated virtual teacher and peer learners to provide students with a high level of immersion. However, since we aim to explore how students respond to the animations of their primary social counterparts (i.e., peer learners), all animations were pre-programmed in the spirit of controllability. However, this may reduce students' sense of immersion since the teacher does not call on them when they raise their hand. Although avatar animations were found to have an impact on students' cognitive and visual attention behaviors, and this impact was related to the number of animations, it is unclear whether students' learning outcome is also associated with and affected by avatar animations. In our future work, we aim to investigate this question as it will provide guidance on the optimal design of VR environments for educational purposes and thus maximize the efficiency of student learning in such VR environments.

### A.2.7   Conclusion

In this paper, we designed an immersive VR classroom with pre-programmed animated virtual avatars to investigate how social interactions between the virtual teacher and peers (i.e., peer hand raising in response to the teacher's questions) affect students' cognitive and visual attention behaviors during a virtual lesson. To this end, eye movements and pupil measures were analyzed. We found that peers' hand raising had an effect on students' behavior in several aspects, including increased cognitive load, attentional shift from instructional contents (i.e., virtual teacher and screen) to peers, and increased information processing time and visual search behavior. The effects of hand raising on various aspects of students reflected in different measures are interrelated. In addition, we found that the magnitude of such effects is related to the number of animations (number of animated peer learners).

   In our study, we developed an immersive VR classroom that closely resembles a real classroom by creating not only a virtual teacher but also a set of animated virtual peer learners with social information (i.e., hand raising) to enhance immersion and authenticity. Our research provides a methodological foundation for investigating students' instantaneous and intuitive responses to virtual human animations (particularly virtual peers) during VR experience using eye movements. Our results imply that the effects of avatar animation should be considered by developers when presenting animated virtual humans to enhance immersion or for inter-action purposes in VR environments. Overall, these findings have important implications for future studies aiming to create more effective, interactive, and authentic VR-based (learning) environments.

### A.3 Exploring Gender Differences in Computational Thinking Learning in a VR Classroom: Developing Machine Learning Models Using Eye-Tracking Data and Explaining the Models

#### A.3.1 Abstract

Understanding existing gender differences in the development of computational thinking skills is increasingly important for gaining valuable insights into bridging the gender gap. However, there are few studies to date that have examined gender differences based on the learning process in a realistic classroom context. In this work, we aim to investigate gender classification using students' eye movements that reflect temporal human behavior during a computational thinking lesson in an immersive VR classroom. We trained several machine learning classifiers and showed that students' eye movements provide discriminative information for gender classification. In addition, we employed a Shapley additive explanation (SHAP) approach for feature selection and further model interpretation. The classification model trained with the selected (best) eye movement feature set using SHAP achieved improved performance with an average accuracy of over 70%. The SHAP values further explained the classification model by identifying important features and their impacts on the model output, namely gender. Our findings provide insights into the use of eye movements for in-depth investigations of gender differences in learning activities in VR classroom setups that are ecologically valid and may provide clues for providing personalized learning support and tutoring in such educational systems or optimizing system design.

#### A.3.2 Introduction

Computational thinking (CT), which refers to the thought processes involved in expressing solutions as computational steps or algorithms that can be carried out by a computer (Wing [231]), is considered as an essential skill that will provide people with a strong competitive edge in the digital future (García-Peñalvo and Mendes [232]). With the growing popularity of user-friendly and open-source programming languages such as Python[9], CT has already been incorporated into K-12 education in many countries such as UK (Sentance and Csizmadia [233]), Singapore (Seow et al. [234]), and New Zealand (Bell et al. [235]) to equip students with this 21st century skills. However, studies show that although the gender gap has been narrowing down in recent years, gender differences in students' interests and attitudes toward CT and in their CT skills are still observed (Kong et al. [163]; Sullivan and Bers [236]). In these works, gender differences are typically examined through the analysis of commonly used self-reports, i.e., measures that do not tap into the learning process, including students' individual characteristics that have been shown to be associated with gender differences, acquired CT skills, or similar learning outcomes. However, the gender differences that might emerge during the process of CT development and that are reflected in real-time human behavior have not yet been investigated due to the lack of advanced methods to measure these differences.

---

[9]https://www.python.org/

In recent years, virtual reality (VR) has become increasingly popular and prevalent in education, offering benefits such as supporting distance learning and teaching (Cryer et al. [237]; Hernández-de-Menéndez et al. [238]; Grodotzki et al. [19]) with increased immersion (Casu et al. [11]). With the growing availability of modern head-mounted displays (HMDs), immersive VR learning experiences can be provided to students in the near future at a reasonable cost and with relatively reduced effort. For instance, immersive VR classrooms in particular, which emulate traditional classrooms, have the potential to provide more flexible and engaging learning contexts for students to develop their CT skills. Furthermore, integrated eye trackers in such setups open up additional opportunities to investigate students' gazing behavior in standardized and yet realistic environments, ultimately offering an in-depth understanding of individual differences in learning, e.g., CT skill development. Indeed, eye-tracking technology has already been used in variety of educational applications for studying learning processes, training, and assessment, such as in mathematics education (Strohmaier et al. [96]), medical education (Ashraf et al. [95]), multimedia learning (Molina et al. [239]).

Compared to commonly used questionnaires and surveys, eye tracking offers the opportunity to obtain objective measurements from subjects in a non-intrusive manner, and these measurements could also be used in real-time for various purposes. Previous studies in this context have mainly focused on investigating various aspects of human behavior using eye movements, such as stress (Hirt et al. [168]), visual attention (Bozkir et al. [76]; Gao et al. [219]), and problem solving (Eivazi and Bednarik [240]), either in VR or conventional setups. In addition, eye movements have been found to provide discriminative information for various psychological behavior related predictions using machine learning methods, such as cognitive load (Appel et al. [128]; Yoshida et al. [241]), personality traits (Hoppe et al. [130]; Berkovsky et al. [131]), and IQ test performance (Kasneci et al. [134]). Closer to our work, several studies in the field of human-computer interaction have shown that gender differences can be inferred from eye movements by analyzing subjects' visual viewing and search behavior with 2D stimuli (Sammaknejad et al. [132]; Hwang and Lee [242]; Mercer Moss et al. [243]). Gender differences were also found to be predictive using classification models developed based on eye-tracking data in reading (Al Zaidawi et al. [133]) and indoor picture viewing tasks (Abdi Sargezeh et al. [244]). However, eye movement information used in these studies was limited to fixation- and saccade-related statistics, and tasks were performed in relatively simple contexts, i.e., screen-based tasks with 2D stimuli. Furthermore, the relationship between eye movements and gender has not yet been fully investigated using explainable machine learning approaches. Therefore, it is an open question whether it is possible to detect gender differences by using eye movement information in learning activities that require more effort in more complex contexts (e.g., in VR-based learning), based on machine learning and explainability approaches.

From an educational perspective, predicting gender differences based on machine learning and explainability approaches to analyzing eye movements information provides great potential: It is widely acknowledged that boys and girls differ in their achievement and interest, especially in STEM subjects, but respective educational research is predominantly based on questionnaires relying on self-reports of students. Hence, little is known about gender differences in the actual learning process. Respective insights into systematic differences of how boys and girls objectively differ in their learning behaviors, not only allow a deeper understanding of gender differences but make it possible to adapt learning environments (especially in VR settings) to the different needs of girls compared to boys to equally foster the skill development of both genders.

Therefore, in this work, we investigated to what extent eye movement data provide discriminative information for gender classification in CT learning in an immersive VR classroom. To this end, we examined a large set of eye movement features that characterize students' real-time visual attention and cognitive behaviors in a VR lesson. Several machine learning models were developed for gender classification, including Support Vector Machine (SVM), Logistic Regression, three ensemble machine learning models, i.e., Random Forest, eXtreme Gradient Boosting (XGBoost), and Light Gradient Boosting Machine (LightGBM). To improve the performance of the model, we performed a feature selection procedure by applying the Shapley additive explanations (SHAP) approach (Shapley [245]). Furthermore, we interpreted the classification model by SHAP approach as well.

In summary, the contributions of this work are four-fold. **(i)** We extracted a large number of eye movement features using different time windows similar to the work of Bulling et al. ([246]) from a lesson on computational thinking in an immersive VR classroom. **(ii)** We developed five machine learning models for gender classification using all extracted eye movement features. We performed feature selection using SHAP and improved the performance of the LightGBM(best) model by training the model with the selected (best) eye movement feature set. Furthermore, **(iii)** we investigated the importance of the eye movement features and explored the effects of high contribution features on classification output, i.e., gender, using the SHAP approach. **(iv)** We provided a fundamental methodology for future studies aimed at investigating gender differences using eye movement information not only in CT but also in other learning activities in immersive VR environments for educational purposes. This could help to gain further insights to optimize the design of educational systems and thus offer personalized tutoring in such educational systems.

### A.3.3 Related Work

As the integration of CT into K-12 STEM education has increased, so has the need for effective teaching and learning methods for CT instruction (Chalmers [247]; Hsu et al. [248]), and an understanding of gender differences can shed light on this. In the literature to date, there are various findings on gender differences in CT development. For instance, it has been found that girls tend to show less interest and self-efficacy in STEM subjects (McGuire et al. [249]; Wang and Degol [250]). Kong et al. ([163]) conducted a study with 287 senior primary school students to investigate their interest in CT and collaboration attitude. It was found that boys showed more interest in programming than girls and thus found programming more meaningful and exhibited higher creative self-efficacy. Similarly, Baser ([251]) observed that in an introductory computer programming course, males had more positive attitude towards programming than females and this attitude was positively correlated with their performance in programming. In addition, Nourbakhsh et al. ([252]) examined gender differences in a robotics course designed to develop CT skills in high school students. Girls were found to have less confidence in their CT skills than boys at the beginning of the course and, according to weekly surveys, girls were more likely to report that they struggled with programming. Notably, these studies measured gender differences using students' individual characteristics in terms of their self-reported interests and attitudes toward CT. In addition, gender differences are often tied to students' achievement test results. Boys are typically found to score higher in CT tests than girls (Polat et al. [253]). Girls in turn have been found to require more training time than boys to achieve the same skill level (Atmatzidou and Demetriadis [254]). Angeli and Valanides ([255]) demonstrated that boys and girls benefit from different scaffolding and learning activities when working on CT-related tasks.

The aforementioned works have highlighted the growing need for research on gender differences in CT-related education, as knowledge of how different gender groups exhibit different attitudes and learning outcomes can inform educators to better support both girls and boys who typically differ in their prerequisites and requirements for CT skill acquisition (Nourbakhsh et al. [252]; Angeli and Valanides [255]). However, most studies on gender differences in STEM (and CT) learning do not take into account the learning process and the respective differences of boys and girls in how they acquire knowledge, especially the differences reflected in their real-time visual attention and cognitive behavior during learning; however, they could offer valuable insights for providing tailored support or developing tailored tutoring systems aimed at reducing gender disparities in STEM subjects (CT skill development) for different gender groups, e.g., tailored VR classrooms that can be easily realized through software. Eye tracking holds such potential.

Recently, several works suggest that eye movement data provide a more intuitive interface for studying conscious and unconscious human behaviors in various tasks, such as visual search pattern recognition (Raptis et al. [256]), web search (Dumais et al. [257]), the n-back task (Appel et al. [128]), decision making measure for intelligent user interfaces (Zhou et al. [258]), and learning in the VR context (Gao et al. [219]; Bozkir et al. [218]). Moreover, eye movements were found to complement self-reports in providing discriminative information for the prediction of subjects' psychological behavior. In terms of eye movements, Hoppe et al. ([130]) utilized a machine learning approach to predict personality traits and perceptual curiosity during an everyday task solely from eye movements. In addition to personality traits, eye movement data have also been used for detecting other individual psychological behaviors. Prediction of cognitive load based on eye movements was investigated by Appel et al. ([198]) and it was found that the models trained based on eye movement data were discriminative in predicting cognitive load in a simulated emergency game. Furthermore, Zhou et al. ([82]) combined eye movements with demographic information and self-reports to predict situational awareness in a take-over task during conditionally automated driving. The results show that the LightGBM model, which was developed based on eye movements alone, performed better than other models in predicting situational awareness with a mean absolute error of 0.096. In addition, Kasneci et al. ([134]) utilized Gradient Boosted Decision Trees (GBDT) model to examine individual differences in IQ tests by using a set of eye movement variables in combination with socio-demographic variables as features. Specifically, eye movements alone were found to be discriminative in predicting participants' intelligence performance on the Cultural-Fair IQ Test 20-R (CFT 20-R). Notably, the eye movement and socio-demographic features together were observed to provide complementary information, indicating that eye movement information is a reliable and effective behavioral measure of learning and performance processes.

Furthermore, eye movements were also found to be connected to gender. Sammaknejad et al. ([132]) used eye movement data to determine gender differences in a face viewing task. It was found that male and female participants exhibited significantly different eye movement transition patterns, indicated by saccades, when viewing facial photographs of male and female subjects that were unknown to them. Similarly, gender differences in eye movement patterns were found in an indoor picture viewing task (Abdi Sargezeh et al. [244]), where females showed more exploratory gaze behavior, as indicated by longer scanpaths and larger saccade amplitudes. In their study, a support vector machine classifier was utilized to predict gender by using ten eye movement features (i.e., features related to fixations, saccades, and scanpaths), achieving an overall accuracy of about 70%. In the work of Hwang and Lee ([242]), gender differences in online shopping were examined using area-of-interest information based on eye-tracking and it was concluded that females paid more attention to shopping content than males. Closer to the CT learning task in this work, different eye movement behaviors were observed in two gender groups during an algorithmic problem solving task (Obaidellah and Haek [259]), with females fixating more on the indicative verbs, while males fixated more on the operational statements. However, in contrast to the present study, the participants were undergraduates and the task was performed in a conventional context with screen-based stimuli. Al Zaidawi et al. ([133]) performed gender classification of children aged 9-10 years, whose ages were comparable to the participants in this work, based on eye movements with reading stimuli using machine learning. Several classifiers were developed based on a group of extracted fixation- and saccade-related features, and accuracies of 63.8% and 60.7% were achieved for the non-dyslexic and dyslexic participants groups, respectively. Furthermore, gender differences in eye movement behavior were found not only in these 2D-based stimulus tasks but also in 2D-based stimulus tasks. In a map direction pointing task (Liao and Dong [260]), males fixated on landmarks significantly longer than females in the 3D map, and a reverse difference between males and females was observed in the 2D map. In addition to these 2D screen-based experimental setups, gender was found to be predictable in a VR-based reading task using only eye movement features and support vector machines, with accuracies near 70% (Steil et al. [261]); however, the reading stimuli in VR were still 2D.

As previous works have examined gender differences in relatively conventional contexts

(i.e., with screen-based stimuli) and in VR contexts (i.e., with 2D stimuli in VR) with limited spatial and temporal characteristics of eye movements, first, it is an open question whether such findings apply to immersive virtual reality environments (e.g., learning environments) for developing computational thinking. It should be mentioned that, to our knowledge, there is no similar research on gender classification based on eye movements with 3D stimuli rendered in VR learning environments. Second, more complex models with multi-modal data and model explanation approach can reveal relationships between gender and the most contributed features, ordered by feature importance, to support computational thinking training rather than analyzing differences separately based on summary statistics (e.g., mean fixation or saccade duration). Gender information is in fact considered protected and should be hidden in the data (Steil et al. [261]; Bozkir et al. [165]), especially when using the commercial application. However, gender recognition is critical for education domain and the development of commercial and noncommercial human-computer interaction applications, and has been studied in depth by a number of researchers (Lin et al. [262]). Particularly in subjects where gender differences typically exist (Reilly et al. [263]), gender prediction can help in providing personalized support during learning and further expand implications for the design of VR and intelligent tutoring systems. Therefore, we investigate gender differences as a proof-of-concept by using eye movements that are obtained in a learning space in an immersive VR environment.

### A.3.4 Dataset

In this section, we give an overview of data acquisition and preprocessing.

#### Participants

Data were collected from 381 sixth-grade volunteer students (179 female, 202 male; average age = 11.5, SD = 0.5), including eye-tracking data and questionnaire data (i.e., demographic data, self-reports of participants' learning background, and VR learning experience). The study was IRB-approved and all participants and their legal guardians provided informed consent in advance.

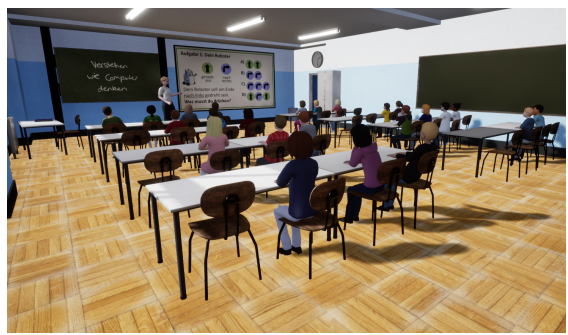
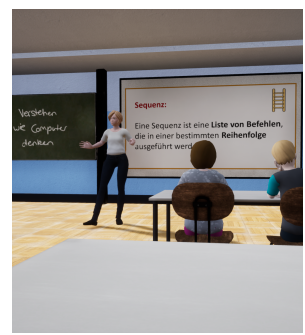

(a) Overall view of the VR classroom       (b) Participant's view

Figure A.10: Computational thinking (CT) learning in an immersive virtual reality classroom.

## Apparatus

We used the HTC Vive Pro Eye with a refresh rate of 90 Hz and a field of view of 110° in our study. Eye-tracking data were recorded using the integrated Tobii eye tracker with a sampling frequency of 120 Hz after a 5-point calibration routine. Virtual environment was rendered using Unreal Game Engine v4.23.1.

## Experimental design and procedure

An immersive VR classroom as depicted in Fig. A.10, similar to conventional classrooms, was used for data collection. Virtual avatars, including the teacher and peer learners, were rendered in one of two visualization styles (i.e., cartoon and realistic). Participants sat in the classroom (i.e., front or back row) and listened to an approximately 15-minute virtual lesson about basic CT principles delivered by the virtual teacher. The lesson *"Understanding how computers think"* consists of four sessions. In the first session, the virtual teacher gives an introduction to CT and asks five simple questions to prompt interaction with learners; in the second session, the teacher explains the terms "loop" and "sequence" to the learners, with four questions following each explanation; in the third session, after the knowledge input by the teacher, the learners are given two exercises to apply the learned content, after a short reflection period, the teacher gives the answers to each question; in the fourth session, the teacher ends the lesson with a summary. During the VR lesson, all relevant learning content, including the CT terms, questions, answers, and exercises, are displayed on the screen on the front wall of the VR classroom. In addition, to mimic a real classroom and increase immersion, a fixed percentage (20%, 35%, 65%, 80%) of virtual peer learners interact with the teacher by raising their hands after each question and by turning around from time to time throughout the lesson.

Each experimental session took about 45 minutes in total, including a paper-based pre-test, the VR lesson, and a post-test. Since the different experimental conditions (i.e., the afore-mentioned visualization styles of the virtual avatar, the seating positions of the participants in the VR classroom, and the percentages of virtual peer learners who were preprogrammed with hand-raising behavior) were not the focus of the present study (see detailed investigation in our previous work, Gao et al., [219]), we trained our classification models based on all eye-tracking data.

## Data preprocessing

We collected raw sensor data including participants' head-poses, gaze vectors, and pupil diameters. Data from participants who experienced sensor-related issues, such as low tracking ratios (less than 90% of eye-tracking signal was recorded), incomplete VR lesson experiences, were excluded. Given that the summary session of the VR lesson ($\approx$ 1.5 minutes) does not include learning activities, we excluded the data from this session. Consequently, data from 280 participants (140 female, 140 male) were used with an average of 13 minutes of head-pose and eye-tracking data. To ensure the quality of the data, we then performed preprocessing of the data for further feature engineering as follows.

Since pupillometry data are affected by noisy sensor readings and blinks, we smoothed and normalized pupil diameter using Savitzky-Golay filter (Savitzky and Golay [154]) and the divisive baseline correction method (Mathôt et al. [155]) with a baseline duration of $\approx$ 1 seconds, respectively. We used a $7°/s$ threshold to detect stationary ($< 7°/s$) and moving ($> 7°/s$) head activities similar to the work of Agtzidis et al. ([122]). In addition, we performed a linear interpolation for the missing gaze vectors. Eye movement events, including fixations and saccades, were detected based on a modified Velocity-Threshold Identification (I-VT) method suitable for the VR setting which takes into account head movements (Agtzidis et al. [122]). In the absence of prior knowledge on how to determine gaze velocity and duration thresholds for fixation and saccade detection in the VR learning context, we set these thresholds based on previous literature (Salvucci and Goldberg [121]; Holmqvist et al. [159]; Agtzidis et al. [122]), but make some adjustments to fit our study. Fixations were detected within stationary head activities using a maximum gaze velocity threshold of $30°/s$, with additional thresholds for a minimum duration of $100ms$ and maximum duration of $500ms$. Saccades were detected by a minimum gaze velocity threshold of $60°/s$ with additional thresholds for a minimum duration of $30ms$ and maximum duration of $80ms$.

### A.3.5  Methods

In this section, we discuss the feature extraction pipeline, machine learning model development for gender classification, model evaluation, and the SHAP explanation approach for feature selection and model interpretation. The procedure of our ML approach for gender prediction is shown in Fig. A.11.

#### Feature extraction

To extract the temporal features from the sensory data, we adopted a sliding-window approach similar to previous studies (Bulling et al. [246]; Hoppe et al. [130]). Since there is no gold standard for the selection of window size in VR learning scenarios, considering different preprogrammed activities (see section A.3.4) occur during the virtual lesson, we initially used a set of window sizes ranging from $10s$ to $100s$ with a step of $10s$. For each window, a vector of 43 features was extracted, most of which related to eye movement information, while one feature was HMD-related; the details of the extracted features and the description of the features are given in Table A.2. For simplicity, we refer to these 43 features collectively as the eye movement features in the following.

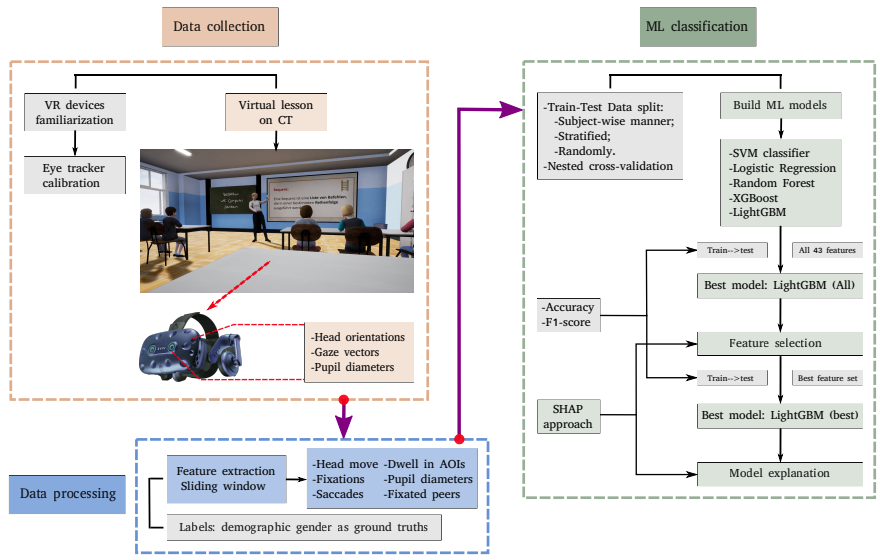

Figure A.11: The ML approach for gender prediction in CT development in the immersive VR classroom.

- HMD-related features: In a previous study, head movements were analyzed and found to be indicative of shifts in social attention during participation in a virtual classroom (Seo et al. [192]). Therefore, we used similar measurement in this study. The number of moving head activities per second, i.e., **hmdMoveRate**, is used as a feature in the classification models (Feature 1 in Table A.2).

- Fixation-related features: Fixations are periods of time when the visual gaze is maintained in a single location. Fixations have been used to understand the learning processes and are considered indicators of attention and cognitive processing activity (Negi and Mitra [202]; Chien et al. [203]). We used features related to fixations, including **fixationRate** and **fixationDuration** (Features 2-7 in Table A.2). Moreover, in our previous study (Bozkir et al., [218]), we found that participants' attention in the VR classroom mainly switches between three virtual objects, also called object-of-interest (OOI), including the virtual peer learners, the virtual teacher, and the screen displaying the instructional content. Therefore, we extracted the number of fixations on these OOIs, i.e., **fixationNumberOnPeer/Teacher/Screen**, as well as their duration, i.e., **peer/teacher/screenFixationDuration** (Features 8-22 in Table A.2). Since dwell time quantifies the time spent looking within an OOI, which includes all fixations and saccades within the OOI as well as revisits (Holmqvist et al. [159]), we additionally extracted **dwellOnPeer/Teacher/Screen** (Features 23-25 in Table A.2). In addition, we extracted the number of peer learners fixated by the participant during the virtual lesson, i.e., **fixatedPeerNumber** as a feature (Feature 43 in Table A.2).

- Saccade-related features: Saccades indicate the rapid shift of the eye from one fixation to another and are also informative eye movements that are highly correlated with visual search behavior (Holmqvist et al. [159]). In a fixation-like manner, we have the number note: Min, max, and SD stand for the minimum, maximum, and standard deviation of the relevant features.

Table A.2: Eye movement features extracted for gender classification model.

| Feature | Description |
|---|---|
| 1. HMD moving rate | Number of moving head activities per second; |
| 2. Fixation rate | Number of fixations per second; |
| 3-7. Fixation duration | Mean, min, max, sum, SD of fixation duration; |
| 8-10. Number of fixation on object of interest | Number of fixations on peer learners, teacher, and screen; |
| 11-22. Duration of fixation on object of interest | Mean, min, max, SD of fixation duration on peer learners, teacher, and screen; |
| 23-25. Dwell time on object of interest | Dwell times on peer learners, teacher, and screen; |
| 26. Saccade rate | Number of saccades per second; |
| 27-31. Saccade duration | Mean, min, max, sum, SD of saccade duration; |
| 32-36. Saccade amplitude | Mean, min, max, sum, SD of saccade amplitude; |
| 37-40. Saccade peak velocity | Mean, min, max, SD of saccade peak velocity; |
| 41-42. Pupil diameter | Mean, SD of pupil diameter; |
| 43. Fixated peer learners | Number of peer learners fixated by the participant. |

of saccades per second, i.e., **saccadeRate**, and their durations, i.e., **saccadeDuration**. Additionally, **saccadeAmplitude** and **saccadePeakVelocity** are employed as features in our models (Features 26-40 in Table A.2).

- Pupil-related features: It is known from previous studies that pupil diameter reflects cognitive load in various human cognitive processes, e.g., visual attention during scene perception (Gao et al. [219]), visual search (Castner et al. [74]), and sustained attention (Appel et al. [128]). Therefore, we extracted features related to pupil diameter (Features 41-42 in Table A.2).

**Classification models**

We used the reported gender demographic data as the ground truth and used discrete variables to represent each gender, i.e., for the purpose of machine learning research only, we set 0 for the female class and 1 for the male class. To be clear, these two numbers have no specific meaning. In this work, we developed five supervised machine learning models to detect participants' gender in a VR lesson on CT learning using eye movement features. Specifically, we used Support Vector Machine (SVM), Logistic Regression, and three ensemble machine learning models, namely Random Forest, eXtreme Gradient Boosting (XGBoost), and Light Gradient Boosting Machine (LightGBM). Thus, a binary classification task was performed on our dataset, which consists of features and the target: $(x_{pw}, y_{pw})$, $x_{pw} = [x_{pw1}, x_{pw2}, ..., x_{pwN}]$, $x_{pw} \in R^d$, where $1 < p < K$, $K = 280$, $p$ represents the sequence number of the participant, $1 < w < M$, $M = 780s / window\_size$, $w$ represents the sequence number of time windows of each participant, and $N$ represents the total number of the eye movement features used, $N = 43$. In short, $x_{pw}$ is the $p * w$-th input vector of all features used for model training; $y_{pw} = \{0, 1\}$ is the target variable (gender), where 0 is class-0 (class-female) and 1 is class-1 (class-male). Before model training, predictor variables were normalized using the maximum-absolute scaling normalization technique.

For training the model based on eye movement features extracted with a specific time window, we performed the nested cross-validation approach to optimize open parameters, i.e., the hyperparameters of the models and window size. A stratified 5-fold cross-validation strategy was applied. For each iteration, we divided the data into a training set, a validation set, and a test set. Particularly, in each iteration, we selected 20% of participants as the test set, 20% of the remaining participants as the validation set, and the rest of the participants as the training set. Thus, for example, with a window size of $60s$, there are more than 3600 data samples, including 2900 data samples for training, and 700 data samples for testing. After 5-fold cross-validation, all participants appeared in the training set and the test set. Note that to avoid overfitting and to generalize our models to unseen data, we performed all data splits in a participant-dependent manner, meaning that all data samples from the same participant should remain in one data set (i.e., either the training, validation, or test set). In addition, participants were randomly assigned without regard to identity. Furthermore, we performed 5-fold cross-validation 10 times, each time selecting different groups of participants as the test set, which further eliminated the participant-group effect on the model. Thus, our models were trained for 50 iterations, and in each iteration, five models were trained on the training set and evaluated on the validation set. We used the F1-score to select the most optimized model hyperparameters. The best hyperparameters were selected based on the validation results. For the final performance evaluation, we trained our models on the unified training and validation set and tested them on the test set to generalize our models to unseen data. Since our dataset is nearly balanced with respect to gender, the chance level is about 50%.

## Model evaluation

Since we have a balanced dataset for binary classification, we evaluated the performance of the models in terms of accuracy and F1-score. TP, TN, FP, and FN stand for True Positive, True Negative, False Positive, and False Negative, respectively. $Accuracy = \frac{TP+TN}{TP+TN+FP+FN}$, is the ratio of correctly predicted observations to total observations. $Precision = \frac{TP}{TP+FP}$, is the ratio of correctly predicted observations to the total predicted positive observations. $Recall = \frac{TP}{TP+FN}$, measures the percentage of true positives that are identified correctly. We measured the F1-score, i.e., $F_1 = 2 \times \frac{Precision * Recall}{Precision + Recall}$, as it accounts for both false positives and false negatives.

**SHAP explanation approach**

Explainability becomes significant in the field of machine learning as it provides insights into how a model can be improved. SHAP (Shapley additive explanation), a game-theoretic approach, is one of the proposed methods to support the interpretation of the prediction results and analyzing the importance of individual features, where the individual feature values are assumed to be in a cooperative game whose payout is the prediction (Shapley [245]). Given that the Shapley value of a feature is its contribution to the payout, weighted and summed across all possible feature value combinations, the Shapley value for a model with a prediction function of $f(x)$ is given as follows. Given $F = \{x_1, x_2, \ldots, x_N\}$ including all the features,

$$\phi_j(f_x) = \sum_{S \subseteq F \setminus \{x_j\}} \frac{|S|!(N - |S| - 1)!}{N!} \left( f_{p=S \cup \{j\}}\left(x_p\right) - f_S(x_S) \right) \tag{A.1}$$

where $S$ is a subset of features and $N$ is the number of features. $\phi_j(f_x) \in R$ stands for the Shapley value of the feature vector $x_j$ (Lundberg and Lee [149]). In our study, we used the local feature attribution method for tree models, TreeExplainer, introduced by Lundberg et al. ([264]). TreeExplainer bridges theory and practice by building on previous model-agnostic research based on classic game-theoretic Shapley values (Shapley [245]; Štrumbelj and Kononenko [265]; Datta et al. [266]; Lundberg and Lee [149]; Sundararajan and Najmi [267]). More details on TreeExplainer can be found in the study by Lundberg et al. ([264]).

SHAP feature importance is measured as mean absolute Shapley values. We used the SHAP approach not only to explain the machine learning models at the feature level (explaining the effects of each predictor variable on the model output, i.e., gender), but also to perform feature selection according to the calculated SHAP feature importance to improve the performance of the classification model. See details in section A.3.6.

## A.3.6 Results

We present gender classification, feature selection by SHAP, and SHAP explainability results as follows.

**Classification results**

To find the optimal window size for eye movement feature extraction in gender classification, we extracted ten sets of eye movement features in ten different time windows. Then we trained five classification models separately with ten extracted feature sets. The hyperparameters of the classifiers were tuned during the training process. Fig. A.12 shows the performance results of five models trained with all eye movement features extracted with different window sizes. As shown, the performance of five models is higher than the chance level (50%) in all time windows. Particularly, we found that a window size of $60s$ provides an optimal trade-off for gender classification compared to time windows of other lengths, as shown by the evaluation results of all five models (see Fig. A.12). Therefore, it can be stated that $60s$ is the optimal window size for eye movement feature extraction in gender classification in this study. In the following analysis, we report the performance of gender classification models trained with eye movement features extracted with a window size of $60s$.

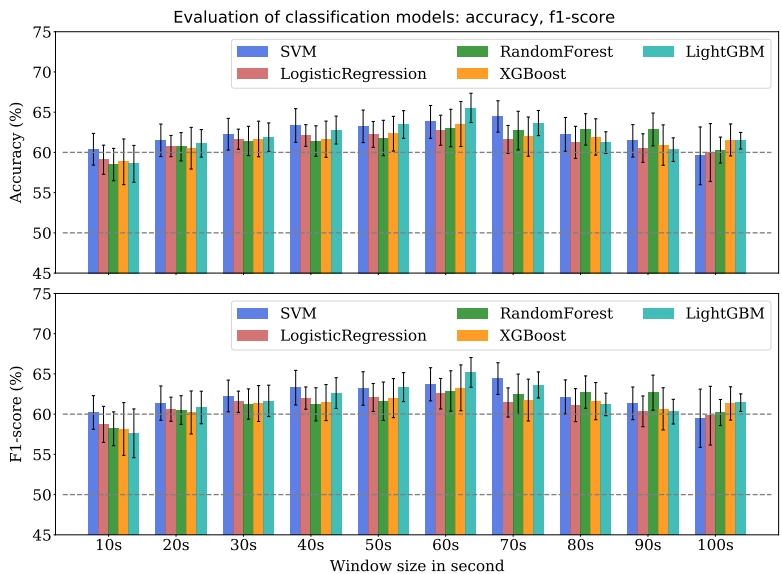

Figure A.12: Performance of all five classification models trained with 43 eye movement features extracted using 10 different time windows; mean values and standard deviations calculated from 50 training iterations.

The comparative performance of all five classification models based on all 43 eye movement features extracted a window size of 60$s$ is shown in the upper part of Table A.3. In particular, the LightGBM classifier performs best with an average accuracy of 65.5% ($SD = 1.8$%), followed by SVM ($M = 63.8$%, $SD = 2.1$%) and XGBoost classifier ($M = 63.5$%, $SD = 2.8$%). The best hyperparameters used for each model trained with all features are listed in the upper part of Table A.4.

Table A.3: Performance of all five classification models trained with all (43) eye movement features and with the selected (best, *feature_number*) eye movement features. The bold font represents the best performance of the model trained with different feature sets.

| Classification model | Accuracy | F1-score |
|---|---|---|
| SVM (all) | 63.8($\pm$2.1) | 63.7($\pm$2.1) |
| Logistic Regression (all) | 62.7($\pm$1.9) | 62.5($\pm$1.9) |
| Random Forest (all) | 63.1($\pm$2.3) | 62.9($\pm$2.5) |
| XGBoost (all) | 63.5($\pm$2.8) | 63.3($\pm$2.7) |
| **LightGBM (all)** | **65.5($\pm$1.8)** | **65.2($\pm$1.8)** |
| SVM (best, 26) | 66.7 ($\pm$1.9) | 66.5 ($\pm$1.9) |
| Logistic Regression (best, 21) | 67.1 ($\pm$2.2) | 66.9 ($\pm$2.2) |
| Random Forest (best, 20) | 67.2($\pm$2.1) | 67.4 ($\pm$2.2) |
| XGBoost (best, 26) | 67.9 ($\pm$2.4) | 67.8 ($\pm$2.3) |
| **LightGBM (best, 24)** | **70.8($\pm$1.7)** | **70.6($\pm$2.5)** |

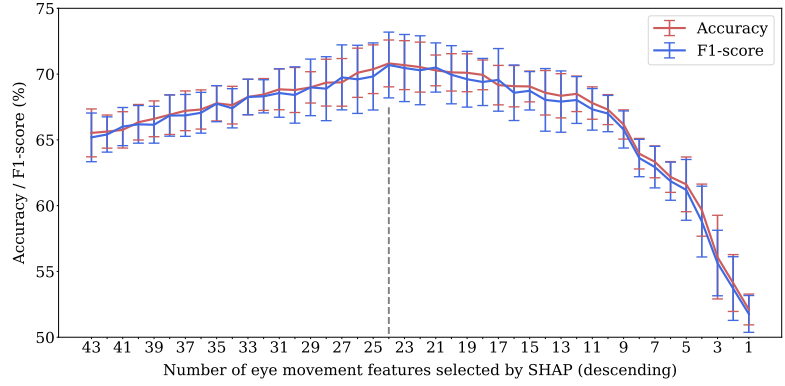

Figure A.13: Performance of the LightGBM model trained with a series of eye movement feature sets obtained by sequentially dropping the least important feature; mean values and standard deviations calculated from 50 training iterations.

Furthermore, we performed feature selection according to the feature importance, represented by Shapley values, to further improve the performance of the machine learning models in gender classification. Specifically, we trained all five models separately with a series of selected feature sets: There are 43 eye movement features in the current feature set, we dropped the least important feature (feature with the lowest feature importance according to the SHAP approach) and trained the model with the remaining features (i.e., the selected feature set); we used the selected feature set as the current feature set for the next loop; in each loop, we dropped the least important feature from the current feature set, we continue the loop until there is only one feature left in the selected feature set. In this way, 43 feature sets with length from 43 to 1 were obtained and used for training.

The comparative performance of all five classification models based on selected best eye movement features extracted with a window size of 60$s$ is shown in the lower part of Table A.3. As can be seen, all models achieved better performance after feature selection by SHAP than the models trained with all 43 features. In particular, the improvement of the LightGBM model is the largest, over 5% improvement in accuracy from 65.5% ($SD = 1.8$%) to 70.8% ($SD = 1.7$%) trained with the top 24 features. In contrast, the SVM classifier shows the least improvement in accuracy, about 3%, trained with the top 26 features. Nevertheless, LightGBM still achieved the best performance in gender prediction among all five models after feature selection. Here, only the feature selection results of the best performed LightGBM model are given, as shown in Fig. A.13. The best hyperparameters used for each model trained with the selected best features are listed in the lower part of Table A.4.

Table A.4: Best hyperparameters of all five classification models trained with all (43) eye movement features and with the selected (best) eye movement features.

| Classification model | Best hyperparameters |
|---|---|
| SVM (all) | 'C': 1, 'kernel': 'RBF', 'gamma': 0.01; |
| Logistic Regression (all) | 'penalty': 'l2', 'C': 0.1, 'solver': 'newton-cg', 'max_iter': 10000; |
| Random Forest (all) | 'n_estimators': 100, 'max_depth': 10, 'min_samples_split': 15, 'min_samples_leaf': 20, 'max_features': 'log2'; |
| XGBoost (all) | 'gamma': 0.1, 'learning_rate': 0.01, 'max_depth': 3, 'min_child_weight': 12, 'subsample": 0.7, 'colsample_bytree': 0.6, 'reg_lambda': 0.5, 'reg_alpha': 0.8; |
| LightGBM (all) | 'n_estimators': 600, 'num_leaves': 150, 'learning_rate': 0.01, 'max_depth': 8, 'min_child_weight': 0.01, 'min_child_samples': 180, 'subsample": 0.7; |
| SVM (best) | 'C': 10, 'kernel': 'RBF', 'gamma': 0.1; |
| Logistic Regression (best) | 'penalty': 'l1', 'C': 0.01, 'solver': 'newton-cg', 'max_iter': 10000; |
| Random Forest (best) | 'n_estimators': 150, 'max_depth': 8, 'min_samples_split': 15, 'min_samples_leaf': 26, 'max_features': 'log2'; |
| XGBoost (best) | 'gamma': 0.1, 'learning_rate': 0.01, 'max_depth': 5, 'min_child_weight': 12, 'subsample": 0.7, 'colsample_bytree': 0.5, 'reg_lambda': 0.5, 'reg_alpha': 0.6; |
| LightGBM (best) | 'n_estimators': 600, 'num_leaves': 200, 'learning_rate': 0.01, 'max_depth': 8, 'min_child_weight': 0.01, 'min_child_samples': 200, 'subsample": 0.8. |

**SHAP explanation**

To understand the contribution of each eye movement feature to the output of the Light-GBM(best) model trained with the top 24 features, we calculated the average SHAP values for each feature in the test data used in the model. Since LightGBM is tree-based model, the Tree-Explainer[10] SHAP was used. A global feature importance and a local explanation summary plot of SHAP values that combines feature importance with feature effects are shown in Fig. A.14 and Fig. A.15, respectively. In the global feature importance plot, a standard bar-chart based on the average magnitude of the SHAP values is illustrated, where the $x$-axis represents the average impact of the features on the model output. In addition, local explanations are plotted in a beeswarm-style SHAP summary plot to examine both the prevalence and magnitude of features' effect. Each point in the summary plot represents the SHAP value for a feature and an observation. The position of each point is determined by the feature on the $y$-axis and by the SHAP value on the $x$-axis. For each feature, blue and red colors indicate low and high feature value, respectively. A change (from left to right) in the color from blue to red along the $x$-axis indicates that the feature has a positive impact on the prediction of the class-1 (class-male) and, on the contrary, a change (from left to right) from red to blue indicates a negative impact of the feature on the prediction of the class-1 (class-male) and thus a positive impact on the prediction of the class-0 (class-female). The greater the impact of a feature on the model output, the more spread out it is on the $x$-axis. In both plots, the features are sorted according to their importance from top to bottom in the $y$-axis.

---

[10]https://shap-lrjball.readthedocs.io/en/latest/index.html

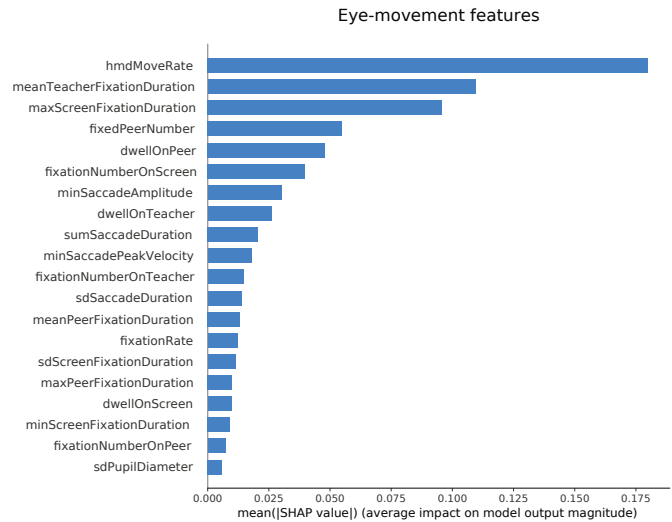

Figure A.14: SHAP global feature importance plot: the top 20 features of the LightGBM(best) model trained with the best eye movement feature set. Bar plot of mean absolute SHAP values of individual features.

As shown in Fig. A.14 and Fig. A.15, features **hmdMoveRate**, **meanTeacherFixationDuration**, and **maxScreenFixationDuration** followed by **fixedPeerNumber** and **dwellOnPeer** provide the maximum information for the LightGBM(best) model in gender classification. Interestingly, we found that pupil- and saccade-related features contribute less to the classification model than HMD- and fixation-related features, and many informative features are features related to objects of interest (i.e., the virtual teacher, virtual peer learners, and the screen) in VR. Furthermore, as can be seen in Fig. A.15, which shows the local explanation of feature importance, features affect the model differently. For instance, the features **hmdMoveRate**, **fixedPeerNumber**, and **dwellOnPeer** have the highest positive influence on classification into class-1 (i.e., class-male); while the features **meanTeacherFixationDuration** and **maxScreenFixationDuration** have the highest negative influence on classification into class-1 (i.e., class-male), in other words, the highest positive influence on classification into class-0 (i.e., class-female).

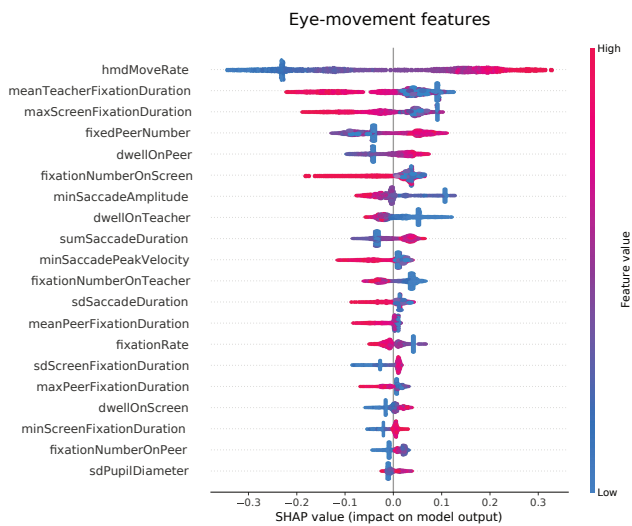

Figure A.15: SHAP local explanation summary plot: the top 20 features of the LightGBM(best) model trained with the best eye movement feature set. The color change in the summary plot (from left to right) of each feature from blue to red indicates a positive influence on classification into class-1; conversely, from red to blue indicates a negative (positive) influence on classification into class-1 (class-0).

Furthermore, before further discussion of the SHAP results in Section A.3.7, we examine the hierarchical relationship between the features to check for redundancy. A dendrogram of the top 24 features used for the best LightGBM was created, as shown in Fig. A.16.

### A.3.7   Discussion

As most gender differences in CT development are assessed by commonly used subjective measures, i.e., questionnaires, acquired CT skills, or similar learning outcomes based on statistical analyses, gender differences assessed by using participants' temporal object behavior with machine learning techniques have not been addressed in previous work. In this work, we investigated the detection of gender differences in CT development process in a VR context using eye movements, which provide a non-intrusive and real-time measure of participants' cognitive and visual behavior during the (VR) learning process. We developed five models using supervised machine learning techniques for gender classification and trained the models with all 43 eye movement features extracted from the recorded eye-tracking data using the selected optimal window size (60$s$, see Fig. A.12). The results show that all classification models perform above chance level (50%), with LightGBM performing best among all models, followed by SVM and XGBoost (see Table. A.3). To improve the performance of the best classification model, i.e., LightGBM, feature selection was implemented using the SHAP approach, and the best eye movement feature set was identified (i.e., top 24 eye movement features according to SHAP feature importance, see Fig. A.13). As a result, an improved average accuracy of 70.8% ($SD = 1.7\%$) for gender classification was achieved after feature selection (see Table. A.3). Therefore, it can be concluded that participants' eye movements provide discriminative information for the LightGBM(best) model in classifying gender. Our gender classification results are in line with previous literature (e.g., 64% in Al Zaidawi et al. [133]; 70% in Abdi Sargezeh et al. [244]), although the stimulus and tasks in our immersive VR setup differ significantly from conventional 2D stimuli used in these works. This is further evidence that even though people's viewing behavior and attention models differ in virtual and real worlds, eye movements that are considered non-intrusive can nevertheless reveal gender differences in learning in VR environments where eye movements and their analysis are more complex. Another previous work (Steil et al. [261]) achieved about 70% accuracy in gender classification, which is similar to the accuracy of our work. As the reading task in their study was performed in a VR context, our results are more comparable; however, the reading task (2D reading stimulus in VR) is much less complex than the CT skills development in an immersive VR classroom in our study. These previous works have further implications for the success of our study regarding the use of eye movement features in predicting gender during more complex learning activities, i.e., CT development, in an immersive VR environment.

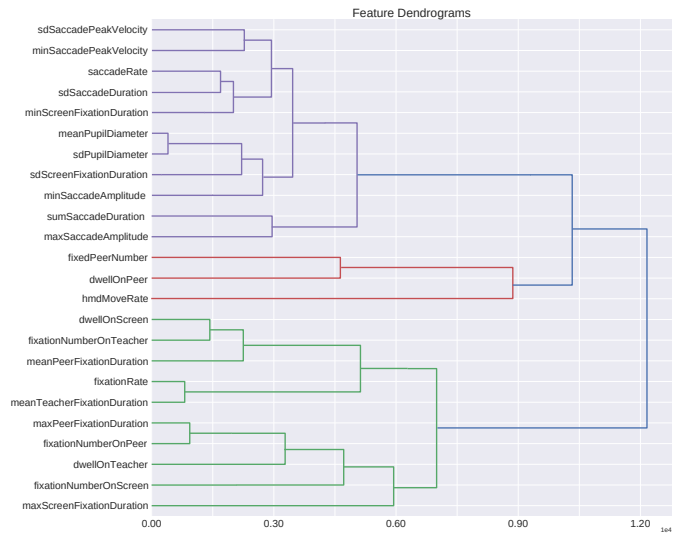

Figure A.16: Hierarchical clustering dendrogram for the features used in the best LightGBM.

In addition to using the SHAP approach for feature selection, the SHAP was used to explain the model, namely to examine the contribution of features to the classification model and the relationships between predictor variables and the target. Specifically, we found that the HMD-related and fixation-related predictor variables influenced the classification model more than the saccade-related and pupillary predictor variables (see Fig. A.14 and Fig. A.15). Notably, the majority of the features in the best eye movement feature set are features related to virtual objects (i.e., the virtual teacher, virtual peer learners, and the screen) in VR, suggesting that participants' eye movement behavior toward these objects in VR provides discriminative information for differentiating gender differences. This finding suggests that the way participants attribute their visual attention to the different classroom content (i.e., the virtual teacher and the screen are instructional content, and virtual peer learners are social comparison information) reflects their gender information. From an educational perspective, this is crucial to know when aiming to design learning environments and instructional support tailored to the different needs of girls compared to boys in STEM subjects (in this case, the development of CT skills). The results imply that, for instance, intelligent tutoring support by virtual peer learners might vary in its positive impact as it is likely to receive different levels of attention depending on the gender of the learners. Similarly, scaffolds and guidance in the instructional material are attended to differently by boys compared to girls, necessitating differentiated implementation (in line with the conclusions drawn by Angeli & Valanides [255]).

Notably, the frequency of participants' head movements was found to be the most informative feature for predicting gender. It was observed that head movement has the greatest influence on classification into class-male, implying that more head movements could be associated with male gender and, in particular, that it is highly likely that boys show more head movement than girls during a VR lesson on CT. This finding is consistent with previous research based on participants' self-reports (Kong et al. [163]; Baser [251]) showing that boys typically exhibit higher levels of interest and self-efficacy in CT and consequently exhibit more exploratory behavior, which is particularly reflected in head movements and further exemplified by eye movements. With regards to the pedagogical design of CT learning environments, this result suggests that girls require more guidance when exploring the VR classroom; girls' interest and self-efficacy in the CT lesson need to be explicitly promoted whereas boys naturally tend to exhibit respective behaviors.

In addition, there are several fixation-related features that have a high impact on the gender classification model, including fixations and dwell on virtual objects, as well as non-specific fixations. This suggests that different gender groups may exhibit different attentional behaviors while participating the VR CT lesson. In particular, participants' attentional behavior toward the virtual teacher has high contribution to the classification model. The features mean fixation duration on the teacher, number of fixations on the teacher, and dwell time on the teacher have high negative impacts on classification into class-male, i.e., a lower feature value than the feature average drives classification into class-male, whereas a higher feature value than the feature average drives classification into class-female. This suggests that different gender groups might display different attentional behavior toward the virtual teacher: Girls were more likely to direct their visual attention to the virtual teacher than boys. Regarding the screen, another instructional object of interest in VR, we found similar finding as for the teacher. In particular, the features maximum fixation duration and number of fixations on the screen have high negative impacts on classification into class-male, implying that a lower feature value than the feature average drives classification into class-male. This may indicate that boys tend to pay less attention to the screen compared to girls. Taken together, our results suggest that the girls' group may pay more visual attention (as indicated by longer and more fixations) to the instructional content (i.e., the virtual teacher and the screen) than the boys' group, which is in line with previous work that girls pay more attention to learning material (Papavlasopoulou et al. [268]).

Moreover, participants' attention to virtual peer learners was also found to provide discriminative information for the model. The features number of peers fixated by participants and dwell time on peers are positively related to the gender classification output; a feature value higher than the feature average leads to classification into class-male. This may indicate that participants' visual attention to virtual peer learners differs between genders, with boys more likely to direct their visual attention to the social comparison information (virtual peer learners) than girls. This could be further supported by the positive impact of feature sum of the saccade duration on classification into class-male, as the teacher, the screen, and especially the peer learners draw attention from different directions around the participants, resulting participants exhibiting longer visual search behavior.

Taken together, all these results further indicate that girls and boys might differ with regards to how they distribute their attention in the VR lesson. Girls, in particular, seem to focus more on the instructional and lesson content. Consistent with previous research (Atmatzidou and Demetriadis [254]; Angeli & Valanides [255]), these results may indicate that girls need more time and more conversation-based instructional support to acquire CT skills compared to boys. Based on the results, specifically instructional guidance provided by the teacher is likely to support girls' CT learning, as they tend to focus more on the teacher compared to virtual peer learners (who could also serve as intelligent tutors in a VR classroom but are more at the focus of boys' visual attention). In addition, the results indicate that boys appear to distribute their attention more across the whole classroom and spend more time decoding the information provided by peer learners in addition to the teacher and the screen. This is consistent with the findings on the differences in head movements between girls and boys (see above) and suggests stronger exploratory behavior in boys, which in turn is likely to indicate greater interest in CT among boys as well as their overall more positive attitudes toward CT development (Baser [251]; Polat et al. [253]; Kong et al. [163]).

Notably, the inferred learning behaviors of boys and girls based on their eye movements in this study reflect relatively high-level and aggregated behavior concerning the overall visual search and attention in the VR classroom rather than responses to specific aspects of the lesson and learning materials. Therefore, implications from an educational perspective concern primarily the overall learning environment design and general instructional support in the VR classroom. Asserting these overall gender differences in learning behaviors in the VR environment provides ground for more fine-grained analyses that further inform the design of tailored learning environments and instructional supports for girls and boys, respectively. SHAP approach provides a valuable way to interpret machine learning models at the feature level by examining the importance of different features that contribute differently to the LightGBM(best) model and by revealing the relationship between eye movement features and gender. Our SHAP results suggest that participants' visual search and visual attention behaviors, particularly attention to different sources of information in the VR environment, such as the virtual teacher (i.e., instruction), the screen (i.e., lesson content), and peer learners (i.e., social orientation), provide different amount of discriminative information for gender classification during a VR lesson on CT. Future studies can build on these findings to examine gender differences in the response to more specific aspects of the learning experience (e.g., critical time points in the lesson, certain presentations of the instructional material) and in other subjects, particularly in those that typically yield gender differences and take place in VR environments.

### A.3.8   Conclusion

To our knowledge, our study is the first to examine gender differences in a VR lesson based on only eye movements using machine learning techniques. Five classification models developed based on all extracted eye movement features were found to be predictive of gender, with LightGBM outperforming the other models. The LightGBM(best) model, which was developed based on the best eye movement feature set selected by SHAP approach, showed an improved performance with an average accuracy of over 70%. In addition, the SHAP approach was used for model interpretation. Our study provides a systematic way to detect gender and explore the relationship between different eye movements (e.g., different attentional, exploratory, search, and cognitive behaviors during CT learning in VR) and gender.

Our findings provide an important foundation for future use of eye movement data to study gender differences in learning in educational contexts, particularly in VR scenarios. Future research building on the findings of this work may offer a promising avenue for improving teaching and learning processes aimed at narrowing gender gaps by gaining a comprehensive understanding of how eye movement features differentially contribute to gender classification: for example, optimizing environmental design in terms of instructional content (i.e., teacher's instructions and screen-based content) and social counterparts (i.e., peer learners) rendered in VR, or optimizing relevant factors for user interface design tailored to different gender groups. These insights offer important implications for the design of future adaptive tutoring systems for educational purposes, particularly in STEM subjects such as CT where gender differences remain pronounced.

### Acknowledgments

The authors thank Jens-Uwe Hahn, Stephan Soller, Sandra Hahn, and Sophie Fink from the Hochschule der Medien Stuttgart for their work and support related to the immersive virtual reality classroom used in this study.

## A.4 Predicting Teacher Expertise Based on Fused Sensor Data from an Immersive VR Classroom and Explainable Machine Learning Models

### A.4.1 Abstract

Currently, VR technology is increasingly used in applications to enable immersive yet controlled research settings. One such area of research is expertise assessment, where novel technological approaches to collecting process data, specifically eye tracking, in combination with explainable models can provide insights into ways to train novices and develop expertise. We present a machine learning approach to predict teacher expertise by leveraging data available from the off-the-shelf VR device collected in a ViRATec study. By fusing eye tracking and controller tracking data, teachers' recognition and handling of disruptive events in the classroom are taken into account. Three classification models were compared, including SVM, Random Forest, and LightGBM, with Random Forest achieving the best ROC-AUC score of 0.768 in predicting teacher expertise. The SHAP approach to the interpretation of models revealed informative features (e.g., fixations on identified disruptive students) for distinguishing teacher expertise. Our study paves the way for evaluating teacher expertise in an interactive virtual environment using eye tracking.

### A.4.2 Introduction

The assessment of individuals' competencies is a central issue across disciplines, such as radiology [269, 270], dentistry [271, 71, 73, 72], surgery [272, 273, 274], sports [135, 275], and computer science [276, 277]. Individual competency is defined as a skill that can be conceived of as a disposition that enables a person to cope with particular situational demands. The disposition may not be directly observable, but actions that result from its existence are. By assessing the expertise of individuals during task performance based on various behavioral data, better exploitation of human resources can be achieved, and moreover, the behavior of experts can be better understood, offering further insights for training novices to become competent. Of particular interest to the education field is the assessment of teacher expertise, especially taking their professional vision while teaching in the classroom into account. In order to ensure a high quality of teaching in the sense of good classroom management, teachers must be able to recognize important cues in students' behavior. The ability to focus attention on important situational features in the classroom, grasp their significance, and use their knowledge to derive effective teaching actions is referred to as teachers' professional vision [278]. However, selectively perceiving visual cues and consciously recognizing their relevance to classroom events is a situation-specific skill and requires practical experience [279]. Previous research has shown that expertise and experience influence teachers' visual perception of the classroom [280].

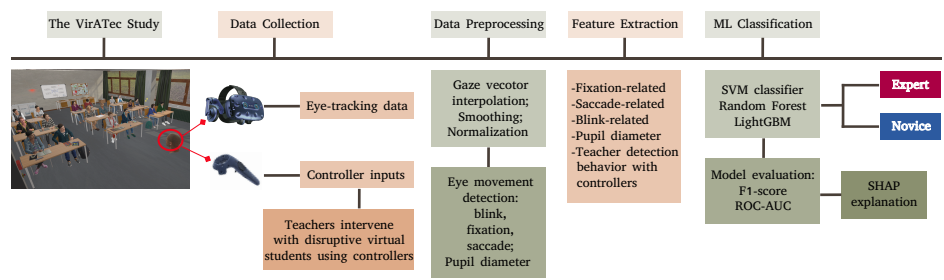

Figure A.17: Steps of the machine learning approach to predict teacher expertise in classroom management based on sensor data collected in the ViRATec study using an immersive VR classroom.

To this end, eye tracking offers a viable and practical avenue to assess teachers' professional vision in the classroom context, as it has been shown to provide an intuitive interface for investigating human visual perception and cognitive processes in a non-intrusive and objective manner [159, 197]. Eye movements have been shown to be capable of revealing underlying individual differences in various tasks [134, 128], and particularly in expertise [71, 135, 157, 281]. Researchers have also made significant advances in using eye tracking technology to study teachers' professional vision in the classroom [173, 282, 283]. Previous studies provided interesting insights into the differences between expert and novice teachers in terms of their visual perceptual behaviors in recognizing disruptive classroom events, using eye movement measures such as fixations and scanpaths [284]. However, studies on teachers' professional vision show discrepancies depending on whether teachers were watching an instructional video (on-action scenario) or were teaching themselves (in-action scenario; [285]). These differences are not surprising, considering that in regular teaching situations, teachers are faced with complex circumstances in which they have to monitor students' behavior, reply to students' questions, and think about how to convey the learning content well understandable – all at the same time and in a fast pace [286]. Thus, in order to provide an authentic context for action while ensuring comparability across study participants, technological advances that have recently been made in areas such as immersive virtual reality can be used [17].

Digital setups combined with eye tracking technology, particularly immersive virtual reality (VR) offer several advantages over real classrooms and provide viable opportunities for exploring teacher expertise in classroom management. With VR, specific experimental conditions can be easily manipulated by pre-programming virtual avatars and virtual environment configurations to create a realistic and immersive environment [40, 218, 219]. In addition, conducting experiments in VR overcomes significant privacy concerns, as no real videos of students are recorded. And, more importantly, a recording of synchronized sensor data (e.g., eye- and controller-tracking data) can be achieved with advanced, off-the-shelf VR devices without too much effort in an accurate way. Considering that digital setups have entered our daily lives, especially during the COVID-19 pandemic and the subsequent push into VR education [287], immersive VR (IVR) classrooms may be the next big step in digital education. In the long run, it may be even possible for students to attend a virtual lecture that takes place in a virtual classroom more regularly to get immersive and authentic experiences. Researchers have already embarked on this line of research [2, 288], particularly from the perspective of students, focusing on examining various classroom configurations [219], or the effectiveness of intervention strategies with distracted students [125], using eye tracking as an assessment tool. While some previous work explored classroom complexity and size from the perspective of novice teachers in VR [42, 44], as virtual reality technology also supports teacher training [29], some previous studies have implemented VR simulations and even an IVR classroom [31] to help novice teachers develop their expertise in classroom management by using VR as a research tool. Particularly, comparisons were made between video and VR as tools to promote interest and self-efficacy in classroom management, with questionnaires [22]. However, these studies have not investigated teachers' in-action behaviors (particularly their eye movements), which could further support the development of teacher expertise. To our best knowledge, there has been no work that considers eye movements in VR from a teacher's perspective.

Considering these research gaps, VR offers a promising avenue to study teachers' expertise in recognizing disruptive events in the classroom and reactions to address them. First, a realistic classroom environment can be easily created to provide teachers with an authentic and immersive teaching experience, and second, not only can teachers' real-time eye movements be recorded, but teachers' reactions to disruptive classroom events performed with controllers can also be recorded simultaneously. In this study, we use data collected from the VirATec study (see Section A.4.4) using an immersive VR classroom. We aim to investigate whether teachers' in-action behaviors, particularly eye movements, contain discriminative information for distinguishing between novice and expert teachers. To do this, we build and train several novice-expert classification models based on a large set of features extracted from sensor data, including eye tracking and controller inputs. Unlike previous studies that used limited eye movements based on an average measure to examine teachers' professional vision in recognizing disruptive events in the classroom [157, 158], this study used all eye movement information along with controller inputs throughout the classroom experience. That is, teachers' pupil diameter, fixations, saccades, and relevant eye movements to area-of-interests (AOIs) were used to capture teachers' visual attention, visual search, as well as underlying cognitive processing behavior [159] during the recognition of disruptive student behavior. With the machine learning approach, analyzing a large number of eye tracking features is no longer a challenge, as this is what machine learning excels at. Based on the fused features, teacher expertise can be successfully predicted with our novice-expert models, with minimum performance above 0.74. Moreover, the Shapley Additive Explanations (SHAP) [149] approach was applied for model explainability to determine what particular features contribute most to the classification models, which provides implications for the development of expert eye movements by novices. We further report basic statistical comparison results of various features between expert and novice teachers. In summary, our main contribution is threefold and can be summarized as follows.

- We explore the task of predicting teacher expertise in an immersive VR setup using fused

sensor features, including eye tracking, along with controller inputs. We build three novice-expert models, including Support Vector Machine (SVM), Random Forest, and Light Gradient Boosting Machine (LightGBM). We demonstrate the predictability of expertise in the educational domain, demonstrating the discriminative power of eye tracking data. In our study, the classification performance of about 0.78 was achieved.

- We analyze summary statistics on features from teachers' behavioral data. We provide statistical evidence for the informativeness of sensor features, further complementing the findings of the model explanation approach.
- We identify the most important features for the novice-expert classification models by providing a post-hoc model explanation using the SHAP approach. Underlying relationships between teachers' behavioral features and their expertise were uncovered, which are supported by the statistical findings. Our study provides a foundation and insights for future studies aimed at recognizing teacher expertise and further developing gaze-based training interfaces for novices.

### A.4.3   Related Work

Since our work benefits from two areas, teachers' professional vision and classification of expertise based on eye tracking, we also discuss the previous work in this way.

#### Teachers' Visual Perception

With the development of pervasive eye tracking technologies that provide easy-to-calibrate procedures without physical constraints on users, a body of research has examined how teachers perceive classroom events based on eye movements and has found that visual perception of classroom events varies, particularly between novice to expert [281, 289, 290, 291]. Previous studies mainly investigated this research question using videotaped classroom observations, which create an on-task assessment setting in which teachers have no agency. For instance, Van den Bogert et al. [157] and Imai-Matsumura [292] investigated differences in teachers' visual attention in recognizing disruptive student behaviors using fixation measures. From these studies, expert teachers generally process visual information more quickly and effectively than novice teachers, as indicated by shorter fixation durations and a greater number of fixations on target students.

In addition to the assessment of visual attention, some other works have examined the differences in teachers' professional vision from the perspective of teachers' viewing strategies. Kosel et al. [158] compared the visual scanning patterns of expert and novice teachers in assessing learning-relevant student characteristics. They found that experts observed all student targets more regularly and exhibited more complex scanpaths that included more frequent revisits of student targets. Similarly, McIntyre and Foulsham's work [293] also analyzed teachers' dynamic viewing behaviors and found significant differences between experts and novices.

These previous works have shown that eye movements are informative for differences between expert and novice teachers in the way they visually perceive classroom events, especially in the way they direct their visual attention to disruptive students and in the visual strategies they use. Therefore, previous findings possess strong evidence that eye movements have a great potential to provide discriminative information for predicting teacher expertise in a machine learning approach.

**Eye-based Expertise Classification**

Expertise classification is one of the most popular topics in the field of machine learning, which helps to identify the expertise of individuals in a particular domain to further facilitate various aspects such as better exploitation of human resources, novice training, and so on. In this research area, eye tracking data, in particular, used as an excellent interface to reveal individual differences in visual perception and cognitive processes during different tasks, has been found to provide discriminative information for machine learning models of expertise classification. In particular, eye tracking has been used extensively to distinguish between experts and novices in various fields, such as radiology [269, 270], dentistry [271, 71, 73, 72], surgery [272, 273, 274], sports [135, 275], and computer sciences [276, 277].

Castner et al. [71] successfully differentiated sixth- through tenth-semester dental students viewing orthopantomograms (OPTs) with classifiers using eye tracking (i.e., scanpath), achieving considerable accuracies above chance level. The authors stated that knowledge of experts' strategies for viewing OPTs offers further potential for facilitating dental student education, particularly for first-year dental student training. The same authors also showed in a follow-up work [73] that using deep semantic embeddings based on eye fixations can achieve prediction accuracies of over 70% to distinguish experts from novices. Another work [72] in the same domain found that employing saccadic features with Long Short-Term Memory (LSTM) achieved comparable performance in predicting expertise as the earlier work [73].

Hosp et al. [274] built a support vector machine (SVM) classifier based solely on eye movement features to distinguish surgeon expertise at three levels. Eye tracking data were collected from a total of fifteen participants (5 for each level). A comprehensive set of eye movement features was used, consisting of a total of 38 related to fixations, saccades, smooth pursuits, and pupil diameters. The results showed that an accuracy of 84.4% was achieved with only the most important features, suggesting that eye movements are highly informative in discriminating surgeons' expertise. In the same domain, Ahmidi et al. [272] employed eye tracking data to predict surgical expertise in the sinus surgery task using Hidden Markov Models and reported an accuracy of over 80%. These results support the argumentation of Eivazi et al. [273, p. 1] that eye gaze has the potential to be further utilized in models of surgical expertise.

Assessment and prediction of expertise on programming tasks in the computer science domain using eye movements have been researched as well. Ahsan and Obaidellah [276] proposed the prediction of programming expertise using k-nearest neighbors and SVMs. The total fixation durations on different AOIs were used as features. Their results showed that it is possible to predict expertise with over 80% accuracy. In another work, Lee et al. [277] predicted developer expertise with an SVM classifier based on biometry, including eye tracking and electroencephalogram (EEG) data. Their classifier could predict developer expertise with over 85% precision using only eye movement data. When eye movements were fused with EEG, the models achieved over 95% precision.

In terms of how the stimuli are presented, the expertise classification based on eye movements has also been studied in a VR-based scenario similar to our current study. Hosp et al. [135] applied a machine learning approach similar to that in [274] in the sports domain to classify the expertise levels of soccer goalkeepers into three levels. Compared with [274], eye movement data were collected from a larger number of participants (i.e., 35 in total, with 13 novices, 10 intermediates, and 12 experts) and more eye movement features (i.e., 46 in total) were extracted for an SVM classifier. An accuracy of 78.2% was achieved, suggesting that eye movements are discriminative features for athlete expertise prediction in VR.

The aforementioned works have demonstrated the great discriminative power of eye movements in distinguishing different expertise classes in different domains using machine learning methods. Expertise prediction in education, and particularly teacher expertise, has not been investigated yet, although previous works on teacher professional vision have shown that eye movements are indicative of teachers' visual expertise, albeit to a limited extent (i.e., using limited eye movement measures in an average fashion). All these research gaps have inspired our proposal of a novel machine learning approach in combination with model explainability that leverages a large amount of eye tracking data along with controller tracking data to predict teacher expertise in professional vision. With such an approach, we not only provide a very suitable machine learning strategy for detecting teacher expertise but also a way of in-depth investigation of the differences between expert and novice teachers in terms of recognizing and intervening in disruptive classroom events.

### A.4.4 Data Collection

To test the feasibility of our proposed machine learning approach for predicting teacher expertise in recognizing disruptive classroom events, we used data from the VirATec study, in which participants gave a presentation to eighteen virtual students and observed and identified those virtual students who exhibited (pre-programmed) disruptive behaviors. Participants' eye tracking data and reactions (indicated by controller inputs) to disruptive events were recorded. The following section describes the design of the VirATec study in detail.

### Participants

In this study, a total of 42 participants, including experienced (i.e., expert) and pre-service (i.e., novice) teachers from diverse subject domains (i.e., chemistry, language, math, sports), were recruited to participate in our study. 17 of them are experienced teachers (8 female; 9 male), aged between 28 and 58 ($M = 40.5$, $SD = 9.7$), teaching in local German high schools and vocational schools; 25 are pre-service teachers (11 female; 14 male) aged between 21 and 32 ($M = 24.5$, $SD = 2.2$), obtaining a teaching qualification for German high schools and vocational schools.

All participants reported their experiences with video games and VR. Of the experienced teachers, two ($\approx 12\%$) played games about once a week, two ($\approx 12\%$) played about once or twice a month, and the rest ($\approx 76\%$) played no video games; two ($\approx 12\%$) used VR more than five times, one ($\approx 6\%$) used VR three to five times, four ($\approx 24\%$) had VR experience once or twice, and the rest ($\approx 58\%$) had no VR experience. Of the pre-service teachers, three (12%) played video games very frequently (more than twice a week), one (4%) played about once a week, nine (36%) played about once or twice a month, and the rest ($\approx 48\%$) played no games; five (20%) used VR more than five times, seven (28%) had VR experience once or twice, and the rest (52%) had no VR experience.

Each novice teacher was compensated with 30 euros for participating in the study. The expert teachers are voluntary participants in the study. The ethics committee of the University of [anonymous] and the regional council approved the study. All participants provided written informed consent prior to the start of the experiment and were informed that they could terminate the experiment at any time if they felt uncomfortable.

### Apparatus

The VR environment was developed in Unity[11] and displayed using the HTC Vive Pro Eye head-mounted display (HMD). The HMD has a refresh rate of 90 Hz and a field of view of $110°$. The screen resolution for each eye was $1440 \times 1600$. For eye tracking, the integrated Tobii eye tracker was used with a sampling rate of 120 Hz and an accuracy of $0.5° - 1.1°$ for default calibration. Vive hand-held controllers supplied with the HMD were used to interact with the VR environment.

### Experimental Design and Material

An immersive VR (IVR) classroom that mimics traditional classrooms was designed and deployed (see Figure A.17). The IVR classroom is configured with three rows of fixed seating, each row with three desks and three pairs of chairs. In this way, eighteen pre-programmed virtual secondary school students sit in the classroom facing the teacher. The teacher's avatar, played by the participant, initially stands at the front of the classroom and gives a presentation on a non-teaching related topic to the virtual students. Teachers explain an illustrated children's story of "Max and Moritz" from Wilhelm Busch [12]. The presenting content is unknown to the teachers so that differences in their behavior are not confounded by their subject knowledge. During the VR immersion, the teacher is free to walk around the podium. Considering our main research goal to investigate the differences in teachers' recognition of disruptive student behaviors, relevant environmental conditions can be implemented in VR. That is, all virtual students are pre-programmed to display specific behaviors during the teacher's presentation. A distinction was made between attentive behavior (i.e., looking ahead) and two types of off-task behavior (see Figure A.18): active (i.e., students look at each other and talk to each other, share learning material with each other) and passive (i.e., students put their head on the desk, look out the window, play with the pen or their hands). These students' attentive and off-task behaviors are programmed into each 30-second time slot of the approximately 10-minute VR presentation. Note that only a certain number of virtual students exhibit off-task behaviors in each 30-second time slot, and that both the virtual students and their off-task behavior are different in these 30-second time slots. To capture the teacher's decision and action regarding students' behavior while increasing authenticity and immersion, teachers can intervene with controllers, i.e., clicking on the disruptive virtual students with the controller stops the disruptive student behavior and the detected virtual students are illuminated as feedback in VR.

---

[11]https://unity.com/
[12]https://www.wilhelm-busch.de/

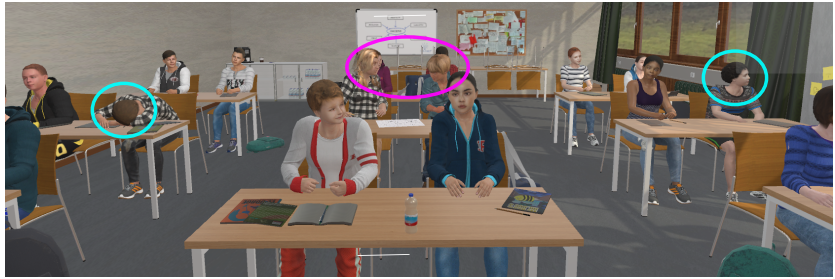

Figure A.18: Passive (blue) and active (red) disruptive behaviors of virtual students during the teacher's presentation in the IVR classroom.

**Procedure**

The experiment consisted of three parts: a paper-based pretest, the virtual classroom experience, and a paper-based posttest. In the first part, before entering the VR environment, participants were asked to complete a questionnaire about demographic information (i.e., age, gender) and their previous experiences with VR and video games. This part took about 30 minutes.

The second part took place in an IVR classroom. Before the actual data collection took place, the experimenter first gave the participants an introduction to the experiment and instructed them on how to use the VR equipment and how to conduct a presentation in VR. Participants were asked to observe the behavior of the virtual students during the presentation and were told that they could use the controller to react to those virtual students who exhibited disruptive behavior. Participants were explicitly informed that their presentation would not be verified, as it was only to increase the immersion and authenticity of the VR experience. After a detailed briefing, participants had time to familiarize themselves with the entire VR experiment process and hardware. This preparation part took about 20-25 minutes. Once participants felt ready, the actual data collection began. Participants wore the HMD and held the controller in their hands. Before entering the IVR classroom, a standard 5-point eye-tracking calibration routine was performed. After successful calibration, participants gave an approximately 10-minute presentation to the virtual secondary school students in the IVR classroom. After completing the presentation, they were asked to take off the HMD. During the actual experimental session, a variety of sensor data were recorded, including eye tracking, HMD tracking, and controller tracking data.

In the last part, participants filled out a questionnaire in which they gave their assessment of the virtual classroom experience. The entire experimental session lasted approximately 2 hours in total.

### Data Preprocessing

Data from two expert teachers were excluded due to major incomplete eye tracking recordings. All remaining participants had eye tracking ratios above 85% (i.e., at least 85% valid eye tracking signal). Therefore, an average of $\approx 10$ minutes of behavioral data from 15 experts (7 female; 8 male) and 25 novices (11 female; 14 male) were used for further data analysis. Since only noisy and unprocessed raw eye tracker data were available, we preprocessed the data to ensure data quality and then detect eye movement events for further feature extraction of the machine learning approach. The preprocessing of the eye tracking data proceeds as follows.

The preprocessing of the data mainly concerns sensor data including pupil diameters, gaze vectors, and head orientations. Given that pupil data are affected by blinks and noise, we smoothed pupil diameters using the Savitzky-Golay filter [154] and applied a divisive baseline correction [155] with a baseline duration of $\approx 1$ second for normalization. In addition, commonly used eye movements such as blinks, fixations, and saccades were detected manually post-experimentally in this study. Specifically, we applied a blink detection algorithm based on the fluctuations that characterize the pupil data, as proposed by Hershman et al. [294]. For fixation and saccade detection, we used the eye movement event detection algorithms from the work of Gao et al. [219], which also employed a VR classroom setup, with parameters adapted to our study. Specifically, eye movement events were detected based on the velocity-threshold identification (I-VT) algorithm [121] with consideration of head movements. Before event detection, linear interpolation was performed for the missing gaze vectors. Fixations were detected with a maximum gaze velocity threshold of $40°/s$ under the condition of relatively stationary head movement (head movement velocity less than $12°/s$). In addition, the minimum ($80ms$) and maximum ($600ms$) duration thresholds were used to filter fixations. Saccades were detected using the normal I-VT algorithm, with the threshold for minimum gaze velocity set at $50°/s$. In addition, the minimum ($30ms$) and maximum ($80ms$) duration thresholds were used to filter saccades. The processed pupil diameters and detected eye movement events can be further used in feature extraction for classification models (see Section A.4.5).

## A.4.5 Expertise Classification

In this session, we present our machine learning approach to predicting teacher expertise in professional vision in an IVR classroom based on teachers' behavioral data recorded in VR, i.e, eye tracking data and reactions towards disruptive virtual students. Our method is presented step by step in Figure A.17. First, we segmented the sensor data according to the 30-second time window used as a unit in the VR program for programming disruptive behaviors (see Section A.4.4). Second, we extracted feature vectors based on segmented data sets. We extracted statistical features from the eye tracking and reaction behavior data to characterize variables in the data (e.g., mean, minimum, maximum, and sum) [130, 147, 148]. Third, we define our task of identifying teacher expertise as a binary classification problem as we have two teacher groups (i.e., experts and novices). Consequently, we labeled all feature vectors according to the teacher's expertise level. We developed and trained several novice-expert classification models based on extracted behavioral features. The details are described below.

### Feature Extraction

Since the teachers' behavioral data consisted of two parts (i.e., eye tracking and controller tracking), the features were extracted accordingly. All extracted features are listed in Table A.5.

With inspiration from earlier works, in this study, we are interested in whether teachers' visual perceptual behavior (measured with eye tracking) in a machine learning approach is informative about teacher expertise. Therefore, we extracted relevant eye movement features that are considered indicators of human visual attention, visual search, and cognitive processing load [159]. Fixations, i.e., periods of time during which the gaze fixates on a fixed point, have been used extensively as a measure to study human visual attention behavior in various tasks, such as visual attention allocation to visual stimuli, processing of visual information, and visual priority [295, 296, 66]. Moreover, previous work has shown that fixations are an indicator of cognitive processing load [297, 298]. Saccades, on the other hand, defined as rapid gaze shifts between fixations, reveal human visual search behavior when exploring visual scenes [159]. Saccadic features such as saccade amplitude and saccade velocity have also been found to correlate with cognitive load [299, 298, 300]. Similarly, pupil diameters and blinks have also been found to correlate with cognitive load [226, 298, 301]. Therefore, we extracted the following eye movement features. Fixation-related features, such as the number of fixations, fixation duration, and fixations in AOIs (i.e., virtual students), were extracted. For saccades, the number of saccades, saccade duration, saccade amplitude, and saccade peak velocity were extracted as features. In addition, we calculated the ratio between saccade duration and fixation duration as a feature indicating the balance between visual search (as indicated by saccades) and visual information processing behavior (as indicated by fixations) [197].

Features indicative of teachers' cognitive processing load during classroom immersion were also extracted, such as the number of blinks, blink duration, and pupil diameter.

We measured teachers' visual perception of disruptive student behavior and whether they identified it as relevant during VR instruction. Thanks to advanced VR technology, teachers' reactions toward disruptive classroom events can be easily and accurately recorded by the controller. We told the participants before the experiment that they can stop the virtual student who exhibits a (pre-programmed) disruptive behavior by pointing the controller at that student. In this way, we can easily observe how many disruptive students were detected (clicked) by the teacher. In addition, we are particularly interested in the teachers' gaze behavior toward virtual students who are perceived as disruptive. Overall, we extracted the features of the teachers' eye tracking in combination with reaction behavior, i.e., the number of reactions (clicks) by the teacher on the AOIs and the fixations in the C-AOIs (i.e., disruptive virtual student clicked by the teacher).

**Model Building**

To investigate whether it is possible to identify teacher expertise based on teachers' behavioral data, mainly eye tracking data, and to further investigate which classification technique performs best on this task, we selected three conventional machine learning classifiers commonly used in binary classification tasks and compared their performance. The classifiers developed and compared are Support Vector Machine (SVM), Random Forest, and Light Gradient Boosting Machine (LightGBM). Thus, we built three different novice-expert models with the following steps.

Since the feature vectors were extracted from segmented data sets of each participant, this results in about 800 observations for training, validation, and testing: $N = P \times W$, where $N$ is the total number of observations, $P$ is the number of participants, $P = 40$, and $W$ is the number of observations for each participant, $W = 10\,minutes/30\,seconds$. Min-max normalization was performed for all feature variables before feeding them into models. Since our binary classification target is the teacher's expertise level, we labeled the feature vectors to one of the classes corresponding to the teachers' expertise (i.e., expert and novice). We used the integers 0 and 1 to represent two classes, i.e., class-0 for novice and class-1 for expert, resulting in 500 observations for the novice class and 300 observations for the expert class. Note that the integers 0 and 1 are used here only for the purpose of machine learning research and have no specific meaning.

After data preparation, we split the data samples into a training set and a test set in a ratio of 80:20 (about 640 samples for training and 160 samples for testing). All three classification models were trained with the training set and tested with the test set. In the training process, 5-fold cross-validation was applied to the training data to tune the hyperparameters of the models to achieve the best prediction performance. In the 5-fold cross-validation, the training set was further split into five subsets, in each iteration with one set serving as the validation note: Min and Max denote the minimum and maximum, respectively; AOI refers to the virtual student; C-AOI refers to disruptive virtual student clicked by the teacher.

Table A.5: Features extracted from teacher behavioral data in the IVR classroom for the novice-expert classification models.

| Categories | Features | Statistical Metrics |
|---|---|---|
| **_a) Eye tracking features_** | | |
| Fixation-related | -Number of fixations | Total number of fixations |
| | -Fixation duration | Mean, Min, Max, Sum of the fixation durations |
| | -AOI number of fixations | Total number of fixations in AOIs |
| | -AOI fixation duration | Mean, Min, Max, Sum of the fixation durations in AOIs |
| Saccade-related | -Number of saccades | Total number of saccades |
| | -Saccade duration | Mean, Min, Max, Sum of the saccade durations |
| | -Saccade amplitude | Mean, Min, Max, Sum of the saccade amplitudes |
| | -Saccade peak velocity | Mean, Min, Max of the saccade peak velocities |
| Fixation-saccade | -Sacc-fixa ratio | Ratio between saccade duration (Sum) and fixation duration (Sum) |
| Blink-related | -Number of blinks | Total number of blinks |
| | -Blink duration | Mean, Min, Max of the blink durations |
| Pupil-related | -Pupil diameter | Mean, Min, Max of the normalized pupil diameters |
| **_b) Teacher's reaction behavior features (combined with eye tracking)_** | | |
| Controller | -Number of clicks on AOIs | Total number of clicks on the disruptive virtual students |
| Controller & Eye | -C-AOI Number of fixations | Total number of fixations in clicked AOIs |
| | -C-AOI Fixation duration | Mean, Min, Max, Sum of the fixation durations in clicked AOIs |

set and the other four sets serving as the sub-training set. The F1-score was used as a measure during hyperparameters tuning. The best hyperparameters of the model were determined by cross-validation.

Notably, to avoid overfitting and to generalize our novice-teacher models to unseen data, all data splits (including the 80:20 split and the splits in 5-fold cross-validation) were participant-dependent, that is, all data samples from the same participant were to remain in one data set (i.e., either the sub-training, validation, or test set). In addition, all data splits were performed using the stratified-sampling approach, that is, data samples were split into different sets while maintaining the percentage of samples for each class. Besides, when splitting the data, participants were randomly assigned without regard to identity. These data splitting policies were performed manually. Furthermore, to reduce the bias caused by data splitting, we performed data splitting 20 times with different random states, that is, data samples were split 20 times with different random seeds so that all data samples (participants) had the opportunity to be used as training and test sets. Since 5-fold cross-validation was applied, our models were trained and validated on $20 \times 5$ different sub-training and validation sets. Therefore, the best model was determined from $20 \times 5$ times training, re-trained on the training set (union of sub-training and validation sets), and tested on the test set. The F1-score and ROC-AUC score were used as measures of the generalizability of the models (see Section A.4.5). The data splitting strategy is described in Figure A.19.

**Model Evaluation and Explainability**

Since our data are not perfectly balanced distributed (i.e., 15 experts vs. 25 novices), we evaluated the performance of the novice-expert models using metrics including F1-score and the area under the Receiver Operating Characteristic Curve (ROC-AUC). The F1-score is the harmonic mean of precision and recall, designed to perform well on data that are not highly imbalanced, where $F_1 = 2 \times \frac{Precision * Recall}{Precision + Recall}$, with $Precision = \frac{TP}{TP + FP}$ and $Recall = \frac{TP}{TP + FN}$ ($TP$, $TN$, $FP$, and $FN$ stand for True Positive, True Negative, False Positive, and False Negative, respectively). The receiver operating characteristic (ROC) curve is a widely accepted method that indicates the True Positive Rate (i.e., $TPR = \frac{TP}{TP + FN}$) and the False Positive Rate (i.e., $FPR = \frac{FP}{FP + TN}$) at different classification thresholds. The area under the ROC curve (AUC) is the value interpreted as the overall performance of a classification model at all classification thresholds.

Model explainability is very important and challenging in the field of machine learning. Many previous machine learning studies focus on improving model performance. However, understanding what particular features and how they contribute to the model performance can not only increase confidence in the validity of models, but also provide clues for improving model performance by knowing exactly what to fine-tune and optimize. In particular, when using interpretable features (e.g., eye movements) that reveal subconscious human behavior during a task, a post-hoc model explanation approach can shed light on the underlying relationships between such features and the specific human task (in this study: teacher expertise).

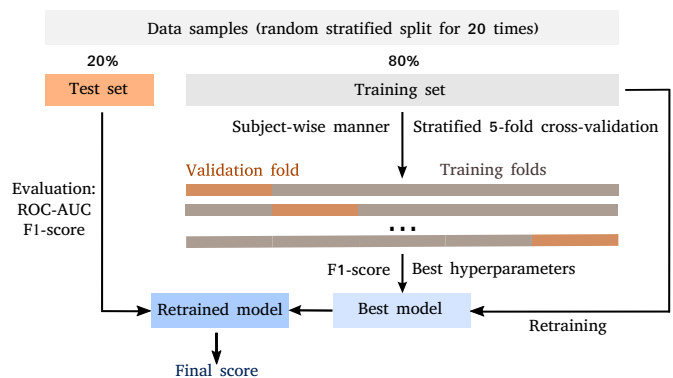

Figure A.19: Data splitting strategy used in the machine learning approach.

SHapley Additive exPlanations (SHAP) by Lundberg and Lee [149] is a novel approach that computes the Shapley value based on a game theory to explain the outputs of machine learning models. It provides a method to gauge and demonstrate how each feature affects the model. Therefore, in this study, we applied the SHAP model explanation approach to explaining our novice-expert models for predicting teacher expertise. The SHAP value for each feature instance and the overall impact of each feature on the model output can be quantified and illustrated. In this way, the most discriminative features can be identified and the underlying relationships between these features and the model output (i.e., novice vs. expert class) can be revealed. The detailed SHAP results are presented in Section A.4.6.

### A.4.6   Results

In this section, we report the results of statistical tests on the sensor features used for expertise prediction, the classification performance of the novice-expert models, and the model explainability by the SHAP approach.

#### Statistical Tests

We performed statistical tests for the extracted features, using the Mann-Whitney U test as a nonparametric test. The results of the statistical tests are presented in Table A.6. We found that both two feature types (i.e., eye tracking and reactions toward disruptive students) were informative of teacher expertise. With regard to eye tracking features, teachers with different levels of expertise were found to show significantly different eye movement behaviors. Specifically, expert teachers were characterized by significantly fewer fixations, shorter fixation duration (sum), and shorter fixation duration (sum) in the AOIs than novice teachers. Furthermore, teachers were also found to show significantly different saccadic behavior, with expert teachers having fewer saccades, shorter saccade duration (max, sum) and saccade amplitude (max, sum), and lower saccade peak velocity (max) than novice teachers. In addition to fixations and saccades, significant differences were found in blink-related and pupil-related features. The expert teachers blinked more frequently with shorter blink duration (mean) and had smaller pupil diameters than novice teachers.

Teachers' reactions toward disruptive virtual students were also found to be informative of teacher expertise. It was found that expert teachers reacted to students showing active note: Min and Max denote the minimum and maximum, respectively; AOI refers to the virtual student; C-AOI refers to disruptive virtual student clicked by the teacher; *, **, ***, and **** represent $p < .05$, $p < .01$, $p < .001$ and $p < .0001$, respectively.

Table A.6: Mann-Whitney U test for features variables between novice and expert groups. Mean value of the variables and p-value of the tests are given.

| Features | Novice | Expert | *P*-value |
|---|---|---|---|
| Number of fixations | 22.40 | 20.26 | .001 ** |
| Fixation duration-Mean[ms] | 186.53 | 184.34 | >.05 |
| Fixation duration-Min[ms] | 86.04 | 85.348 | >.05 |
| Fixation duration-Max[ms] | 440.65 | 433.79 | >.05 |
| Fixation duration-Sum[ms] | 4280.72 | 3785.91 | .002** |
| AOI number of fixations | 10.55 | 9.46 | >.05 |
| AOI fixation duration–Mean[ms] | 175.71 | 175.26 | >.05 |
| AOI fixation duration-Min[ms] | 83.33 | 82.16 | >.05 |
| AOI fixation duration-Max[ms] | 370.26 | 356.80 | >.05 |
| AOI fixation duration-Sum[ms] | 1944.49 | 1723.17 | .027 * |
| Number of saccades | 17.12 | 14.43 | <.0001 **** |
| Saccade duration-Mean[ms] | 33.63 | 32.91 | >.05 |
| Saccade duration-Min[ms] | 24.20 | 24.01 | >.05 |
| Saccade duration-Max[ms] | 50.86 | 47.49 | .011 * |
| Saccade duration-Sum[ms] | 562.39 | 471.02 | <.0001 **** |
| Saccade amplitude-Mean[°] | 5.25 | 5.15 | >.05 |
| Saccade amplitude-Min[°] | 2.05 | 2.17 | >.05 |
| Saccade amplitude-Max[°] | 11.15 | 10.10 | .013 * |
| Saccade amplitude-Sum[°] | 39.37 | 29.55 | <.0001 **** |
| Saccade peak velocity-Mean[°/$s$] | 201.79 | 199.71 | >.05 |
| Saccade peak velocity-Min[°/$s$] | 92.07 | 99.34 | .016 * |
| Saccade peak velocity-Max[°/$s$] | 355.69 | 329.54 | .012 * |
| Sacc-fixa ratio | 0.131 | 0.124 | <.001 *** |
| Number of blinks | 10.25 | 11.68 | <.001 *** |
| Blink duration-Mean[ms] | 208.32 | 202.71 | .001 ** |
| Blink duration-Min[ms] | 121.87 | 119.99 | >.05 |
| Blink duration-Max[ms] | 321.16 | 328.09 | >.05 |
| Pupil diameter-Mean | 1.08 | 0.95 | .002 ** |
| Pupil diameter-Min | 0.75 | 0.71 | >.05 |
| Pupil diameter-Max | 1.37 | 1.21 | <.0001 **** |
| Number of clicks on AOIs | 1.28 | 1.53 | .003 ** |
| C-AOI number of fixations | 3.21 | 4.15 | .005 ** |
| C-AOI fixation duration-Mean[ms] | 123.59 | 127.73 | >.05 |
| C-AOI fixation duration-Min[ms] | 54.78 | 50.74 | >.05 |
| C-AOI fixation duration-Max[ms] | 217.33 | 239.28 | >.05 |
| C-AOI fixation duration-Sum[ms] | 625.48 | 805.38 | .014 * |

Table A.7: Performance comparison of the classification models. F1-score and ROC-AUC for each model are given. Best classifier in bold.

| Classifier | F1-score | ROC-AUC |
|---|---|---|
| SVM | 0.739 | 0.741 |
| **Random Forest** | **0.779** | **0.768** |
| LightGBM | 0.755 | 0.745 |

off-task behavior more often than novices, as indicated by more clicks on disruptive students. In addition, expert teachers were found to have significantly more fixations and longer fixation duration (sum) on these disruptive virtual students detected by teachers using the controller.

Taken together, these significant differences in behavioral features between novice and expert teachers suggest that eye tracking and reaction behavior data (i.e., reaction to disruptive students by the controller) are informative of teacher expertise and have great potential to provide discriminative information for novice-expert models. These statistical results were further discussed in conjunction with the findings of the SHAP model explanation approach.

**Model Performance**

We compared the performance of three different models, i.e, SVM, Random Forest, and LightGBM, by reporting the F1-score, ROC-AUC score, and ROC curve used in previous work for performing an imbalanced classification task [302]. Our prediction results show that the best performing algorithm on the test set is Random Forest, with an F1-score of 0.779 and a ROC-AUC score of 0.768. The SVM classifier and LightGBM achieved slightly lower but still predictive performance, with an F1-score of 0.739 and a ROC-AUC score of 0.741, and an F1-score of 0.755 and a ROC-AUC score of 0.745, respectively. Model performance results are shown in Table A.7 and Figure A.20.

**SHAP Model Explanation**

In this work, only the SHAP results for the best performing classification model, i.e., Random Forest, are presented. The SHAP summary plots are shown in Figure A.21, where Figure A.21 (a) and Figure A.21 (b) show the global and local SHAP value (impact) of each feature of the model, respectively. The position of the feature on the $y$-axis from top to bottom is determined by the feature importance. In the global summary plot, the bar indicates the average impact of each feature on the model; the longer the bar, the greater the impact of the feature on the model. In the local summary plot, each point represents a SHAP value for a feature instance and is positioned along the $x$-axis corresponding to the SHAP value. The color represents the value of a feature instance (blue-low; red-high). The color change of a feature along the $x$-axis (from left to right) indicates the relationship between the feature value and the impact of the feature on the model output. That is, a color change from blue to red (feature value from low to high) indicates that this feature has a positive impact on the model output, i.e., pushes the model prediction to a higher value (class-1, i.e., expert), and vice versa.

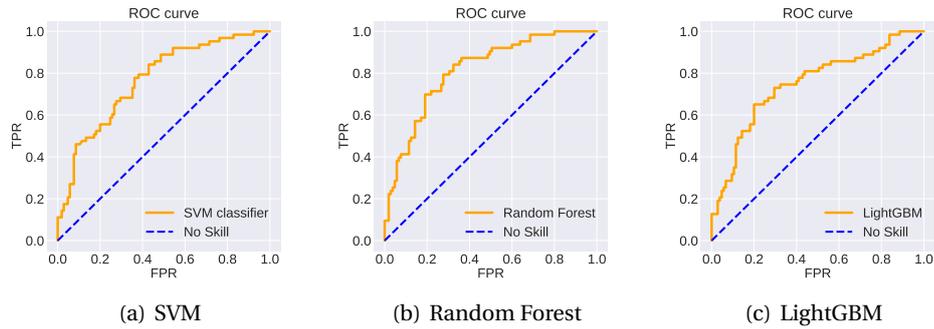

(a) SVM    (b) Random Forest    (c) LightGBM

Figure A.20: ROC curves of classification models trained on features extracted from teacher behavioral data in the IVR classroom: (a) SVM (AUC=0.741), (b) Random Forest (AUC=0.768), and (c) LightGBM (AUC=0.745).

As can be seen, both types of features, i.e., teachers' gaze behavior and their reaction behavior towards disruptive virtual students, provide discriminative information to the novice-expert model. In particular, the feature *ClickedAOINumberOfFixations* contributed the most to the model, followed by two other features related to fixation in C-AOIs (i.e., disruptive virtual students clicked by the teacher), namely *ClickedAOIFixationDurationSum* and *ClickedAOIFixationDurationMax*. Note that the three most important features are related to teachers' fixation behaviors toward C-AOIs. The feature *PupilDiameterMean*, which could indicate teachers' cognitive processing load, also provided discriminative information for predicting expertise. In addition, the feature *NumberOfClicks*, which provides intuitive information about the teacher's classroom management activities, was also found to provide important information for the model.

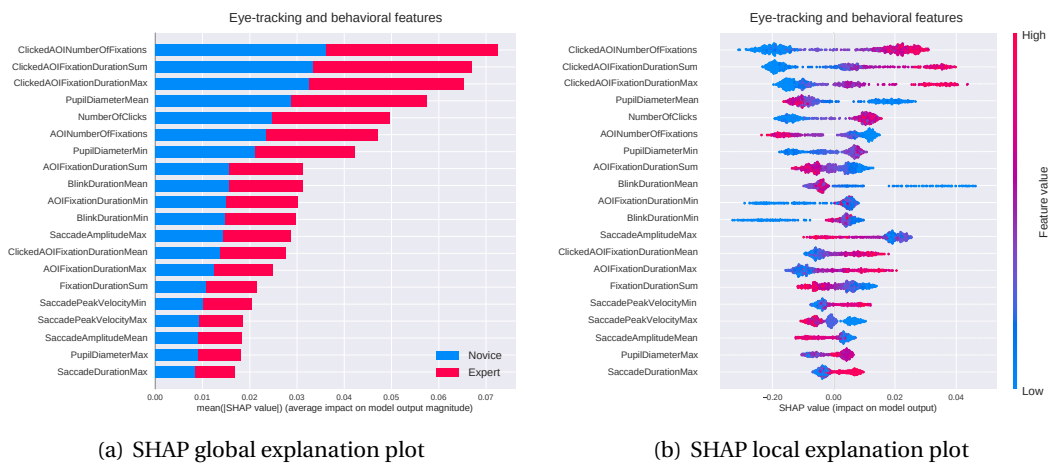

(a) SHAP global explanation plot    (b) SHAP local explanation plot

Figure A.21: SHAP summary plots of the top 20 features in the best Random Forest model. Features are ranked along the $y$-axis in descending order of feature importance. (a) Mean absolute SHAP values of each feature for each class. (b) SHAP values for all feature instances: Each dot represents a SHAP value of a feature instance, where the color represents the value of the feature instance (blue-low; red-high); the more spread out the dots of a feature are on the $x$-axis, the greater the impact of the feature on the model; a color change from blue to red (from left to right) indicates a positive impact of a feature on the model and vice versa.

Notably, of these five most important features, only the feature *PupilDiameterMean* has a negative impact on the novice-expert model output (See Figure A.21 (b), a color change from red to blue), i.e., a pupil diameter larger than its average contributes to the model output of class-0 (novice). The other four most important features all have a positive impact on the model output, i.e., a feature value higher than the feature average contributes to the model output of class-1 (expert). In addition, it is observed that other eye movement features such as blinks (*BlinkDurationMean,BlinkDurationMin*) and saccades (*SaccadeAmplitudeMax*, *SaccadePeakVelocity*) also show their discriminative power in expertise prediction; however, compared to pupil diameter and fixation features, the blink and saccade features play a less important role in prediction. It is noteworthy that of the top twenty features, almost half (i.e., nine) are accounted for by fixation-related features, suggesting that teachers' visual perceptual behaviors, especially toward the AOIs and C-AOIs, contain the most information about the teachers' expertise in terms of classroom management.

### A.4.7   Discussion

**General Discussion of Results**

We presented an investigation on predicting teacher expertise with machine learning techniques and explainable models based on ecologically valid sensor data, including eye tracking and controller tracking, in an immersive VR classroom scenario. In the following, we discuss our findings from different aspects that contribute to this research field.

As a contribution to the research field of expertise prediction, we demonstrate that machine learning models can be trained to predict expertise in educational domains based solely on various sensor data, particularly eye tracking, collected in an immersive VR classroom. All three novice-expert classification models (i.e., SVM, Random Forest, and LightGBM) are predictive, with Random Forest in particular performing best, with an F1-score of 0.779 and a ROC-AUC of 0.768 (see Table A.7 and Figure A.20). Although there are no previous studies on predicting teacher expertise in professional vision, our results are comparable to the model performance of previous similar work on predicting other types of expertise (novice vs. expert) based on eye trackings, such as athlete expertise (78.2% accuracy, in [135]), surgeon expertise (84.4% accuracy, in [274]), and programming expertise (80% accuracy, in [276]), especially considering that a VR setup was used in our study. Our approach shows that eye tracking features are highly informative and well suited to distinguish between novice and expert teachers in terms of how teachers observe and intervene in disruptive classroom events. Our results demonstrate the practicality of predicting teacher expertise in VR-based systems and fill the research gap in detecting expertise in the educational domain.

Certainly, the success of our machine learning approach to predicting teacher expertise is the result of a combination of several efforts. In particular, we built the novice-expert classification models based on a large number of features extracted from different types of sensor data collected in VR. With advanced VR technologies, an immersive VR classroom with pre-programmed animated avatars (i.e., virtual students exhibiting disruptive behaviors) was created and presented in advanced HMDs to provide teachers with a more realistic classroom experience than videotapes used in previous studies [289, 292]. In this case, teachers might exhibit more natural gaze behavior while immersed in a virtual 3-D classroom environment. This further contributes to the availability of ecologically valid eye tracking data, which may be difficult to obtain in situations with video stimuli because teachers' eye tracking data can be affected by head movements in remote eye-trackers. In this study, we included more eye tracking measurements in our machine learning approach compared to previous studies that investigated teachers' visual perceptual behavior using limited measures [157, 158]. Common eye tracking measures such as pupil diameter, blink, fixation, and saccade were all calculated in this study. Based on this, a large number of eye tracking features were extracted from data segments of the entire VR experience data to capture the various conscious and unconscious behaviors of teachers during the teaching experience. Additional sensor data were recorded that explicitly related to teachers' classroom management performance, namely teachers' handling of disruptive events as indicated by controller input. Due to the limited experimental setting, such information has not been investigated in previous work. Therefore, teachers' implicit and explicit behaviors during classroom management that could be indicative of teacher expertise were recorded. Since machine learning techniques are data-driven and therefore learn more from more data, a large number of features we used increases the success of our machine learning approach to predicting teacher expertise. Our approach is an important contribution to advancing future research that uses VR as a tool to explore expertise in education using machine learning techniques and explainable models.

Model explainability is a priority in today's data science community. In addition to training machine learning models to predict teacher expertise and achieve the best model performance, it is equally important to interpret the model to understand how each feature contributes to the model, especially when interpretable features are used (e.g., eye movements). These are also the two main focuses of our current study. To achieve this, we applied the SHAP approach to better understand how novice-expert models make decisions about expertise classes and what particular features and how they contribute to the model outputs. Our SHAP results show that both gaze behaviors and teachers' handling of disruptive events provide discriminative information for the novice-expert model, but to varying degrees (see Figure A.21). Such an approach provides information on possible ways to further improve model performance and, moreover, on the underlying relationships between teacher expertise and (in particular) classroom perceptual behaviors, which could provide a viable way to assess teacher competencies.

In particular, three eye tracking features related to fixations in C-AOIs (including the number of fixations and the total and maximum fixation durations in C-AOIs) provide the most information for the novice-expert model, with all having a positive impact on the model output. That is, a feature value that is higher than the feature average contributes to the model output of class-1 (expert). These findings are consistent with the statistical results showing that experts have significantly more fixations and longer total fixation duration in C-AOIs than novices (see Table A.6). Conversely, another highly informative feature -mean pupil diameter, which correlates positively with cognitive load- has a negative impact on the model output. That is, a higher value of mean pupil diameter than its average contributes to the model output of class-0 (novice). This result is also consistent with the statistical results showing that novices have significantly larger mean pupil diameters than experts. It is noteworthy that fixation-related features (nine of the twenty most informative features) provide the most discriminative information for the model predicting teacher expertise. This suggests that teachers' visual attention behaviors, especially toward these disruptive students in the classroom, which were found to differ between experts and novices in previous studies [289, 157, 291], also show high discriminative power in predicting teacher expertise in a VR scenario. It is also interesting to note that pupil diameter also has high predictive power for teacher expertise, which has not been examined before. Pupil diameter is widely considered to be indicative of cognitive processing load (positive correlation) [303]. Our result suggests that teachers' expertise in classroom management may also be strongly reflected in their cognitive processing behavior, which is second only to the importance of teachers' attentional behaviors toward disruptive students. Compare with fixation and pupil diameter, saccadic features indicative of visual search behavior, which differed significantly between novices and experts according to statistical tests (see Table A.6), were less discriminative in the novice-expert model. Notably, the feature of teachers' handling of disruptive events, as measured by the number of controller clicks they made toward disruptive students, was also highly informative and contributed positively to the model. This is also consistent with the statistical results showing that the experts detected more disruptive students than the novices.

Overall, our SHAP results help to identify the most informative features for the novice-expert model and, moreover, to reveal the relationships between the features and the model outputs, i.e., teacher expertise classes. This contribution of SHAP is very important for machine learning studies. First, if we know which particular features contribute most to the models and which do not, we know what to fine-tune and how to optimize to improve model performance. Second, model explanations shed light on expert behavior, which in turn provides new insights for designing systems for training novices and developing expertise. This is the case in our study. On the one hand, in the statistical analysis, we found some significant differences between experts and novices in their general saccade and fixation behavior across the classroom (here referring to those fixation behaviors that are not directed toward the AOIs, i.e., disruptive students), but these features were not very informative in the model. This may suggest that teachers' visual attention behaviors toward disruptive students play a more important role in predicting teacher expertise than teachers' general visual exploration and attention behaviors in the IVR classroom. On the other hand, we found that teachers' pupil diameter, which is indicative of cognitive load, contributes considerably to the model; these features have not been previously explored in the literature, but their informativeness was revealed by our model explanation approach. Such findings provide various insights for future studies exploring teacher expertise. Furthermore, the role of individual features in predicting expertise and the underlying relationships between features and expertise may provide clues for the design and development of VR-based training systems that utilize gaze-based interfaces to help novices develop visual strategies employed by experts for classroom management.

**Future Work**

The purpose of our study is to test the feasibility of our machine learning approach in predicting teacher expertise and to explore the effectiveness of eye tracking along with teachers' reactions in the classroom to provide foundations for future studies assessing teacher expertise. Improving model performance, e.g., by further feature selection based on the SHAP approach, is beyond the scope of our current study and can be investigated in future work. Furthermore, for the purpose of model explainability, we used only those features that can be interpreted. Future work aiming for higher accuracy could investigate features that are difficult to interpret but may also provide information about expertise, such as unprocessed eye, head, and controller orientations [136]. The use of these features will result in a loss of explanatory power of the model but may improve the performance of the model if accuracy is the primary goal.

In addition, in our work, we used SVM, Random Forest, and LightGBM, which have been commonly used in many previous studies. In a further study, deep learning approaches using neural networks such as Long Short-Term Memory (LSTM), which are good at handling temporal data, could be explored with the aim of achieving better performance. However, such approaches require more data compared to the classifiers we use, since there are many more parameters to optimize. In the future, we may look to increase our sample size for an investigation of deep learning approaches, although expert recruitment is always a challenge in the field of expert research.

### A.4.8 Conclusion

In this work, we present a novel machine learning approach for predicting teacher expertise based on teachers' behavioral data (especially eye tracking) collected in an immersive VR classroom. We evaluated the performance of three novice-expert classification models created based on different classifiers, including SVM, Random Forest, and LightGBM in the binary classification task. Our evaluation showed that Random Forest performed best with an F1-score of 0.779 and a ROC-AUC score of 0.768. The most discriminative features for predicting teacher expertise were identified by the SHAP model explanation approach. Consistent with previous literature, we found that teachers' visual attention behaviors toward disruptive classroom events contributed the most to the model. In addition, teachers' blinking behaviors and pupil diameters, indicative of cognitive load, as well as teachers' reaction behaviors toward disruptive events, were also highly discriminative of teacher expertise, although these features are rarely considered in this area of research. Overall, our computational framework based on machine learning techniques and model explainability approach demonstrates the predictability of educational expertise based on a rich set of sensor features in a VR setting. Our findings provide valuable insights for future studies assessing teacher expertise in VR-based systems and offer further insight into the design of gaze-based training interfaces.

# B Human Behavior in VR Locomotion

This chapter is based on the following publications:

1. **Hong Gao**, Lasse Frommelt, and Enkelejda Kasneci. The Evaluation of Gait-Free Locomotion Methods with Eye Movement in Virtual Reality. In *2022 IEEE International Symposium on Mixed and Augmented Reality (ISMAR-Adjunct'22)*. Singapore. 2022. doi:10.1109/ISMAR-Adjunct57072.2022.00112.

2. **Hong Gao** and Enkelejda Kasneci. Eye-Tracking-Based Prediction of User Experience in VR Locomotion Using Machine Learning. In *Computer Graphics Forum*. 2022. doi: 10.1111/cgf.14703.

---

For formatting considerations, the papers in the appendix are attached with minor changes to the publication template.

## B.1 The Evaluation of Gait-Free Locomotion Methods with Eye Movement in Virtual Reality

### B.1.1 Abstract

As VR becomes increasingly popular in the entertainment industry, VR locomotion, a technique that allows users to navigate virtual environments beyond the spatial confines of the real world, is being increasingly studied by developers and researchers. Previous work has examined the effects of locomotion methods on various aspects of users, such as user experience, motion sickness, and task performance. However, how locomotion methods affect users' eye movements that might indicate cognitive load has not yet been investigated, although several relevant works have addressed these effects as being important to study. To contribute to this area of research, in this work we investigate the evaluation of five common gait-free locomotion methods using eye movements during VR navigation. The results show that locomotion methods significantly affect participants' eye movement behavior (i.e., blinks, fixations, and saccades), suggesting that different cognitive responses were elicited with different locomotion methods. Our research provides a viable tool for future studies evaluating locomotion methods, thus providing further in-depth insights for developing more effective and enjoyable VR locomotion methods.

### B.1.2 Introduction

With the increasing development of commodity-level virtual reality (VR) head-mounted displays (HMDs), VR is being widely used in entertainment and education [3, 2], leading thus to a growing interest in the design and development of VR applications. Navigation is one of the most important features in VR applications (especially VR games), allowing users to efficiently and infinitely navigate large virtual environments (VEs) while remaining confined in a room-scale real-world environment. To date, a variety of locomotion methods have been developed and employed in VR applications, such as arm swinging, dash, joystick, and teleportation-like locomotion methods [55, 60, 304, 305, 306]. Since locomotion methods are only tools to help the user move around in the VE, they should not interfere with the user's main task during VR navigation. Therefore, one of the major challenges for VR developers and researchers is to develop advanced locomotion methods that offer an enjoyable experience while imposing a low cognitive load on users and causing less or no VR sickness.

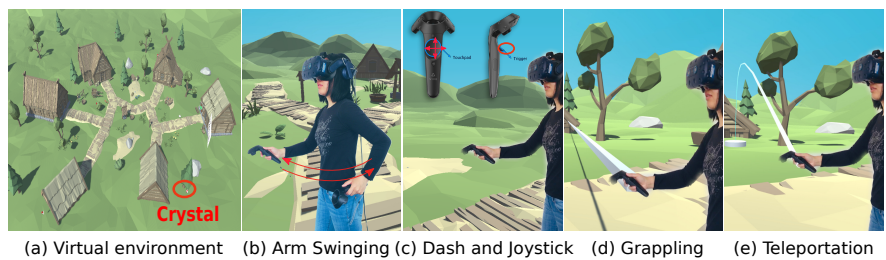

(a) Virtual environment  (b) Arm Swinging  (c) Dash and Joystick  (d) Grappling  (e) Teleportation

Figure B.1: The virtual environment for locomotion (a) and the illustration of the five evaluated gait-free locomotion methods (b)-(e).

A body of previous work has evaluated currently popular locomotion methods and compared their effectiveness in VR and games. This involved quantifying assessments such as user experience (usability of locomotion methods), sense of presence, motion sickness, and other post-hoc surveys such as user preferences that were usually quantified with questionnaires [60, 61]. However, the real-time user cognitive load induced by the locomotion method and the visual behavior during VR navigation have rarely been studied and compared between different locomotion methods. Some previous studies have examined the cognitive load of users while experiencing different 3D travel techniques (e.g., real walking, steering, and joystick), but used either a post-task or dual-task paradigm to measure cognitive load [62, 63]. The cognitive load induced by the travel techniques alone was not directly measured as additional cognitive tasks were involved. In our study, however, we aim to measure the cognitive load induced purely by the locomotion method. With this in mind, an eye-tracking-based analysis could provide a great opportunity for such evaluation from a new perspective and should attract the attention of researchers. In the literature, eye tracking has been widely used as a non-intrusive and objective measurement of human conscious and unconscious temporal cognitive and visual behavior in various tasks [204, 64, 219, 307].

The aim of this work is to evaluate and compare five commonly used gait-free locomotion methods, namely arm swinging, dash, grappling, joystick, and teleportation. For this purpose, we introduced eye movements as objective measurements of cognitive load for the first time in this research domain. Specifically, we conducted a VR user study using a within-subjects design. Participants performed a simple search & collection task by navigating the VE using five different locomotion methods. The eye-tracking data providing information about participants' underlying cognitive processes and visual behavior during navigation were collected and analyzed. Our results show that locomotion methods have significant effects on participants' eye movements.

### B.1.3   Related Work

Previous studies have evaluated a variety of locomotion methods and compared them with respect to different aspects. Coomer et al. [61], for example, examined the effects of four locomotion methods on users, including joystick, arm-cycling, teleportation, and point-tugging. User experience, simulator sickness, and participants' task performance were evaluated. Results showed that arm-cycling was the best out of the four methods, as it performed best on the search task and had better simulator sickness scores than joystick and point-tugging. teleportation, on the other hand, performed worst in the search task but also had better simulator sickness scores than joystick and point-tugging.

Similar to [61], Paris et al. [60] also evaluated four commonly used locomotion methods,

including skiing, magic carpet, grappling, and teleporting. Post-hoc questionnaires on user experience, presence, simulator sickness, and path integration performance in the absence of external landmarks were evaluated. Results showed that continuous methods had advantages over discontinuous methods in path integration performance, consistent with the work of Coomer et al. [61]. However, the advantage of discontinuous methods over continuous methods with respect to simulator sickness, generally reported in previous work [57], was not present in [60].

Frommel et al. [308] conducted a user study to evaluate four controller-based locomotion methods, including free teleportation, fixpoint teleportation, touchpad-based (joystick-like), and guided automatic locomotion. Participants' discomfort, presence, enjoyment, affective state, and simulator sickness during VR locomotion were evaluated. Results showed that free teleportation was superior to the other three locomotion methods, eliciting the least discomfort and scoring highest on enjoyment, presence, and affective state. In addition, free teleportation was found to have significantly lower simulator sickness scores than touchpad-based locomotion, which is inconsistent with the literature [57].

Based on different evaluation purposes, the aforementioned works have demonstrated the effects of locomotion methods on different aspects of navigation in VEs. However, the measures used are similar and limited, mostly based on subjective self-reports by users. This inspires our study to explore the feasibility of using eye movements as objective metrics to assess participants' cognitive processing responses during VR navigation.

The finding that eye-tracking measurements correlate with cognitive load is no longer new and has been widely used in previous studies. In particular, it is well known that pupil diameter correlates positively with human cognitive load [303]. In addition, eye movements such as blinks, fixations, and saccades have also been found to be indicative of cognitive load. Specifically, blink and fixation rates were found to be negatively correlated with cognitive load [299]. Conversely, fixation duration was found to be positively correlated with cognitive load [299, 298]. In addition, saccadic measures were also found to be correlated with cognitive load, e.g., saccade amplitude was found to be negatively correlated with cognitive load [309]. These previous works strongly support our current study, which examines the cognitive response of users during VR locomotion using eye movements.

### B.1.4 Locomotion Methods

Since our goal in this study is to investigate the feasibility of using eye-tracking technology to measure users' cognitive processing load during VR locomotion, we selected the five most commonly used and representative gait-free locomotion methods as test subjects rather than proposing new locomotion techniques. The HTC Vive Pro Eye HMD was used to display the VE. All scripts of the locomotion methods were written in C#. The user navigates the VE using Vive controllers that come with the HMD. The models of the controllers were rendered and updated as the user moved around the VE. In the Unity coordinate system, all movements occur in the ($x$, $z$) plane, and the eye height (i.e, $y$ coordinate) remains unchanged. The distance in Unity is in units (by default, 1 Unity unit is 1 meter). The default scale between the virtual world and the real world is 1:1. Detailed information about the VE design can be found in Section B.1.5. Five locomotion methods are illustrated in Figure B.1.

### Arm Swinging

Arm swinging converts physical arm movements into player movements in the VE. To navigate, the user holds controllers and swings their arms simultaneously. Once the user holds down the grip button (of right controller), the coordinates of the controllers in the current frame ($Left\_x1$, $Left\_z1$, $Right\_x1$, $Right\_z1$) and in the next frame ($Left\_x2$, $Left\_z2$, $Right\_x2$, $Right\_z2$) are read and stored. Then the differences in the controller positions between two consecutive frames are calculated. The player moves forward in each frame by the sum of the absolute differences in the yaw direction of the HMD. The locomotion stops when the grip button is released or the arm stops swinging.

### Dash

To perform the dash locomotion, the user touches the trackpad on one of the controllers to select the desired direction (thumb on trackpad: up is forward, down is backward, the left side is left, and the right side is right direction). The $2D$ coordinates of the contact point on the trackpad are read and converted to a direction in the Unity coordinate system. Then the user taps the trigger button to perform a dash in the selected direction. The player is then moved 2 units distance in the VE within 0.2 seconds. To perform a continuous movement over a longer distance, the user can hold down the trigger button instead of tapping it. Multiple dashes are then performed continuously, which makes dash a continuous locomotion method. The user can release the trigger button to stop the locomotion.

### Grappling

In grappling, the user uses a grappling hook to pull the player to a desired location in the VE. To navigate, the user presses the trigger button, whereupon a ray is projected from the controller in the direction the controller is pointing. Once the ray hits the desired location ($x$, $z$), the grappling hook is extended from the controller to the desired location and the user is then moved to the desired location. The movement speed of the player is automatically adjusted according to the movement state within a minimum of 1 unit per second and a maximum of 4 units per second. Once the player has moved to within one meter of the desired location, locomotion is stopped. Then the visualization of the grappling hook is reset and the movement is unlocked. Locomotion can be ended prematurely by pressing the grip button.

**Joystick**

Joystick is a locomotion method that simply relies on the controller's trackpad inputs. Similar to dash locomotion, the user touches the trackpad with the thumb to select the desired direction (see above Section B.1.4). The $2D$ coordinates of the contact point are read and converted to a direction in the Unity coordinate system. The player moves at a smoothed speed of 1 unit per second in the VE towards the desired location. Locomotion stops when the user removes the thumb from the trackpad button.

**Teleportation**

For teleportation, we directly used the teleportation prefab provided by SteamVR[1] for Unity3D. To teleport, the user holds down the top of the trackpad on one of the controllers. A parabolic projection is then displayed, emanating from the controller and meeting the ground of the VE, visible to the user. The user points the end of the projection to the desired location $(x, z)$. As soon as the trackpad button is released, the player is instantly transported to the desired location in the VE.

### B.1.5   Experimental Evaluation

We designed an experiment in which participants perform a very simple search & collection task by navigating a VE using five different locomotion methods. We assessed participants' eye movements that might indicate cognitive load during VR locomotion.

**Participants**

Fifteen volunteers (10 male, 5 female) between the ages of 23 to 31 participated in our experiment. They were university students, from which six participants reported no experience with video games, four played video games 0 to 5 hours per week, and five played video games for more than 5 hours per week. Eight have no experience with VR, six reported to have some VR experience, and one used VR headsets regularly. All participants provided informed consent. Since all fifteen participants successfully completed the task and the eye-tracking data were valid, no participant was excluded from the study.

**Apparatus and Materials**

The HTC Vive Pro Eye HMD was used to display the VE, which has a resolution of $1440 \times 1600$ per eye, a refresh rate of 90 Hz, and a field of view of $110°$. The HTC Vive Pro Eye is seamlessly integrated with the Tobii eye tracker with a sampling rate of 120 Hz and an accuracy of $0.5° - 1.1°$, which can be used to record eye-tracking data. The HTC Vive controllers served as the input device for locomotion. We used two HTC Vive Base Stations to track a $2m \times 2m$ area. The VE was rendered with Unity engine[2] (version 2020.03.23f) on a computer with a 3.5GHz Core i7 processor and 16GB RAM.

---

[1]https://store.steampowered.com/

To avoid additional effects on participants caused by the VE and the task, we created a simple VE based on a package from the Unity asset store. The VE was an outer ground plane with several houses and paths and trees in between. In addition, to encourage participants to experience more VR locomotion, we placed the five crystals in different locations (outside the house) in the VE to make these targets easy to find. The top-down view of the VE is shown in Figure B.1 (a).

**Experimental Procedure**

We used a within-subjects design, where the independent variable of locomotion methods had five levels, i.e., arm swinging, dash, grappling, joystick, and teleportation. Therefore, all participants took part in five trials (locomotion methods). The entire experiment lasted approximately 50 minutes per participant. All participants were informed before the experiment that they could abort the experiment at any time if they felt uncomfortable.

After participants signed an informed consent form, they were asked to complete a demographic questionnaire (e.g., age, gender) and a questionnaire about their previous experiences with video games and VR. Then, the experimenter gave instructions on the experimental task. Next, participants were assigned to one of the five locomotion conditions in a counter-balanced order created using a Latin square. In each trial, participants first practiced the current locomotion method with text instructions in the VE and, if necessary, with the help of the experimenter until they were familiar with it. A standard 5-point eye-tracking calibration routine was then performed before starting the actual data collection. Participants performed the crystal collection task with controllers and they were asked to collect all (five) crystals (see Figure B.1 (a)). Note that to avoid additional cognitive load from the task, there was no time limit for performing and completing the task. Eye-tracking data was recorded during the task. After completing the task, participants took off the headset and took a rest. To avoid additional cognitive load from previous experiences, participants were asked to rest for about four to five minutes after each exercise and trial until they felt comfortable to start again. The entire experiment ended after the participant completed all five trials.

**Measures**

To evaluate and compare different locomotion methods, different eye movement measures were used. Eye tracking has already been considered a valuable tool for assessing users' cognitive process and visual behavior during interaction with a system [64, 298]. Therefore, in this work, we measured commonly used eye movements such as blinks, fixations, and saccades.

---

[2]https://unity.com/

Since the eye tracker integrated into the HMD only outputs pupil size and gaze vectors, and no standard method or software is available for eye movement event detection in VR, eye movements such as blinks, fixations, and saccades had to be detected manually post-experimentally. A blink detection algorithm based on the fluctuations that characterize pupil data as proposed by Hershman et al. [294] was applied in our study.

Before fixation and saccade detection, linear interpolation was performed for the missing gaze vectors. A modified velocity-threshold identification (I-VT) algorithm [122, 219] was used for fixation detection. Specifically, fixations were detected with a maximum gaze velocity threshold of $40°/s$ under the condition of relatively stationary head movement (head moving velocity lower than $12°/s$). In addition, the minimum ($100ms$) and maximum ($500ms$) duration thresholds were used to filter fixation. Saccades were detected using the normal I-VT algorithm [121], with a minimum gaze velocity threshold of $80°/s$, and minimum ($30ms$) and maximum ($80ms$) duration thresholds for the follow-up filtering. Based on those detected eye movement events, measures such as blink rate, fixation rate and duration, and saccade amplitude were calculated.

### B.1.6 Results

Dependent variables of eye movements were calculated and compared. The Shapiro-Wilk test was performed to test for data normality. For normally distributed data, the dependent variables across the five locomotion conditions were compared using repeated-measures ANOVA and paired t-test as post-hoc tests for the pairwise comparisons. Friedman test was used as a non-parametric test with Nemenyi as the post-hoc test. Bonferroni correction was applied in the post-hoc tests. The significance level was set at $\alpha = 0.05$ for all tests. The statistical test results are summarized in Table B.1 and plotted in Figure B.2 to B.5, with asterisks in the results indicating significant differences (*, **, *** and n.s. for $p < .05$, $p < .01$, $p < .001$, and no statistical significance, respectively). All statistical analyses were performed using Pingouin[3], an open-source Python[4] package.

Table B.1: Statistical comparison results of eye movement metrics between different locomotion methods. Significant differences are highlighted with *, **, and *** for $p < .05$, $p < .01$, and $p < .001$, respectively.

| Metrics | Con_1 | Mean (SD) | Con_2 | Mean (SD) | Significance |
|---|---|---|---|---|---|
| Blink rate | ArmSwing | 6.85 (5.35) | Joystick | 9.53 (3.68) | $p = .045$, * |
| Blink rate | Dash | 5.05 (3.09) | Joystick | 9.53 (3.68) | $p = .004$, ** |
| Blink rate | Dash | 5.05 (3.09) | Teleport | 9.06 (6.68) | $p = .018$, * |
| Fixation rate | ArmSwing | 34.16 (15.67) | Joystick | 59.53 (11.85) | $p = .002$, ** |
| Fixation rate | ArmSwing | 34.16 (15.67) | Teleport | 56.17 (14.20) | $p = .012$, * |
| Fixation duration | ArmSwing | 239.86 (19.54) | Joystick | 205.04 (23.67) | $p < .001$, *** |
| Fixation duration | ArmSwing | 239.86 (19.54) | Teleport | 219.54 (22.92) | $p = .047$, * |
| Saccade amplitude | ArmSwing | 8.60 (2.21) | Joystick | 10.12 (1.53) | $p = .021$, * |
| Saccade amplitude | ArmSwing | 8.60 (2.21) | Teleport | 10.28 (1.48) | $p = .010$, * |
| Saccade amplitude | Grappling | 9.47 (1.58) | Teleport | 10.28 (1.48) | $p = .017$, * |

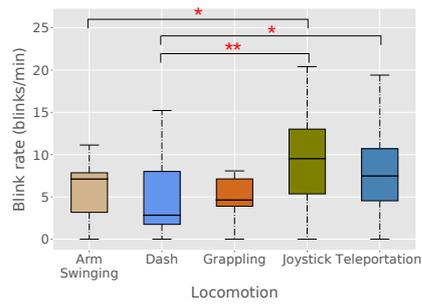

Figure B.2: Blink rates in each locomotion condition. Significant differences are highlighted with * and ** for $p < .05$ and $p < .01$, respectively.

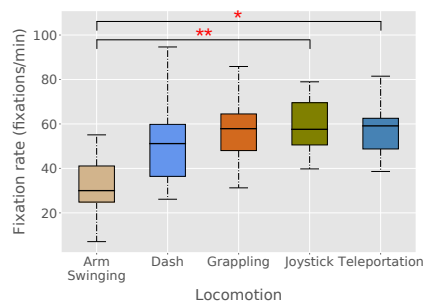

Figure B.3: Fixation rates in each locomotion condition. Significant differences are highlighted with * and ** for $p < .05$ and $p < .01$, respectively.

### Blink rate

Statistical tests revealed a significant effect of locomotion methods on participants' blink rate, with $F(4, 56) = 2.65$, $p = .042$. As shown in Figure B.2, the blink rate ($blinks/min$, abbreviated as $b/m$) differed significantly between some of the locomotion conditions, with the blink rate in the joystick condition ($M = 9.53 b/m$, $SD = 3.68 b/m$) is significantly higher than in the dash ($M = 5.05 b/m$, $SD = 3.09 b/m$) and arm swinging ($M = 6.85 b/m$, $SD = 5.35 b/m$) conditions, with $p = .004$ and $p = .045$, respectively. In addition, the blink rate in the teleportation condition ($M = 9.06 b/m$, $SD = 6.68 b/m$) is also significantly higher than in the dash condition, with $p = .018$.

---

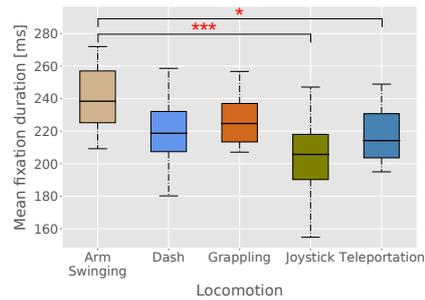

Figure B.4: Mean fixation durations in each locomotion condition. Significant differences are highlighted with * and *** for $p < .05$ and $p < .001$, respectively.

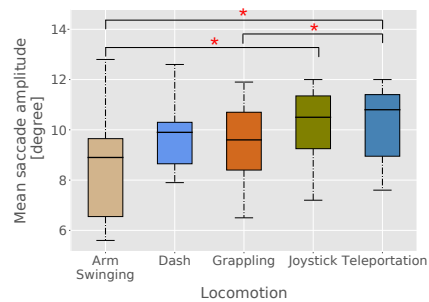

Figure B.5: Mean saccade amplitudes in each locomotion condition. Significant differences are highlighted with * for $p < .05$.

### Fixation rate

With regard to fixation, statistical tests showed a significant effect of locomotion methods on participants' fixation rate, with $F(4, 56) = 8.78$, $p < .001$. As shown in Figure B.3, the fixation rate ($fixations/min$, abbreviated as $f/m$) in the arm swinging condition ($M = 34.16 f/m$, $SD = 15.67 f/m$) is significantly lower than in the joystick condition ($M = 59.53 f/m$, $SD = 11.85 f/m$), with $p = .002$. In addition, the fixation rate in the arm swinging condition is also significantly lower than in the teleportation condition ($M = 56.17 f/m$, $SD = 14.20 f/m$), with $p = .012$.

### Fixation duration

Not only fixation rate but also the mean fixation duration was found to be significantly affected by the locomotion method. The mean fixation duration in the arm swinging condition ($M = 239.86 ms$, $SD = 19.54 ms$) is significantly longer than in the joystick condition ($M = 205.04 ms$, $SD = 23.67 ms$), with $p < .001$. In addition, the mean fixation duration in the arm swinging condition is significantly longer than in the teleportation condition ($M = 219.54 ms$, $SD = 22.92 ms$), with $p = .047$.

**Saccade amplitude**

Furthermore, statistical tests revealed a significant effect of locomotion methods on participants' mean saccade amplitude, with $F(4, 56) = 7.59$, $p < .001$. As shown in Figure B.5, the mean saccade amplitude in the teleportation condition ($M = 10.28°$, $SD = 1.48°$) is significantly larger than in the arm swinging ($M = 8.60°$, $SD = 2.21°$) and grappling ($M = 9.47°$, $SD = 1.58°$) conditions, with $p = .010$ and $p = .017$, respectively. In addition, the mean saccade amplitude in the joystick condition ($M = 10.12°$, $SD = 1.53°$) is also significantly larger than in the arm swinging condition, with $p = .021$.

### B.1.7    Discussion

We found that eye movements, reflecting participants' temporal cognitive processing load, were significantly affected by locomotion methods (see Figure B.2 to Figure B.5). In particular, the mean blink rate was higher in the joystick condition than in other locomotion conditions and significantly higher than in the arm swinging and dash conditions. In addition, the mean blink rate was significantly higher in the teleportation condition than in the dash condition. Previous work has shown that the blink rate is negatively correlated with cognitive load [299, 298]. Therefore, the blink rate results in this study may suggest that participants had lower cognitive load when navigating the VE using the joystick and teleportation locomotion methods than when using the arm swinging and dash locomotion methods. The finding that the joystick elicited less cognitive load in participants during VR locomotion is consistent with the previous study showing that the joystick is less fatiguing than the arm-cycling and point-tugging, providing a greater sense of control, and is more enjoyable than teleportation [61].

In addition to blinks, our results showed that the mean fixation duration in the joystick and teleportation conditions was significantly shorter than in the arm swinging condition. Previous work has shown that fixation duration is indicative of users' cognitive processing load. It was found that higher cognitive load was associated with longer fixation duration [299, 298]. This may indicate that participants had lower cognitive load and shorter processing time for visual information when using the joystick and teleportation compared with other locomotion methods. In addition, fixation rate was also found to correlate with cognitive load [298]: the higher the load, the lower the fixation rate. This further supports the results on fixation duration, which state that joystick and teleportation elicit lower cognitive load than arm swinging. This is not surprising, as fixation duration was negatively correlated with the number of fixations. The fixation results are consistent with the blink rate results discussed above.

Furthermore, we found that the mean saccade amplitude in the teleportation condition was significantly larger than in the arm swinging and grappling conditions, and the mean saccade amplitude in the joystick condition was significantly larger than in the arm swinging condition. Similar to blink and fixation rate, saccade amplitude was also found to be negatively correlated with cognitive load in previous work [309]. Therefore, these results again suggest that participants had a lower cognitive load when using the joystick and teleportation methods than when using the arm swinging and grappling.

Overall, our results on eye movements (i.e., blinks, fixations, and saccades) are interrelated, suggesting that joystick and teleportation cause less cognitive processing load in participants than other locomotion methods. Although no previous locomotion studies have evaluated eye movements as a measure of cognitive load, our findings were validated by a number of previous eye movement studies suggesting that eye movements are a valuable tool for assessing cognitive load. Furthermore, our findings that joystick and teleportation elicit less cognitive load are also consistent with previous work showing that joystick is an enjoyable locomotion method and teleportation elicits less motion sickness [308, 305, 61]. Typically, participants are assumed to have a lower cognitive load when they have a more pleasant VR experience and less motion sickness. Taken together, our findings provide strong evidence for the effectiveness of eye movements as a proxy for assessing cognitive load in interaction tasks with VR systems (e.g., VR navigation); moreover, eye movements as objective measures can compensate for or even substitute questionnaires. This argues for the use of eye-tracking methods in future studies, as cognitive processing load during VR locomotion has rarely been investigated, providing important avenues for the evaluation of VR locomotion methods.

### B.1.8   Conclusion

In this study, we evaluated five common gait-free locomotion methods using eye movements to assess participants' cognitive processing load during navigation in VEs. Our results showed that the locomotion methods significantly affected participants' cognitive state, with joystick and teleportation locomotion methods found to elicit less cognitive load in participants. However, no significant differences in eye movements were found between other three locomotion methods, i.e., arm swinging, dash, and grappling.

Our results demonstrate the effectiveness and the feasibility of using eye movements as a proxy to study human cognitive behavior during VR locomotion, which can be used as compensation or potential replacement for questionnaires. Our study offers profound implications for future studies in this research domain that use eye movements as a superior objective measure of users' cognitive load, and thus can be further used as a tool to assess VR locomotion techniques.

## B.2 Eye-Tracking-Based Prediction of User Experience in VR Locomotion Using Machine Learning

### B.2.1 Abstract

VR locomotion is one of the most important design features of VR applications and is widely studied. When evaluating locomotion techniques, user experience is usually the first consideration, as it provides direct insights into the usability of the locomotion technique and users' thoughts about it. In the literature, user experience is typically measured with post-hoc questionnaires or surveys, while users' behavioral (i.e., eye-tracking) data during locomotion, which can reveal deeper subconscious thoughts of users, has rarely been considered and thus remains to be explored. To this end, we investigate the feasibility of classifying users experiencing VR locomotion into **L-UE** and **H-UE** (i.e., low- and high-user-experience groups) based on eye-tracking data alone. To collect data, a user study was conducted in which participants navigated a virtual environment using five locomotion techniques and their eye-tracking data was recorded. A standard questionnaire assessing the usability and participants' perception of the locomotion technique was used to establish the ground truth of the user experience. We trained our machine learning models on the eye-tracking features extracted from the time-series data using a sliding window approach. The best random forest model achieved an average accuracy of over 0.7 in 50 runs. Moreover, the SHapley Additive exPlanations (SHAP) approach uncovered the underlying relationships between eye-tracking features and user experience, and these findings were further supported by the statistical results. Our research provides a viable tool for assessing user experience with VR locomotion, which can further drive the improvement of locomotion techniques. Moreover, our research benefits not only VR locomotion, but also VR systems whose design needs to be improved to provide a good user experience.

### B.2.2 Introduction

In recent years, with the proliferation of consumer-grade head-mounted displays (HMDs), virtual reality (VR) has become increasingly integrated into entertainment and education, and thus into people's everyday lives [3, 2]. VR applications, especially those for entertainment, should not only provide users with an immersive user experience but also enable the exploration of large and unlimited virtual environments in a limited physical space, such as the popular VR theme and amusement parks on Oculus Rift and Steam. This is made possible by VR locomotion techniques. VR locomotion is an essential technique that enables users to move effectively in virtual environments. It is one of the pillars of a great VR experience and has been studied extensively by designers and researchers recently due to the increasing popularity of VR-based entertainment [50]. To date, a wide variety of VR locomotion techniques have been developed for different purposes and scenarios in the VR domain, and can be categorized into controller-based (e.g., joystick [305], teleportation [55], dash [306]) and motion-based (e.g., walking-in-place (WIP) [310], arm swing [311]) locomotion. A locomotion technique that performs the intended locomotion function should also provide a good experience for users, i.e., it should be easy to use and not impose much additional cognitive load on users performing primary tasks in VR. Therefore, when evaluating a VR locomotion technique, it is important to evaluate the user experience.

Previous literature has primarily focused on the development of new locomotion techniques for various use cases in VR applications [312, 313], but in recent years more emphasis has been placed on the evaluation of locomotion techniques [61, 308]. Various aspects of locomotion techniques have been evaluated, such as locomotion effectiveness and self-reported post-hoc user experience, presence, motion sickness, etc [61, 305, 60]. In terms of the locomotion technique itself, user experience is most valued in this research area as it provides direct feedback on the usability of the locomotion technique and how users feel about using it. This provides the designer and researcher with concrete advice on how to improve the locomotion technique. To date, most studies on VR locomotion (or other VR systems design) have assessed user experience using methods such as observational data during VR locomotion (or during interaction with VR systems), post-hoc surveys, and questionnaires [310]. However, such methods can be time-consuming, especially for a large number of trials, and furthermore may lack deeper insights into users' subconscious thoughts and behaviors during locomotion.

With this in mind, we are thinking about evaluating the user experience with VR locomotion in alternative ways, to which eye tracking can contribute. With the increasing popularity of eye-tracking studies in various research areas such as education [219], entertainment [135], daily activities [314, 76], and of course with the development of HMDs that allow easy acquisition of eye-tracking data via integrated eye trackers in the HMDs, eye tracking holds great potential to facilitate VR studies. Eye tracking, widely used as a reliable tool to study human behaviors in real-time (e.g., cognitive processing load and visual attention behavior) [159, 91], has already demonstrated its informality in classification tasks, using machine learning to predict various human- and task-related goals such as personality traits [131], learning gains [136], cognitive load [307], and task performance [134]. Although studies on the use of eye-tracking in VR scenarios are limited, there is still literature that uses eye movements to study user behavior during immersion in VR and provide insights to improve VR systems [315, 219]. However, eye tracking is rarely used for evaluation purposes in VR locomotion studies. Overall, we consider that eye tracking has the potential to be a viable tool for evaluating user experience in VR locomotion research.

Therefore, in this study, we aim to propose a new approach to detecting user experience with VR locomotion. In more detail, the user experience evaluated in this study refers to the usability of the VR locomotion technique and the user's subjective perception and thoughts about the locomotion technique they are using. Specifically, we explore the possibility of classifying participants experiencing VR locomotion techniques into **L-EU** (i.e., low user experience) and **H-UE** (i.e., high user experience) using machine learning methods based on eye-tracking data alone. To this end, we designed a user study involving a navigation task with different locomotion techniques in a virtual environment for data collection. We collected participants' post-hoc user experiences of the usability of VR locomotion techniques and their feelings and enjoyment of using VR locomotion techniques as ground truth with a questionnaire. We applied a sliding window approach [130, 136] to extract features from time-series data, i.e., eye-tracking data; the extracted features include pupil diameter, fixations, and saccades. We developed classification models using the random forest (RF) algorithm based on the extracted eye-tracking features. The trained RF model achieved an average accuracy of 0.71 in our binary classification task, indicating that the user experience level is predictable from eye-tracking data alone. In addition, we applied the SHapley Additive exPlanations (SHAP) approach to further investigate how eye-tracking features contribute to the model outputs. Our SHAP results show that there are underlying explainable relationships between eye-tracking features and user experience (i.e., **L-UE** and **H-UE**) and that these relationships can be further supported by the statistical results of the eye-tracking metrics.

To our knowledge, this is the first study to predict user experiences with VR locomotion using eye-tracking data, which can reveal human underlying cognitive and visual perceptual behaviors that might be difficult to capture with questionnaires. Our study provides a potential avenue for future studies to detect user experience using eye-tracking data not only in VR locomotion but also in other VR contexts where user experience needs to be evaluated to improve VR systems. Moreover, our results provide deep insights into real-time user experience prediction that may be relevant and important for interactive VR systems or intelligent user interfaces for educational and entertainment purposes. By obtaining real-time feedback from users regarding their experiences with the systems they interact with, it is possible for such systems to provide users with tailored and optimal experiences by adjusting system settings accordingly.

### B.2.3 Related Work

**VR Locomotion**

Previous work has evaluated and compared locomotion techniques from various aspects of the user experience, such as the locomotion usability, users' subjective thoughts about the locomotion technique, preference, etc. For instance, Frommel et al. [308] evaluated the impact of controller-based locomotion methods on user experience during a VR exploration task. Participants navigated a virtual zoo using four locomotion methods, namely free teleport, fixpoint teleport, touchpad-based and guided automatic locomotion. User experience was measured using a questionnaire that recorded discomfort and enjoyment. Similarly, Coomer et al. [61] compared four commonly used locomotion methods, including joystick, point-tugging, teleportation, and arm-cycling. Post-hoc questionnaires were used to determine the usability of each locomotion method and the participants' opinions about it. Compared to [308], user experience was assessed in more aspects and detail. Questions were asked about the difficulty of understanding and operating the locomotion method, the feeling of being in control, fatigue, and the feeling of enjoyment. Paris et al. [60] compared two joystick-like (i.e., skiing and magic carpet) and two teleportation-like (i.e., grappling and teleportation) locomotion methods in a navigation task. Results of task performance and post-hoc questionnaires on simulator sickness, presence, and system usability were reported. However, no significant difference was found in system usability, which measures the ease of use of locomotion methods.

In addition to quantitative questionnaires, qualitative surveys were also conducted to measure user experience. Funk et al. [55] presented and evaluated three point & teleport locomotion methods that differed in the way how their teleportation trajectories were rendered in the virtual environment. Participants reported their qualitative thoughts on these locomotion methods after VR locomotion. The post-hoc survey allows researchers to directly learn how users feel about the locomotion method they used, which gives researchers further insight into how to improve locomotion methods, however, this could also be time-consuming.

In the aforementioned literature on VR locomotion research, user experience has been assessed post-hoc with either quantitative questionnaires or qualitative surveys. However, users' real-time behavioral data during locomotion, which can provide direct insights into their experiences, has rarely been considered and thus remains to be explored. As such, our study aims to propose a novel way to gain a deeper understanding of the user experience using time-series data (i.e., eye-tracking). If the effectiveness of eye tracking in predicting user experiences with VR locomotion can be demonstrated, this will provide deep insights into predicting user experience in other VR applications (e.g., VR for training and education) and thus provide clues for improving VR systems.

### Eye Tracking

Eye tracking has long been used as a tool to study and improve user experience [92], such as smartphone app development [316], marketing [314], etc. Eye-tracking data contains rich information about how users process visual scenes and what cognitive processing load is simultaneously triggered during information processing, and such information reveals the user's deep subconscious behavior and thoughts about the system they are interacting with. For this motivation, many researchers are considering incorporating eye-tracking technology into machine learning studies, i.e., training machine learning models based on eye-tracking features to predict various human- or task-related goals. In fact, eye tracking has already made its way into the machine learning field and has proven its effectiveness as an informative feature.

For instance, in the work of Conati et al. [147], eye movements were identified as informative features in machine learning models for predicting binary labels of different cognitive abilities during a visualization task [317]. Eye-tracking features related to pupil, fixation, saccade, and area-of-interest (AOI) were calculated using various descriptive statistics. Random forest classifiers trained on eye-tracking features alone achieved accuracies over 0.63 for various prediction targets (i.e., cognitive abilities). With respect to cognitive behavior, Appel et al. [307] trained classification models to predict cognitive load in an emergency simulation game using eye-tracking features of pupils, fixations, blinks, and microsaccades. The trained models achieved accuracies of 0.63 and 0.69 across participants and tasks in the binary classification task. In addition, eye movements also proved informative in predicting personality traits [130, 131] and expertise [135]. Apart from these previous works, eye-tracking data can also be used in conjunction with other task performance data or questionnaire data to improve model accuracy [317, 82]. Kasneci et al. [134, 318], for example, used eye-tracking and socio-demographic data to predict participants' performance in solving an IQ task. Gradient boosting decision trees (GBDT) models developed on eye movements alone were found to be discriminative with a ROC-AUC of 0.63 and could be improved to 0.65 with socio-demographic features. These previous works have demonstrated the feasibility and effectiveness of eye movements in revealing various underlying human behaviors, and have shown that the underlying features of eye-tracking data can be efficiently learned by machine learning models.

However, eye-tracking research in VR is still limited currently due to hardware limitations and the lack of fine-grained data analysis tools but is increasing with the advent of more and more HMDs with integrated eye trackers. Although no research has yet used eye tracking in VR locomotion for evaluation purposes, let alone to evaluate the user experience with VR locomotion, there are few studies that have used eye tracking to investigate user behavior in VR scenarios under different environmental conditions. For instance, Gao et al. [219] investigated students' cognitive and visual attention behaviors during a virtual lesson in an immersive VR classroom. Several VR environment design factors were evaluated to improve the design of the VR learning system. An approach for detecting eye movement events, i.e., fixations and saccades, suited for VR was proposed. The results showed that students' eye movements were significantly affected by the environmental factors of the VR classroom, and the underlying meaning of such effects can be interpreted to provide further guidance for improving the VR system. This study provides compelling evidence that eye movements shed light on students' perceptions of different VR environmental configurations. Although VR classroom learning is different from VR locomotion, this study reinforces our belief that eye movements provide insight into how users perceive different VR locomotion techniques and can further contribute to classification models for predicting the user experience.

### B.2.4 VR Locomotion User Study

In this section, we provide an overview of our VR locomotion user study designed for data collection. Since our research goal is to assess the feasibility of classifying participants who experienced VR locomotion into **L-UE** and **H-UE** based on eye-tracking data alone, we used five different VR locomotion techniques in our study to obtain as diverse user experiences as possible. Considering that our study is an initial exploration of our research question, we used well-studied and widely used controller-based locomotion techniques as the experimental setup rather than using a niche or introducing new locomotion techniques. The locomotion techniques used in this study are arm swing, dash, grapple, joystick, and teleportation, which reportedly provide different experiences to users [61, 60, 306]. Our user study involved partici-pants navigating a designed virtual environment using different locomotion techniques. They then provided feedback in a questionnaire about the usability of the locomotion techniques they were currently using, as well as their personal thoughts, such as how they felt and enjoyed the locomotion techniques. Details on the participants, materials, experimental procedure, and data acquisition are given below.

### Participants

Fifteen university students (10 male, 5 female) with an average age of 24.93 ($SD$ = 2.84) participated in our experiment as volunteers. All participants have normal or corrected-to-normal (with glasses) vision. They all reported their experience with video games and VR, with five reporting playing video games for more than 5 hours per week, four playing video games for 0 to 5 hours per week, and six reporting no experience with video games. In addition, one participant regularly used VR HMDs, six had some VR experience, and eight had no VR experience. All participants provided informed consent prior to data collection. Our study was approved by the institutional review board (IRB).

### Materials

In order to investigate participants' experiences with different locomotion techniques without the virtual environment exerting additional effects on participants, we designed a very simple virtual environment that simultaneously meets the navigation requirement. The virtual environment consists of a green lawn area bordered by houses and trees. To encourage locomotion, we developed a search and collect task that is typically used in previous studies investigating VR locomotion [61, 319]. Five crystals were placed at different locations in the virtual environment and were easy to find. The virtual environment was created and rendered using the Unity engine (version 2020.03.23f) on a computer with a 3.5GHz Core i7 processor and 16GB RAM. The HTC Vive Pro Eye HMD was used to display the virtual environment, which has a resolution of 1440 × 1600 per eye, a refresh rate of 90 Hz, and a field of view of 110°. In addition, the HMD is seamlessly integrated with the Tobii eye tracker with a sampling rate of 120 Hz and an accuracy of $0.5° - 1.1°$. The Vive controllers served as the input device for VR locomotion. Two HTC Vive base stations were used to track a $2m × 2m$ area for the locomotion study. All five selected locomotion techniques are controller-based, meaning participants navigate the virtual environment using Vive controllers that come with the HMD. The locomotion techniques were implemented with SteamVR based on previous work [61, 60, 306] and the scripts were written in C#.

### Experimental Procedure

Our study used a within-subjects design with five levels of the independent variable of locomotion conditions, namely arm swing, dash, grapple, joystick, and teleportation.

After providing informed consent, all participants completed a demographic questionnaire (e.g., age, gender) and reported their previous experiences with video games and VR. The experimenter then gave instructions about the experiment. All participants took part in a total of five trials (locomotion conditions), with the order counterbalanced using a Latin square to offset order effects. Each trial consisted of three sessions, i.e., the practice session, the actual experiment (recording of eye-tracking data), and the post-hoc user experience questionnaire. In the practice session, participants were asked to practice the current locomotion technique in a clean virtual environment and no task was given. Once they became familiar with the locomotion technique, they were asked to take off the VR headset and take a break until they were ready for the experiment. The pause is compulsory to avoid the practice session exerting additional influence on the actual data collection session. In the actual experimental session, participants entered the task virtual environment after a successful 5-point eye tracker calibration routine. They were asked to search for five crystals placed in different directions of the virtual environment and collect them with controllers. After completing the task, participants were asked to take off the headset and fill out questionnaires about their locomotion experience. This marked the end of one trial. Before starting the next trial, participants were asked to take a break until they felt comfortable for the next trial.

The experiment ended after the participant completed all five trials. The entire experiment lasted approximately 50 minutes for each participant, with each participant wearing the HMD for approximately 20 minutes (including the practice session). All participants were informed before the experiment that they could terminate the experiment at any time if they felt uncomfortable.

### Data Acquisition

All fifteen participants had valid and complete eye-tracking and questionnaire data, so all their data were used in our study. An average of 2.5 minutes of time-series data (i.e., eye-tracking) was collected for each trial. Thus, in total, more than three hours of eye-tracking data were collected, including raw pupil and gaze vector data, which can be further used in our machine learning experiment.

In addition to eye-tracking data, participants reported their experiences after each VR locomotion trial. We used a modified version of Loewenthal's [320] core elements of the game experience questionnaire. In accordance with our study, six subcategories were selected and measured in our study, including difficulty in understanding and operating the locomotion technique, effort required, feeling of control over the locomotion technique, tiredness, and enjoyment. Responses were on a 5-point Likert scale, with 1 representing "not at all" and 5 representing "very much". For convenience in data analysis, we reversed the scores of some items and then calculated the mean of all six items as the final user experience score, where 1 represents the worst user experience and 5 represents the best. In this case, the best user experience means that participants report that the locomotion technique is easy to use with little effort and no fatigue and that they have a strong sense of control when using it and feel

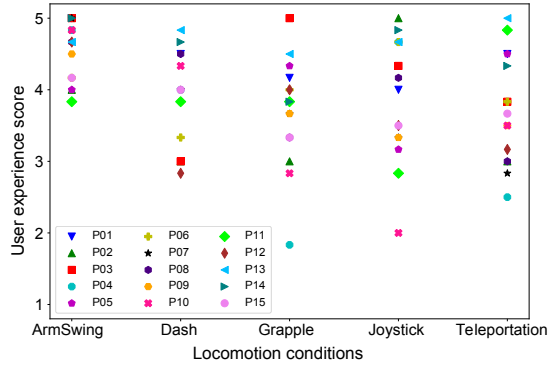

Figure B.6: Distribution of user experience scores for all participants in all locomotion conditions.

very comfortable with it.

The mean user experience score of all trials was 3.85 ($SD = 0.74$, $Median = 4.0$, ranging from 1 to 5). This indicates that participants had a good experience (above the middle level, i.e., 2.5) with different locomotion techniques. As can be seen in Figure B.6, participants showed slightly larger variance in their user experience in some conditions, e.g., grapple and joystick (scores range from 2 to 5). We consider this to be a normal bias between participants, as different individuals perceive the locomotion technique differently. In addition, participants reported different user experiences across the five locomotion conditions, which is what we expected. And this supports our strategy of training our classification models in a trial-dependent manner (see Section B.2.5 for details).

### B.2.5   Machine Learning Method

In this section, we present our machine learning method to investigate our research question: *Is it possible to predict participants' user experience with VR locomotion by building machine learning models based solely on eye-tracking data?*

In the following, we first describe how we fit the time-series and questionnaire data into machine learning models. This includes the preprocessing of both types of data and the extraction of eye-tracking features. Next, we present details on building classification models, including partitioning training and test datasets in a trial-dependent manner to minimize overfitting, model building and training, and model evaluation and explanation.

#### Data Preprocessing

We obtained the ground truths of user experiences from the post-hoc questionnaires. To label the data samples ($75 trials$) as low level of user experience (i.e., **L-UE**) and high level of user experience (i.e, **H-UE**), we binned the user experience scores in a data-driven manner considering the median value and equal frequency. Therefore, we labeled all user experience scores below the median, i.e., $< 4.0$, as **L-UE** and all other scores as **H-UE**. This resulted in 33 and 42 data samples in **L-UE** and **H-UE** groups, respectively.

Table B.2: Thresholds for detection of eye movement events.

| Threshold | Fixation | Saccade |
|-----------|----------|---------|
| *head_velocity* | $< 12°/s$ | / |
| *gaze_velocity* | $< 40°/s$ | $> 80°/s$ |
| *event_duration* | $100 < dur < 500 (ms)$ | $30 < dur < 80 (ms)$ |

In the literature, pupil size (e.g., pupil diameter) and eye movements (e.g., fixation, saccade) are commonly used to examine subjects' cognitive processing load and visual perceptual behavior [91]. We also used these measures in our study. Since only raw sensor data, including raw pupil data and gaze vectors recorded by the eye tracker and head orientations recorded by the HMD tracking system, are available, we should preprocess the data to obtain the above eye-tracking measures [100]. Pupillometric data can be extremely noisy due to blinks and noisy sensor readings. Therefore, smoothing and normalization are usually performed before calculating pupil diameter. The Savitzky-Golay filter [154] was applied to smooth the raw pupil diameters; the divisive baseline correction method [155] with a baseline duration of $\approx 1.5$ seconds was applied for normalization. Detection of eye movement events in VR remains challenging due to head movements and 3D stimulus, and there is no standard method or software to solve this problem. Fixations and saccades had to be detected manually post-experimentally. In our study, a modified velocity-threshold identification (I-VT) algorithm proposed by Gao et al. [219] that takes head movement into account was used to detect fixations, with parameters adapted to our study. Since saccades are not affected by head movements, the normal I-VT algorithm was used for saccade detection. Before detecting such eye movement events, linear interpolation was performed for the missing gaze vectors. Specifically, fixations were detected with a maximum gaze velocity threshold of $40°/s$ under the condition that head moving velocity was less than $12°/s$; saccades were detected with a minimum gaze velocity threshold of $80°/s$. Additional duration thresholds were used for filtering. All parameters used for eye movement event detection were listed in Table B.2.

### Feature Extraction

Considering that each trial averaged 2.5 minutes of eye-tracking data, we applied the sliding window approach to extract eye-tracking features rather than averaging over the entire trial to avoid eliminating too much temporal information from the data. Since there is no gold standard for determining window size, our study examined time windows ranging from 5s to 30s (with a step of 5s) based on previous literature [130, 136, 65]. The window size was considered as a hyperparameter during model training (see Section B.2.5), and the best window size was determined based on the training results (see Section B.2.6).

Table B.3: Extracted eye-tracking features for machine learning models.

| Features | Descriptive statistics |
|---|---|
| Pupil diameter | Mean, Std.dev, Min, Max of the normalized pupil diameter |
| Pupil diameter (fixation) | Mean, Std.dev, Min, Max of the normalized pupil diameter during fixation |
| Fixation rate | Number of fixations per minute |
| Fixation duration | Mean, Std.dev, Min, Max, Sum of the fixation durations |
| Saccade rate | Number of saccades per minute |
| Saccade duration | Mean, Std.dev, Min, Max, Sum of the saccade durations |
| Saccade amplitude | Mean, Std.dev, Min, Max, Sum of the saccade amplitudes |
| Saccade velocity | Mean, Std.dev, Min, Max of the saccade velocities |
| Saccade peak velocity | Mean, Std.dev, Min, Max of the saccade peak velocities |

note: Std.dev, Min, and Max denote the standard deviation, minimum, and maximum, respectively.

For each of the sliding time windows, we calculated and extracted eye-tracking metrics that have been commonly used as features in previous machine learning studies, i.e., metrics related to pupil diameter, fixations (number and duration), and saccades (number, duration, amplitude, and velocity) [147, 134, 318]. For some of the features, we calculated not only the mean, but also the standard deviation, the minimum, the maximum, and the sum of the values to characterize variables throughout the data [130, 148]. Specifically, the feature of pupil diameter was extracted since pupil diameter has been considered an indicator of cognitive processing load during various tasks in previous literature [227]. Fixations have been shown to reveal visual attention behavior and are also an indicator of cognitive processing load [202]. We calculated fixation rate and fixation duration and used them as features. In addition, pupil diameter during fixation is an indicator of cognitive processing load (pupil diameter) during visual information processing (fixation) [321], so we extracted it as a feature. Saccades are informative eye movements that correlate highly with visual search behavior and also with cognitive processing load [197, 309]. Several features can be extracted from saccades, including saccade rate, saccade duration, saccade amplitude, and saccade velocity. We extracted these features and used them for model training. All extracted 33 eye-tracking features were listed in Table B.3.

## Model Building

In this work, we developed random forest (RF) models to classify participants into low and high user experience groups. We used the integers 0 and 1 to represent two prediction targets, i.e., class-0 for **H-UE** and class-1 for **L-UE**. For each sliding window, a $1 \times 33$ feature vector was generated. Thus, with a window size of $ws$ seconds, there are approximate $N = 75 \times 150 / ws$ samples for machine learning training, where 75 is the number of trials and 150 seconds (i.e., 2.5 minutes) is the averaged duration of each trial. Min-max normalization was performed for all feature variables. Then, data samples were randomly split into the training set (80%, about $N \times 0.8$ samples) and the test set (20%, about $N \times 0.2$ samples). For example, with a window size of 10s, we have about 1100 samples, of which about 900 samples are used as the training set and about 200 samples are used as the test set. The classification models were trained on the training set and tested on the test set. For model training, we performed 5-fold cross-validation to tune the hyperparameters of the random forest models on the training set, which means we further split the training set into the sub-training set and the validation set.

To avoid overfitting and to generalize our models to unseen data, all data splits in the training and testing processes were trial-dependent, that is, all feature vectors from the same trial were to remain in the same data sets (i.e., either sub-training, validation, or test set). Here, we split the data in a trial-dependent manner rather than a participant-dependent manner, which may risk overfitting the model because the models might be tested on the seen data (if data samples from a participant exist in both the training and test sets). Actually, such concern can be eliminated as participants had different user experiences across different conditions (See Figure B.6) and these differences are reflected in the eye-tracking behavior as well (eye-tracking feature), which means that we can consider each participant's trial as an independent data sample. Thus, we assumed that the models were still tested on the unseen data. Moreover, we performed stratified data split, that is, data samples were split into different sets while maintaining the percentage of samples for each class. These data split policies were performed manually. Notice that participants were randomly assigned during all data splits without regard to identity. Furthermore, to reduce the bias caused by data splitting, we split the data 50 times with different random seeds so that all trials had the opportunity to be used as training and test sets. Thus, in a 5-fold cross-validation, our model was trained and validated on $50 \times 5$ different sub-training and validation sets. The best model identified through the training process was then tested 50 times on unseen data, i.e., on the 50 test sets. The final model test results reported in our study are the average of the 50 runs.

**Model Evaluation and Explainability**

In this work, the random forest models solved a binary classification problem. Considering that our dataset is not fully balanced (33 **L-UE**: 42 **H-UE**), metrics including accuracy, precision, recall, and F1-score were used to evaluate model performance. These four metrics were calculated based on True Positive (TP), False Positive (FP), True Negative (TN), and False Negative (FN). We report only the model test results using these metrics.

Moreover, post-hoc model explainability becomes significant in the field of machine learning as it provides insights into how a model can be improved. As in our study, a large set of eye-tracking features indicative of different human behaviors were extracted and used. In addition to training the models to achieve the best performance in predicting user experiences, we also emphasize explaining the model at the feature level, i.e., how individual features contribute to the model outputs. To this end, we applied the state-of-the-art method for model explainability, namely SHapley Additive exPlanations (SHAP) [149]. First, SHAP can help gain insight into the feature importance and identify the most informative eye-tracking features for the models. Second, SHAP can reveal the underlying relationships between eye-tracking features and model outputs (i.e, **L-UE**, **H-UE**) by demonstrating what impact an individual feature has on model outputs.

Table B.4: The test performance of RF models trained on features extracted with different window sizes. Best window size and model performance are in bold.

| $WindowSize[s]$ | Accuracy | Precision | Recall | F1-score |
|:---:|:---:|:---:|:---:|:---:|
| 5 | 0.62 | 0.64 | 0.62 | 0.63 |
| 10 | 0.68 | 0.70 | 0.68 | 0.69 |
| 15 | 0.67 | 0.69 | 0.67 | 0.67 |
| **20** | **0.71** | **0.72** | **0.71** | **0.71** |
| 25 | 0.67 | 0.69 | 0.67 | 0.67 |
| 30 | 0.68 | 0.70 | 0.68 | 0.68 |
| 35 | 0.63 | 0.65 | 0.63 | 0.63 |
| 40 | 0.64 | 0.65 | 0.64 | 0.64 |

### B.2.6 Results

In this section, we report our results in three parts. First, we report the performance of our RF models on the binary classification task of predicting user experience level. Second, we report the results of SHAP explainability. We identified the most informative eye-tracking features and explained the classification model at the feature level to reveal how these informative eye-tracking features contribute to the model outputs. Third, to further support the SHAP results, we also report the statistical results of the main eye-tracking metrics.

**Model Performance**

As can be seen in Table B.4, the RF models trained only on eye-tracking features extracted with different window sizes were able to classify participants into **L-UE** and **H-UE**, with average accuracies above 0.62. Of all eight window sizes examined, the RF model trained on features extracted with a 20-second time window performed best, with an average accuracy of 0.71 ($SD = 0.019$), precision of 0.72 ($SD = 0.020$), recall of 0.71 ($SD = 0.019$), and F1-score of 0.71 ($SD = 0.018$), as shown in Figure B.7. The best RF model was created with 300 decision trees. The maximum depth of the tree used was 10, the minimum number of samples for splitting an internal node was 4, and the minimum number of samples at a leaf node was 1.

**SHAP Explanation**

We applied the SHAP TreeExplainer [264] for tree-based classifiers, i.e., random forest. The local explanations of the best RF model are shown in a beeswarm-style summary plot of SHAP values, as shown in Figure B.8. The 20 most informative features are displayed and sorted by feature importance from top to bottom along the $y$-axis. SHAP values are located along the $x$-axis. Each dot in the summary plot represents the SHAP value of one feature observation. The more spread out the dots of a feature are on the $x$-axis, the greater the impact of the feature on the model. The color of the dots in the summary plot indicates the value of the feature, with blue color indicating a low feature value and red color indicating a high feature value. A color change from blue to red along the $x$-axis from left to right indicates that a feature has a positive (negative) impact on the prediction of class-1 (class 0), i.e., **L-UE** (**H-UE**), in contrast, a color change from red to blue along the $x$-axis from left to right indicates that a feature has a negative (positive) impact on the prediction of class-1 (class-0), i.e., **L-UE** (**H-UE**).

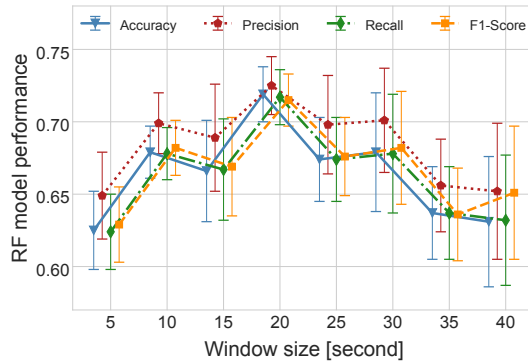

Figure B.7: The test performance of RF models trained on features extracted with different time windows. The best performance was obtained with a 20-second time window.

As shown in Figure B.8, the feature *meanPupilDiameterOfFixation* followed by the features *maxPupilDiameterOfFixation* and *meanPupilDiameter* are the three most informative features for the RF model in classifying the user experience into low and high level. Moreover, it is worth noting that these three most important features have a positive impact on the prediction output into class-1 (i.e., **L-UE**), which means that the feature value higher than the feature average drives the classification into the prediction output of **L-UE**. The feature *maxSaccadeAmplitude* also contributes greatly to the classification model. Unlike the above three most informative features, *maxSaccadeAmplitude* was observed to have a positive impact on the prediction output into class-0 (i.e., **H-UE**), which means that the feature value higher than the feature average drives the classification into prediction output of **H-UE**. Notably, four of the five most informative features are pupil-related. However, compared to pupil- and saccade-related features, fixation-related features tend to be less informative in classifying user experience levels.

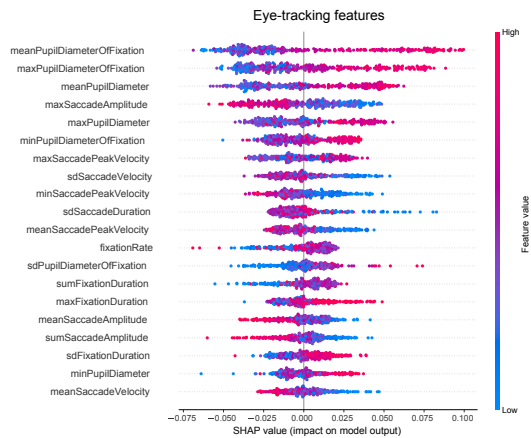

Figure B.8: The top 20 eye-tracking features in the best RF model, ranked by feature importance from top to bottom.

**Statistical Test**

To further validate the SHAP explanation results, we applied statistical tests to the eye-tracking metrics typically analyzed in eye-tracking studies, i.e., these features can be interpreted as associated with various human behaviors (e.g., cognitive load, visual attention, and visual search). For this reason, we did not apply statistical tests to these eye-tracking metrics calculated as minimum, maximum, standard deviation, and sum values as the features of such metrics can be easily learned by machine learning models, but their statistical differences are hard to interpret. The Shapiro-Wilk test was used to test for normality and the Mann-Whitney U test was used as a non-parametric test. The significance level was set at $\alpha = 0.05$ for all the tests. The statistical results for the comparison between **L-UE** and **H-UE** groups are shown in Table B.5.

As can be seen, there was no significant difference between the two groups for fixation-related features that were found to be of low informativeness in the classification model. Conversely, significant differences were found between the two groups for pupil diameters, note: *, **, ****, and n.s. represent $p < .05$, $p < .01$, $p < .0001$, and no statistical significance, respectively.

Table B.5: Statistical comparison of eye-tracking metrics between L-UE and H-UE groups.

| Eye-tracking metrics | L-UE group Mean (SD) | H-UE group Mean (SD) | $p$−value | Sig. |
|---|---|---|---|---|
| Mean pupil diameter | 1.09 (0.11) | 1.05 (0.12) | $1.79 \times 10^{-14}$ | **** |
| Mean pupil diameter (fixation) | 1.08 (0.16) | 1.01 (0.19) | $2.03 \times 10^{-17}$ | **** |
| Fixation rate | 55.01 (27.83) | 52.91 (29.57) | .096 | n.s. |
| Mean of fixation duration | 216.12 (53.96) | 210.53 (62.11) | .079 | n.s. |
| Saccade rate | 46.11 (25.14) | 49.04 (23.49) | .014 | * |
| Mean of saccade duration | 47.48 (8.82) | 48.93 (6.41) | .008 | ** |
| Mean of saccade amplitude | 8.97 (3.05) | 10.13 (2.85) | $2.95 \times 10^{-10}$ | **** |
| Mean of saccade velocity | 187.21 (63.59) | 206.68 (53.62) | $9.98 \times 10^{-9}$ | **** |
| Mean of saccade peak velocity | 306.49 (96.26) | 327.77(79.30) | $4.65 \times 10^{-6}$ | **** |

which are highly informative for the classification model. In particular, the mean pupil diameters in the **L-UE** group are significantly larger than the values in the **H-UE** group, with $p < .0001$. Similar to the mean pupil diameter, a significant difference was also found between the groups for the most informative pupil feature, the mean pupil diameter during fixation, with the mean pupil diameters during fixation in the **L-UE** group being significantly larger than in the **H-UE** group, with $p < .0001$. In addition to pupil diameter and fixations, significant differences were also found in saccade metrics. It was found that the saccade rates in the **H-UE** group are higher than in the **L-UE** group, with $p = .014$. Also, the mean saccade duration, amplitude, velocity, and peak velocity were found to be higher in the **H-UE** group than in the **L-UE** group, with $p < .001$, $p < .0001$, $p < .0001$, and $p < .0001$, respectively.

### B.2.7 Discussion

In this section, the results are discussed according to the structure of the results section.

First, the sliding window approach used for feature extraction proved successful, as our results show that the user experience with VR locomotion is predictable using machine learning models trained solely on eye-tracking features. In particular, the RF model trained on features extracted with a sliding window of 20 seconds performed best with an average accuracy of 0.71 (see Table B.4). On the other hand, the models performed worse when the window sizes were too short or too long. This could be because when the sliding window is too short, the features behind the eye movements that could be indicative of the users' perception of VR locomotion are truncated, whereas when the sliding window is too long, these eye movement features can be easily smoothed out by the averaging effects. Previous literature on the use of the sliding window also shows that the optimal window size is dependent on the task [130, 136]. Second, although no previous research has predicted user experiences with VR locomotion using eye tracking, the best performance we obtained is comparable to eye-tracking-based classification results for other behavioral targets in other scenarios (see Section B.2.3). Our results demonstrate the robustness of our approach to predicting user experiences with VR locomotion based solely on eye-tracking data. This fills a research gap in the field of VR locomotion.

Furthermore, we applied the SHAP approach to explore how eye-tracking features contribute to the classification models and to further uncover the underlying relationships between features and model outputs (see Figure B.8). The results show that pupil- and saccade-related features are highly informative for the RF model in predicting user experiences. Fixation-related features, on the other hand, are less informative compared to the other two types of features. In addition, we found that most pupil-related features, such as mean, maximum, and minimum pupil diameter during fixations, as well as mean and maximum pupil diameter, have a positive influence on the classification model into the prediction output of **L-UE** (class-1), which means that a higher value of these features than their respective average value drives the classification into the prediction output of **L-UE**. This may suggest that a higher value of these pupil-related features correlates with a lower user experience, as evidenced also by the statistical test results of pupil diameter (see Table B.5). Specifically, mean pupil diameter and mean pupil diameter during fixation were found to be significantly larger in the **L-UE** group than in the **H-UE** group, supporting the SHAP results showing that pupil-related features correlate negatively with user experience level. We found similar consistent results for another type of informative feature, namely saccade-related features. The SHAP results show that features such as maximum saccade amplitude, minimum and mean saccade peak velocity have a negative influence on the classification model into the prediction output of **L-UE**, which means, that a lower value of these features than their respective average value drives the classification into the prediction output of **L-UE**, in contrast, a higher value of these features than their respective average value drives the classification into the prediction output of **H-UE** (class-0). This may suggest that a higher (lower) value of these saccade features correlates with a higher (lower) user experience. It is worth noting that these SHAP results for saccade features are also consistent with the statistical results where saccade-related metrics have a higher value in the **H-UE** group than in the **L-UE** group, further suggesting that these saccade-related features correlate positively with user experience level. However, no significant difference was found in fixation rate and fixation duration in the statistical tests, but we found in the SHAP results that these fixation features affected the classification models differently. This suggests that the model can learn the deeper characteristics of eye movements that are difficult to detect with statistical tests.

Although there is no research investigating correlations between eye movements and user experiences with VR locomotion, our findings from the SHAP approach and statistical tests can still be interpreted based on previous literature. As in our study, participants' user experience was evaluated as to their feelings about using the locomotion methods, i.e., whether the locomotion method was easy to use, whether they felt tired, whether it was fun, that is, a higher user experience indicates that the participant can navigate the virtual environment easily and with little fatigue using the locomotion methods. Thus, in our case, we consider that a higher user experience implies a lower cognitive load, while a lower user experience implies a higher cognitive load. Our results show that there might be a negative relationship between pupil diameter and user experience, which means that participants with high user experience have small pupil diameter. This can be supported by previous literature, stating that pupil diameter correlates positively with cognitive load [322]. That is, participants who have a high user experience (i.e., little fatigue, high enjoyment) have low cognitive load as indicated by pupil diameter. Similarly, our results suggest that saccade amplitude and saccade peak velocity are positively correlated with user experience, implying that participants with high user experience have large saccade amplitude and saccade peak velocity. This can be supported by previous literature as well, stating that saccade amplitude and saccade peak velocity are negatively correlated with cognitive load [323, 309, 300]. Thus, our model explainability results, revealing the underlying relationships between eye-tracking features and user experience, can be supported by previous literature.

### B.2.8   Conclusion and Implication

In this work, we investigated the feasibility of predicting user experiences with VR locomotion (i.e., the usability of locomotion methods and users' feelings about them during navigation) based on eye-tracking data alone. We conducted a user study in which participants performed a navigation task in a virtual environment with five different locomotion methods. We collected participants' experience data using a standard questionnaire and binned them into low and high user experience groups as the ground truth. We extracted a variety of eye-tracking features from time-series data using a sliding window approach. We built classification models using the random forest algorithm based solely on the extracted eye-tracking features. Our best model achieved an average accuracy of over 0.7 in 50 runs, demonstrating the feasibility of predicting user experiences with VR locomotion based on eye-tracking data alone and the robustness of our research. By applying the SHAP approach, we identified the most contributed features of the classification model. In addition, the SHAP explainability results revealed some relationships between eye movements and user experience, i.e., **L-UE** and **H-UE**, which can be further supported by the statistical results.

To the best of our knowledge, our study is the first to use eye tracking as a tool to investigate user experience in the research field of VR locomotion. User experience is one of the most important aspects in evaluating or comparing locomotion methods, as it is closely related to the improvement of locomotion methods. As such, our research provides a viable user experience assessment tool for future studies, especially when new locomotion techniques are proposed, and can be extended to other VR research that aims to provide a good experience to users or requires system assessment and improvement. The ultimate goal would be a system that can detect the user experience level in real-time to offer users a tailored and optimal experience in real-world applications.